\documentclass[aos,preprint]{imsart}

\RequirePackage{amsthm,amsmath,amsfonts,amssymb}
\RequirePackage[numbers]{natbib}
\usepackage{amsgen,amsmath,amstext,amsbsy,amsopn,amssymb,latexsym}
\usepackage[usenames]{color}
\usepackage{hyperref}
\usepackage{array}
\usepackage{cite}
\usepackage{float}
\usepackage{multirow}
\usepackage{graphicx}
\usepackage{amssymb}
\usepackage{mathtools}
\usepackage{tikz}
\usepackage{subfigure} 
\usepackage{microtype}
\usepackage{enumitem}
\usepackage{natbib}
\usepackage{pgfplots}
\pgfplotsset{width=10cm,compat=1.9}
\usepackage{algorithm, algorithmicx, algpseudocode}
\arxiv{2004.03758}

\startlocaldefs

\newtheorem{Corollary}{Corollary}
\newtheorem{Proposition}{Proposition}
\newtheorem{Lemma}{Lemma}

\newtheorem{Theorem}{Theorem}

\newtheorem{Remark}{Remark}

\newcommand{\R}{\mathbb{R}}

\newcommand{\cip}{\overset{p}{\to}}
\newcommand{\cid}{\overset{d}{\to}}
\newcommand{\RE}{{\rm RE}}
\newcommand{\err}{e}
\newcommand{\Q}{\mathcal{Q}}
\newcommand{\NT}{\mathcal{P}^{(j)} }
\newcommand{\BM}{B}
\newcommand{\Sb}{\mathcal{T}}
\newcommand{\Sg}{\mathcal{T}_j}

\newcommand{\betainit}{\widehat{\beta}^{init}}
\newcommand{\T}{\mathcal{O}}

\newcommand{\PP}{\mathbb{P}} 
\renewcommand{\P}{\mathbb{P}} 
\newcommand{\E}{\mathbb{E}} 
\newcommand{\Mq}{M}
\newcommand{\Cq}{CM}
\DeclareMathOperator*{\argmin}{arg\,min} 
\DeclareMathOperator*{\argmax}{arg\,max} 

\DeclarePairedDelimiter\floor{\lfloor}{\rfloor} 
\DeclarePairedDelimiter{\norm}{\lVert}{\rVert} 

\graphicspath{ {figures/} }

\endlocaldefs

\begin{document}
\begin{frontmatter}
\title{Doubly Debiased Lasso:\\
High-Dimensional Inference under Hidden Confounding\protect\thanksref{T1}}
\runtitle{Doubly Debiased Lasso}
\thankstext{T1}{Z. Guo and D. \'{C}evid contributed equally to this work. The research of Z. Guo was supported in part by the NSF-DMS 1811857, 2015373 and NIH-1R01GM140463-01; {Z. Guo also acknowledges financial support for visiting the Institute of Mathematical Research (FIM) at ETH Zurich}. D. \'Cevid and P. B\"{u}hlmann received funding from the European Research Council (ERC) under the European Union's Horizon 2020 research and innovation programme (grant agreement No. 786461).}
\begin{aug}
\author[A]{\fnms{Zijian} \snm{Guo}\ead[label=e1]{zijguo@stat.rutgers.edu}},
\author[B]{\fnms{Domagoj} \snm{\'{C}evid}\ead[label=e2]{cevid@stat.math.ethz.ch}}
\and
\author[B]{\fnms{Peter} \snm{B\"{u}hlmann}\ead[label=e3]{buhlmann@stat.math.ethz.ch}}\\
\address[A]{Rutgers University, Piscataway, USA, \printead{e1}}
\address[B]{ETH Z\"{u}rich, Switzerland, \printead{e2,e3}}
\runauthor{Z. Guo, D. \'{C}evid, P. B\"{u}hlmann}

\end{aug}

\begin{abstract}
Inferring causal relationships or related associations from observational data can be invalidated by the existence of hidden confounding. We focus on a high-dimensional linear regression setting, where the measured covariates are affected by hidden confounding and propose the {\em Doubly Debiased Lasso} estimator for individual components of the regression coefficient vector. Our advocated method simultaneously corrects both the bias due to estimation of high-dimensional parameters as well as the bias caused by the hidden confounding. We establish its asymptotic normality and also prove that it is efficient in the Gauss-Markov sense. The validity of our methodology relies on a dense confounding assumption, i.e. that every confounding variable affects many covariates. The finite sample performance is illustrated with an extensive simulation study and a genomic application. 
\end{abstract}



\end{frontmatter}

\section{Introduction}

Observational studies are often used to infer causal relationship in fields such as genetics, medicine, economics or finance.
A major concern for confirmatory conclusions is the existence of hidden confounding \citep{guertin2016performance, manghnani2018metcc}. In this case, standard statistical methods can be severely biased, particularly for large-scale observational studies, where many measured covariates are possibly confounded. 

To better address this problem, let us consider first the following linear Structural Equation Model (SEM) with a response $Y_{i}$, high-dimensional measured covariates $X_{i, \cdot}\in \R^{p}$ and hidden confounders $H_{i, \cdot} \in \R^{q}$:
\begin{equation}
Y_{i} \leftarrow \beta^{\intercal}X_{i, \cdot}+\phi^{\intercal}H_{i, \cdot}+\err_i, \quad \text{and} \quad X_{i, \cdot}\leftarrow\Psi^{\intercal} H_{i, \cdot}+E_{i, \cdot} \quad \text{for}\; 1\leq i\leq n,
\label{eq: hidden SEM}
\end{equation}
where the random error $\err_i\in \R$ is independent of $X_{i, \cdot}\in \R^{p}$, $H_{i, \cdot}\in \R^{q}$ {and $E_{i, \cdot} \in \R^p$} and the components of $E_{i, \cdot}\in \R^{p}$ are uncorrelated with the components of $H_{i, \cdot}\in \R^{q}$. The focus on a SEM as in \eqref{eq: hidden SEM} is not necessary and we relax this restriction in model \eqref{eq: confounder model} below.
Such kind of models are used for e.g. biological studies to explore the effects of measured genetic variants on the disease risk factor, and the hidden confounders can be geographic information \citep{novembre2008genes}, data sources in mental analysis \citep{price2006principal} or general population stratification in GWAS \citep{mccarthy2008genome}.

Our aim is to perform statistical inference for individual components $\beta_j$, $1\leq j\leq p$, of the coefficient vector, where $p$ can be large, in terms of obtaining confidence intervals {or statistical tests}. This inference problem is challenging due to high dimensionality of the model and the existence of hidden confounders. As a side remark, we mention that our proposed methodology can also be used for certain measurement error models, an important general topic in statistics and economics \citep{carroll2006measurement,wooldridge2010econometric}. 

\subsection{Our Results and Contributions}
We focus on a dense confounding model, where the hidden confounders $H_{i, \cdot}$ in \eqref{eq: hidden SEM} are associated with many measured covariates $X_{i, \cdot}$.
Such dense confounding model seems reasonable in quite many practical applications, e.g. for addressing the problem of batch effects in biological studies \citep{haghverdi2018batch,johnson2007adjusting,leek2010tackling}.

We propose a two-step estimator for the regression coefficient $\beta_{j}$ for $1\leq j\leq p$ in the high-dimensional dense confounding setting, where a large number of covariates has possibly been affected by hidden confounding. In the first step,  we construct a penalized spectral deconfounding estimator $\betainit$ as in \citep{cevid2018spectral}, where the standard squared error loss is replaced by a squared error loss after applying a certain spectral transformation to the design matrix $X$ and the response $Y$. In the second step, for the regression coefficient of interest $\beta_j$, we estimate the high-dimensional nuisance parameters $\beta_{-j} = \{\beta_l;\ l \neq j\}$ by $\betainit_{-j}$ and construct an approximately unbiased estimator $\widehat{\beta}_j$. 

The main idea of the second step is to correct the bias from two sources, one from estimating the high-dimensional nuisance vector $\beta_{-j}$ by $\betainit_{-j}$ and the other arising from hidden confounding. In the standard high-dimensional regression setting with no hidden confounding, debiasing, desparsifying or Neyman's Orthogonalization were proposed for inference for $\beta_j$ \citep{zhang2014confidence,van2014asymptotically, javanmard2014confidence,belloni2014inference,chernozhukov2015valid,farrell2015robust,chernozhukov2018double}. 
However, these methods, or some of its direct extensions, do not account for the bias arising from hidden confounding. In order to address this issue, we introduce a {\em Doubly Debiased Lasso} estimator which corrects both biases simultaneously. Specifically, we construct a spectral transformation $\mathcal{P}^{(j)}\in \R^{n\times n}$, which is applied to the nuisance design matrix $X_{-j}$ when the parameter of interest is $\beta_j$. This spectral transformation is crucial to simultaneously correcting the two sources of bias.

We establish the asymptotic normality of the proposed {\em Doubly Debiased Lasso}  estimator in Theorem \ref{thm: limiting dist}.  
An efficiency result is also provided in Theorem \ref{thm: efficiency} of Section \ref{sec: normal and efficient}, showing that the {\em Doubly Debiased Lasso} estimator retains the same Gauss-Markov efficiency bound as in standard high-dimensional linear regression with no hidden confounding \citep{van2014asymptotically,jankova2018semiparametric}. Our result is in sharp contrast to Instrumental Variables (IV) based methods, see Section \ref{sec: related}, whose inflated variance is often of concern, especially with a limited amount of data \citep{wooldridge2010econometric,boef2014sample}. This remarkable efficiency result is possible by assuming denseness of confounding. Various intermediary results of independent interest are also derived in Section \ref{sec: key results} of the supplementary material. Finally, the performance of the proposed estimator is illustrated on simulated and real genomic data in Section \ref{sec: empirical}. 

To summarize, our main contribution is two-fold:
\begin{enumerate}
\item We propose a novel Doubly Debiased Lasso estimator for individual coefficients $\beta_j$ and estimation of the corresponding standard error in a high-dimensional linear SEM with hidden confounding.
\item We show that the proposed estimator is asymptotically Gaussian and efficient in the Gauss-Markov sense. This implies the construction of asymptotically optimal confidence intervals for individual coefficients $\beta_j.$
\end{enumerate}

\subsection{Related Work}
\label{sec: related}

In econometrics, hidden confounding and measurement errors are unified under the framework of endogenous variables. 
Inference for treatment effects or corresponding regression parameters in presence of hidden confounders or measurement errors has been extensively studied in the literature with Instrumental Variables (IV) regression.
The construction of IVs typically requires a lot of domain knowledge, and obtained IVs are often suspected of violating the main underlying assumptions 
\citep{han2008detecting,wooldridge2010econometric,kang2016instrumental,burgess2017review,guo2018confidence,windmeijer2019use}. In high dimensions, the construction of IVs is even more challenging, since for identification of the causal effect, one has to construct as many IVs as the number of confounded covariates, which is the so-called ``rank condition" \citep{wooldridge2010econometric}. Some recent work on the high-dimensional hidden confounding problem relying on the construction of IVs includes 
\citep{gautier2011high,fan2014endogeneity,lin2015regularization,belloni2017program,zhu2018sparse,neykov2018unified,gold2019inference}. Another approach builds on directly estimating and adjusting with respect to latent factors \citep{wang2019blessings}.

A major distinction of the current work from the contributions above is that we consider a confounding model with a denseness assumption
\citep{chandrasekaran2012latent,cevid2018spectral,shah2018rsvp}.
\citep{cevid2018spectral} consider point estimation of $\beta$ in the high-dimensional hidden confounding model \eqref{eq: hidden SEM}, whereas \citep{shah2018rsvp} deal with point estimation of the precision and covariance matrix of high-dimensional covariates, which are possibly confounded. The current paper is different in that it considers the challenging problem of confidence interval construction, which requires novel ideas for both methodology and theory.

The dense confounding model is also connected to the high-dimensional factor models  \citep{fan2008high,lam2011estimation,lam2012factor,fan2016projected,wang2017asymptotics}. 
The main difference is that the factor model literature focuses on accurately extracting the factors, while our method is essentially filtering them out in order to provide consistent estimators of regression coefficients, under much weaker requirements than for the identification of factors. 

{Another line of research \citep{gagnon2012using,sun2012multiple,wang2017confounder} studies the latent confounder adjustment models but focuses on a  different setting where many outcome variables can be possibly associated with a small number of observed covariates and several hidden confounders.}

\medskip\noindent
\noindent{\bf Notation.} We use $X_{j}\in \R^{n}$ and $X_{-j}\in \R^{n\times (p-1)}$ to denote the $j-$th column of the matrix $X$ and the sub-matrix of $X$ excluding the $j-$th column, respectively; $X_{i, \cdot}\in \R^{p}$ is used to denote the $i-$th row of the matrix $X$ (as a column vector); $X_{i,j}$ and $X_{i,-j}$ denote respectively the $(i,j)$ entry of the matrix $X$ and the sub-row of $X_{i, \cdot}$ excluding the $j$-th entry.  Let $[p]=\{1,2,\ldots,p\}$. For a subset $J\subseteq[p]$ and a vector $x\in \R^{p}$, $x_{J}$ is the sub-vector of $x$ with indices in $J$ and $x_{-J}$ is the sub-vector with indices in $J^{c}$. For a set $S$, $\left|S\right|$ denotes the cardinality of $S$. {For a vector $x\in \R^{p}$, the $\ell_q$ norm of $x$ is defined as $\|x\|_{q}=\left(\sum_{l=1}^{p}|x_l|^q\right)^{\frac{1}{q}}$ for $q \geq 0$ with $\|x\|_0=\left|\{1\leq l\leq p: x_l \neq 0\}\right|$ and $\|x\|_{\infty}=\max_{1\leq l \leq p}|x_l|$.} We use $e_i$ to denote the $i$-th standard basis vector in $\R^p$ and ${\rm I}_{p}$ to denote the identity matrix of size $p\times p$.
We use $c$ and $C$ to denote generic positive constants that may vary from place to place. {For a sub-Gaussian random variable $X$, we use $\|X\|_{\psi_2}$ to denote its sub-Gaussian norm; see definitions 5.7 and 5.22 in \citep{vershynin2010introduction}.}
For a sequence of random variables $X_n$ indexed by $n$, we use $X_n \cip X$ and $X_{n} \cid X$ to represent that $X_n$ converges to $X$ in probability and in distribution, respectively. For a sequence of random variables $X_n$ and numbers $a_n$, we define $X_{n}=o_{p}(a_n)$ if $X_n/a_n$ converges to zero in probability. 
For two positive sequences $a_n$ and $b_n$,  $a_n \lesssim b_n$ means that $\exists C > 0$ such that $a_n \leq C b_n$ for all $n$;
$a_n \asymp b_n $ if $a_n \lesssim b_n$ and $b_n \lesssim a_n$, and $a_n \ll b_n$ if $\limsup_{n\rightarrow\infty} {a_n}/{b_n}=0$.
For a matrix $M$, we use $\|M\|_{F}$, $\|M\|_2$ and $\|M\|_{\infty}$ to denote its Frobenius norm, spectral norm and element-wise maximum norm, respectively. We use $\lambda_{j}(M)$ to denote the $j$-th largest singular value of some matrix $M$, that is, $\lambda_{1}(M)\geq \lambda_{2}(M) \geq \ldots \geq \lambda_{q}(M) \geq 0$. For a symmetric matrix $A$, we use $\lambda_{\max}(A)$ and $\lambda_{\min}(A)$ to denote its maximum and minimum eigenvalues, respectively.

\section{Hidden Confounding Model}
\label{sec: HCM}
We consider the Hidden Confounding Model for i.i.d. data $\{X_{i, \cdot}, Y_i\}_{1\leq i\leq n}$ and unobserved i.i.d. confounders $\{H_{i, \cdot}\}_{1\leq i\leq n}$, given by:
\begin{equation}
\begin{aligned}
Y_{i}= \beta^{\intercal}X_{i, \cdot}+\phi^{\intercal}H_{i, \cdot}+\err_i \quad \text{and} \quad X_{i, \cdot}=\Psi^{\intercal} H_{i, \cdot}+E_{i, \cdot}, 
\end{aligned}
\label{eq: confounder model}
\end{equation}
where $Y_{i}\in \R$ and $X_{i, \cdot}\in \R^{p}$ respectively denote the response and the measured covariates and $H_{i, \cdot}\in \R^{q}$ represents the hidden confounders. 
We assume that the random error $\err_i\in \R$ is independent of $X_{i, \cdot}\in \R^{p}$, $H_{i, \cdot}\in \R^{q}$ {and $E_{i, \cdot} \in \R^p$} and the components of $E_{i, \cdot}\in \R^{p}$ are uncorrelated with the components of $H_{i, \cdot}\in \R^{q}.$

The coefficient matrices $\Psi\in \R^{q\times p}$ and $\phi \in \R^{q \times 1}$ encode the linear effect of the hidden confounders $H_{i, \cdot}$ on the measured covariates $X_{i, \cdot}$ and the response $Y_i$. 
We consider the high-dimensional setting where $p$ might be much larger than $n$. 
Throughout the paper it is assumed that the regression vector $\beta \in \R^p$ is sparse, with a small number $k$ of nonzero components, and that the number $q$ of confounding variables {is a small positive integer}. However, both $k$ and $q$ are allowed to grow with $n$ and $p$. We write $\Sigma_E$ or $\Sigma_X$ for the covariance matrices of $E_{i, \cdot}$ or $X_{i, \cdot}$, respectively.  Without loss of generality, it is assumed that {$\E X_{i, \cdot}=0$, $\E H_{i, \cdot}=0$,} ${\rm Cov}(H_{i, \cdot})= {\rm I}_{q}$ and hence $\Sigma_{X}=\Psi^{\intercal}\Psi+\Sigma_{E}$. 

The probability model \eqref{eq: confounder model} is more general than the Structural Equation Model in \eqref{eq: hidden SEM}.
It only describes the observational distribution of the latent variable $H_{i, \cdot}$ and the observed data $(X_{i, \cdot}, Y_i)$, which possibly may be generated from the hidden confounding SEM \eqref{eq: hidden SEM}.

Our goal is to construct confidence intervals for the components of $\beta$, {which in the model \eqref{eq: hidden SEM} describes the causal effect of $X$ on the response $Y$.} The problem is challenging due to the presence of unobserved confounding. In fact, the regression parameter $\beta$ can not even be identified without additional assumptions. Our main condition addressing this issue is a denseness assumption that the rows $\Psi_{j,\cdot} \in \R^{p}$ are dense in a certain sense {(see Condition (A2) in Section \ref{sec: theory})}, i.e., many covariates of $X_{i, \cdot}\in \R^{p}$ are simultaneously affected by hidden confounders $H_{i, \cdot}\in\R^{q}$.

\subsection{Representation as a Linear Model} 
\label{sec: equi-model}
The Hidden Confounding Model \eqref{eq: confounder model} 
{can be represented as a linear model} for the observed data $\{X_{i, \cdot}, Y_i\}_{1\leq i\leq n}$:
\begin{equation}
\begin{aligned}
Y_i = (\beta + b)^{\intercal}X_{i, \cdot} + \epsilon_i \quad \text{and}\quad 
X_{i, \cdot} = \Psi^{\intercal} H_{i, \cdot} + E_{i, \cdot}, 
\end{aligned}
\label{eq: perturb}
\end{equation}
by writing
$$\epsilon_i=\err_{i}+\phi^{\intercal}H_{i, \cdot}-b^{\intercal} X_{i, \cdot} \quad \text{and} \quad b=\Sigma_{X}^{-1}\Psi^{\intercal}\phi.$$
As in \eqref{eq: confounder model} we assume that $E_{i, \cdot}$ 
is uncorrelated with $H_{i, \cdot}$ and, by construction of $b$, $\epsilon_i$ is uncorrelated with $X_{i, \cdot}$.  
With $\sigma_{\err}^2$ denoting the variance of $\err_i$, the variance of the error $\epsilon_i$ equals $\sigma_{\epsilon}^2=\sigma_{\err}^{2}+\phi^{\intercal}\left({\rm I}_q-\Psi\Sigma_{X}^{-1}\Psi^{\intercal}\right)\phi.$ In model \eqref{eq: perturb}, the response is generated from a linear model where the sparse coefficient vector $\beta$ has been perturbed by some perturbation vector $b \in \R^p$. This representation reveals how the parameter of interest $\beta$ is not in general identifiable from observational data, where one can not easily differentiate it from the perturbed coefficient vector $\beta + b$, where the perturbation vector $b$ is induced by hidden confounding. However, as shown in Lemma \ref{lem: confounding error} in the supplement, $b$ is dense and $\|b\|_2$ is small for large $p$ under the assumption of dense confounding, which enables us to identify $\beta$ asymptotically. {It is important to note that 
the term $b ^{\intercal}X_{i, \cdot}$ induced by hidden confounders $H_{i, \cdot}$ is not necessarily small and hence cannot be simply ignored in model \eqref{eq: perturb}, but requires novel methodological approach.} 

\paragraph*{Connection to measurement errors}
We briefly relate certain measurement error models to the Hidden Confounding Model \eqref{eq: confounder model}. Consider a linear model for the outcome $Y_i$ and covariates $X^{0}_{i\cdot}\in \R^{p}$, where we only observe $X_{i, \cdot}\in \R^{p}$ with measurement error $W_{i, \cdot} \in \R^{p}$:
\begin{equation}
Y_i = \beta^{\intercal}X^{0}_{i\cdot}+e_{i} \quad \text{and} \quad X_{i, \cdot}=X^0_{i, \cdot}+ W_{i, \cdot}  \quad \text{for}\; 1\leq i\leq n.
\label{eq: measurement-error}
\end{equation}
Here, $e_{i}$ is a random error independent of $X_{i, \cdot}^0$ and $W_{i, \cdot}$, and $W_{i, \cdot}$ is the measurement error independent of $X^{0}_{i}$. We can then express a linear dependence of $Y_i$ on the observed $X_{i, \cdot}$,
\begin{equation*}
Y_i=\beta^{\intercal}X_{i, \cdot}+(e_{i}-\beta^{\intercal}W_{i, \cdot}) \quad \text{and} \quad X_{i, \cdot}= W_{i, \cdot}+ X^0_{i, \cdot}
\label{eq: obs model}
\end{equation*}
We further assume the following structure of the measurement error: 
\begin{equation*}
W_{i, \cdot}= \Psi^{\intercal}H_{i, \cdot},
\label{eq: latent ME}
\end{equation*}
i.e. there exist certain latent variables $H_{i, \cdot}\in \R^{q}$ that contribute independently and linearly to the measurement error, a conceivable assumption in some practical applications. Combining this with the equation above we get
\begin{equation}
\begin{aligned}
Y_i=\beta^{\intercal} X_{i, \cdot}+(e_{i}-\phi^{\intercal}H_{i, \cdot}) \quad \text{and} \quad
X_{i, \cdot}=\Psi^{\intercal} H_{i, \cdot} +X^{0}_{i\cdot}, \end{aligned}
\label{eq: latent ME model}
\end{equation}
where $\phi=\Psi \beta\in \R^{q}$. Therefore, the model \eqref{eq: latent ME model} can be seen as a special case of the model \eqref{eq: confounder model}, by identifying $X^{0}_{i\cdot}$ in \eqref{eq: latent ME model} with $E_{i, \cdot}$ in \eqref{eq: confounder model}.

\section{Doubly Debiased Lasso Estimator}
\label{sec: method}
In this section, for a fixed index $j \in \{1,\ldots ,p\}$, we propose an inference method for the regression coefficient $\beta_j$ of the Hidden Confounding Model \eqref{eq: confounder model}. The validity of the method is demonstrated by considering the equivalent model \eqref{eq: perturb}.

\subsection{Double Debiasing}
\label{sec: definition}
We denote by $\betainit$ an initial estimator of $\beta$. We will use the spectral deconfounding estimator proposed in \citep{cevid2018spectral}, described in detail in Section \ref{sec: initial betahat}. We start from the following decomposition: 
\begin{equation}
Y-X_{-j}\betainit_{-j}=X_{j}\left(\beta_j+b_j\right)+X_{-j}(\beta_{-j}-\betainit_{-j})+X_{-j}b_{-j}+\epsilon \quad \text{for}\quad j \in \{1,\ldots ,p\}.
\label{eq: decomp}
\end{equation}
The above decomposition reveals two sources of bias: the bias $X_{-j}(\beta_{-j}-\betainit_{-j})$ due to the error of the initial estimator $\betainit$ and the bias $X_{-j} b_{-j}$ induced by the perturbation vector $b$ in the model \eqref{eq: perturb}, arising by marginalizing out the hidden confounding in \eqref{eq: confounder model}. Note that the bias $b_j$ is negligible in the dense confounding setting, see Lemma \ref{lem: confounding error} in the supplement. The first bias, due to penalization, appears in the standard high-dimensional linear regression as well, and can be corrected with the debiasing methods proposed in \citep{zhang2014confidence,van2014asymptotically,javanmard2014confidence} {when assuming no hidden confounding. However, in presence of hidden confounders, methodological innovation is required for correcting both bias terms and conducting the resulting statistical inference.}
We propose a novel Doubly Debiased Lasso estimator for correcting both sources of bias simultaneously. 

Denote by $\NT \in \R^{n\times n}$ a symmetric spectral transformation matrix, which shrinks the singular values of the sub-design $X_{-j}\in \R^{n\times (p-1)}$. The detailed construction, together with some examples, is given in Section \ref{sec: spectral construction}. {We shall point out that the construction of the transformation matrix $\NT $ depends on which coefficient $\beta_j$ is our target and hence refer to $\NT $ as the nuisance spectral transformation with respect to the coefficient $\beta_j$.}  Multiplying both sides of the decomposition \eqref{eq: decomp} with the transformation $\NT $ gives:
\begin{equation}
\mathcal{P}^{(j)}(Y-X_{-j}\betainit_{-j})=\NT X_{j}\left(\beta_j+b_j\right)+\NT X_{-j}(\beta_{-j}-\betainit_{-j})+\NT X_{-j}b_{-j}+\NT  \epsilon. \label{eq: transformed decomposition}
\end{equation}

The quantity of interest $\beta_j$ appears on the RHS of the equation \eqref{eq: transformed decomposition} next to the vector $\NT X_j$, whereas the additional bias lies in the span of the columns of $\NT X_{-j}$. For this reason, we construct a projection direction vector $\NT Z_j\in \R^{n}$ as the transformed residuals of regressing $X_j$ on $X_{-j}$:
\begin{equation}
Z_j=X_j-X_{-j}\widehat{\gamma},
\label{eq: residue}
\end{equation} 
where the coefficients $\widehat{\gamma}$ are estimated with the Lasso for the transformed covariates using $\NT$:
\begin{equation}
\widehat{\gamma}=\argmin_{\gamma\in \R^{p-1}}\left\{\frac{1}{2n}\|\NT X_j-\NT X_{-j}\gamma\|_2^2+\lambda_j \sum_{l\neq j}\frac{\|\NT  X_{\cdot,l}\|_2}{\sqrt{n}} |\gamma_l|\right\},
\label{eq: est-gamma}
\end{equation}
with $\lambda_j = A \sigma_{j}\sqrt{\log p/n}$ for some positive constant $A>\sqrt{2}$ {(for $\sigma_{j}$, see Section 4.1)}.

Finally, motivated by the equation \eqref{eq: transformed decomposition},
we propose the following estimator for $\beta_j$:
\begin{equation}
\widehat{\beta}_j =\frac{(\NT Z_{j})^{\intercal}\NT (Y-X_{-j}\betainit_{-j})}{(\NT Z_{j})^{\intercal}\NT X_j}. 
\label{eq: proposed estimator}
\end{equation}
We refer to this estimator as the Doubly Debiased Lasso estimator as it simultaneously corrects the bias induced by $\betainit$ and the confounding bias $X_{-j}b_{-j}$ by using the spectral transformation $\mathcal{P}^{(j)}.$ 

In the following, we briefly explain why the proposed estimator estimates $\beta_j$ well. We start with the following error decomposition of $\widehat{\beta}_j$, derived from \eqref{eq: transformed decomposition}
\begin{equation}
\widehat{\beta}_j -\beta_j=\underbrace{\frac{(\NT Z_{j})^{\intercal}\NT  \epsilon}{(\NT Z_{j})^{\intercal}\NT X_j}}_{\rm Variance}+\underbrace{\frac{(\NT Z_{j})^{\intercal}\NT X_{-j}(\beta_{-j}-\betainit_{-j})}{(\NT Z_{j})^{\intercal}\NT X_j}+\frac{(\NT Z_{j})^{\intercal}\NT X_{-j}b_{-j}}{(\NT Z_{j})^{\intercal}\NT X_j}+b_j}_{\rm Remaining \; Bias}.
\label{eq: error decomp}
\end{equation}
In the above equation, the bias after correction consists of two components: the remaining bias due to the estimation error of $\betainit_{-j}$ and the remaining confounding bias due to $X_{-j}b_{-j}$ and $b_j$. These two components can be shown to be negligible in comparison to the variance component under certain model assumptions, see Theorem \ref{thm: limiting dist} and its proof for details.
Intuitively, the construction of the spectral transformation matrix $\NT $ is essential for reducing the bias due to the hidden confounding. The term $\frac{(\NT Z_{j})^{\intercal}\NT X_{-j}b_{-j}}{(\NT Z_{j})^{\intercal}\NT X_j}$ in equation \eqref{eq: error decomp} is of a small order because $\NT $ shrinks the leading singular values of $X_{-j}$ and hence $\NT X_{-j}b_{-j}$ is significantly smaller than $X_{-j}b_{-j}$. 
The induced bias $X_{-j}b_{-j}$ is not negligible since $b_{-j}$ points in the direction of leading right singular vectors of $X_{-j}$, thus leading to  $\|\tfrac{1}{\sqrt{n}}X_{-j}b_{-j}\|_2$ being of constant order. By applying a spectral transformation to shrink the leading singular values, one can show that $\|\tfrac{1}{\sqrt{n}}\NT X_{-j}b_{-j}\|_2=O_{p}(1/\sqrt{\min\{n,p\}})$. 

Furthermore, the other remaining bias term  $\frac{(\NT Z_{j})^{\intercal}\NT X_{-j}(\beta_{-j}-\betainit_{-j})}{(\NT Z_{j})^{\intercal}\NT X_j}$ in \eqref{eq: error decomp} is small since the initial estimator $\betainit$ is close to $\beta$ in $\ell_1$ norm and $\NT Z_{j}$ and $\NT X_{-j}$ are nearly orthogonal due to the construction of $\widehat{\gamma}$ in \eqref{eq: est-gamma}. This bias correction idea is analogous to the Debiased Lasso estimator introduced in \citep{zhang2014confidence} for the standard high-dimensional linear regression:
\begin{equation}
\begin{aligned}
\widehat{\beta}^{DB}_j = \frac{(Z^{DB}_{j})^{\intercal}(Y-X_{-j}\betainit_{-j})}{(Z^{DB}_{j})^{\intercal}X_j},
\end{aligned}
\label{eq: plain estimator}
\end{equation}
where $Z^{DB}_{j}$ is constructed similarly as in \eqref{eq: residue} and \eqref{eq: est-gamma}, but where $\NT $ is the identity matrix.
Therefore, the main difference between the estimator in \eqref{eq: plain estimator} and our proposed estimator \eqref{eq: proposed estimator} is that for its construction we additionally apply the nuisance spectral transformation $\NT $. 

We emphasize that the additional spectral transformation $\NT$ is necessary even for just correcting the bias of $\betainit_{-j}$ in presence of confounding (i.e., it is also needed for the first besides the second bias term in \eqref{eq: error decomp}).  To see this, we define the best linear projection of $X_{1,j}$ to all other variables $X_{1,-j}\in \R^{p-1}$ {with the coefficient vector} 
$\gamma=[\E(X_{i,-j}X_{i,-j}^{\intercal})]^{-1}\E(X_{i,-j} X_{i,j}) \in \R^{p-1}$ {(which is then estimated by the Lasso in the standard construction of $Z_j^{DB}$). We
} 
notice that $\gamma$ need not be sparse due to the fact that all covariates are affected by a common set of hidden confounders yielding spurious associations. Hence, the standard construction of $Z_j^{DB}$ in \eqref{eq: plain estimator} is not favorable in the current setting.  {In contrast, the proposed method with $\NT $ works for two reasons:} 
first, the application of $\NT$ in \eqref{eq: est-gamma} leads to a consistent estimator of the sparse component of $\gamma$, denoted as $\gamma^{E}$ (see the expression of $\gamma^{E}$ in Lemma \ref{lem: decomposition lemma}); second, the application of $\NT$ leads to a small prediction error $\NT X_{-j}(\widehat{\gamma}-\gamma^{E})$. We illustrate in Section \ref{sec: empirical} how the application of $\NT$ corrects the bias due to $\betainit_{-j}$ and observe a better empirical coverage after applying $\NT$ in comparison to the standard debiased Lasso in \eqref{eq: plain estimator}; see Figure \ref{fig: no_bias}.

\subsection{Confidence Interval Construction}
\label{sec: CI construction}
In Section \ref{sec: theory}, we establish the asymptotic normal limiting distribution of the proposed estimator $\widehat{\beta}_j$ under certain regularity conditions. Its standard deviation can be estimated by $\sqrt{\frac{\widehat{\sigma}_{e}^2 \cdot Z_{j}^{\intercal}(\NT) ^4 Z_{j}}{[Z_{j}^{\intercal}(\NT) ^2 X_j]^2}}$ with $\widehat{\sigma}_{e}$ denoting a consistent estimator of ${\sigma}_{e}$. The detailed construction of $\widehat{\sigma}_{e}$ is described in Section \ref{sec: noise level}. Therefore, a confidence interval (CI) with asymptotic coverage $1-\alpha$ can be obtained as
\begin{equation}
{\rm CI}(\beta_j)=\left(\widehat{\beta}_j-z_{1-\frac{\alpha}{2}}\sqrt{\frac{\widehat{\sigma}_{e}^2 \cdot Z_{j}^{\intercal}(\NT) ^4 Z_{j}}{[Z_{j}^{\intercal}(\NT) ^2 X_j]^2}}, \widehat{\beta}_j+z_{1-\frac{\alpha}{2}}\sqrt{\frac{\widehat{\sigma}_{e}^2 \cdot Z_{j}^{\intercal}(\NT) ^4 Z_{j}}{[Z_{j}^{\intercal}(\NT) ^2 X_j]^2}}\right),
\label{eq: CI}
\end{equation}
where $z_{1-\frac{\alpha}{2}}$ is the $1-\tfrac{\alpha}{2}$ quantile of a standard normal random variable.

\subsection{Construction of Spectral Transformations}
\label{sec: spectral construction}
Construction of the spectral transformation $\NT \in \R^{n\times n}$ is an essential step for the Doubly Debiased Lasso estimator \eqref{eq: proposed estimator}. The transformation $\NT \in \R^{n\times n}$ is a symmetric matrix shrinking the leading singular values of the design matrix $X_{-j}\in \R^{n\times (p-1)}$. Denote by $m=\min\{n,p-1\}$ and the SVD of the matrix $X_{-j}$ by
$X_{-j}=U(X_{-j}) \Lambda(X_{-j}) [V(X_{-j})]^{\intercal},$
where $U(X_{-j}) \in \R^{n\times m}$ and $V(X_{-j}) \in \R^{(p-1) \times m}$ have orthonormal columns and $\Lambda(X_{-j}) \in \R^{m \times m}$ is a diagonal matrix of singular values which are sorted in a decreasing order
$\Lambda_{1,1}(X_{-j}) \geq \Lambda_{2,2}(X_{-j}) \geq \ldots \geq \Lambda_{m,m}(X_{-j}) \geq 0$. We then define the spectral transformation $\NT $ for $X_{-j}$ as 
$\NT =U(X_{-j}) S(X_{-j}) [U(X_{-j})]^{\intercal},$
where $S(X_{-j})\in \R^{m\times m}$ is a diagonal shrinkage matrix with $0 \leq S_{l,l}(X_{-j}) \leq 1$ for $1\leq l\leq m$. The SVD for the complete design matrix $X$ is defined analogously. We highlight the dependence of the SVD decomposition on $X_{-j}$, but for simplicity it will be omitted when there is no confusion. 
Note that $\NT  X_{-j}=U \left(S\Lambda\right) V^{\intercal}$, so the spectral transformation shrinks the singular values $\left\{\Lambda_{l,l}\right\}_{1\leq l\leq m}$ to $\left\{S_{l,l}\Lambda_{l,l}\right\}_{1\leq l\leq m}$, {where $\Lambda_{l,l} = \Lambda_{l,l}(X_{-j})$.}

\paragraph*{Trim transform} For the rest of this paper, the spectral transformation that is used is the Trim transform \citep{cevid2018spectral}. It limits any singular value to be at most some threshold $\tau$. This means that the shrinkage matrix $S$ is given as: for $1\leq l\leq m,$
\begin{equation*}
S_{l,l}=\begin{cases} {\tau}/{\Lambda_{l,l}} & \quad \text{if} \quad \Lambda_{l,l} > \tau\\
1 & \quad \text{otherwise}\\
\end{cases}.
\end{equation*}

A good default choice for the threshold $\tau$ is the median singular value $\Lambda_{\floor{m/2}, \floor{m/2}}$, so only the top half of the singular values is shrunk to the bulk value $\Lambda_{\floor{m/2}, \floor{m/2}}$ and the bottom half is left intact. 
More generally, one can use any percentile {$\rho_j\in (0,1)$} to shrink the top $(100\rho_j)\%$ singular values to the corresponding $\rho_j$-quantile $\Lambda_{\lfloor \rho_j m \rfloor,\lfloor \rho_j m \rfloor}$. We define the $\rho_j$-Trim {transform} $\NT$ as 
\begin{equation}
\NT =U(X_{-j}) S(X_{-j}) [U(X_{-j})]^{\intercal} \; \text{with} \;
S_{l,l}(X_{-j})=\begin{cases} \frac{{\Lambda_{\lfloor \rho_j m \rfloor,\lfloor \rho_j m \rfloor}}(X_{-j})}{{\Lambda_{l,l}}(X_{-j})} & \; \text{if} \;\; l\leq \lfloor \rho_j m\rfloor\\
1 & \;\; \text{otherwise}\\
\end{cases}
\label{eq: rho shrinkage}
\end{equation}

In Section \ref{sec: theory} we investigate the dependence of the asymptotic efficiency of the resulting Doubly Debiased Lasso $\widehat{\beta}_j$ on the percentile choice $\rho_j=\rho_j(n)$. There is a certain trade-off in choosing $\rho_j$: a smaller value of $\rho_j$ leads to a more efficient estimator, but one needs to be careful to keep $\rho_j m$ sufficiently large compared to the number of hidden confounders $q$, in order to ensure reduction of the confounding bias. In Section \ref{sec: valid transformations} of the supplementary material, we describe the general conditions that the used spectral transformations need to satisfy to ensure good performance of the resulting estimator.

\subsection{Initial Estimator $\betainit$}
\label{sec: initial betahat}
For the Doubly Debiased Lasso \eqref{eq: proposed estimator}, we use the spectral deconfounding estimator proposed in \citep{cevid2018spectral} as our initial estimator $\betainit$. It uses a spectral transformation $\Q = \Q(X)$, constructed similarly as the transformation $\NT $ described in Section \ref{sec: spectral construction}, with the difference that instead of shrinking the singular values of $X_{-j}$, $\Q$ shrinks the leading singular values of the whole design matrix $X\in \R^{n\times p}$.
Specifically, for any percentile {$\rho\in (0,1)$}, the $\rho$-Trim transform $\Q$ is given by
\begin{equation}
\mathcal{Q}=U(X) S(X) [U(X)]^{\intercal} \; \text{with} \;
S_{l,l}(X)=\begin{cases} \frac{{\Lambda_{\lfloor \rho m \rfloor,\lfloor \rho m \rfloor}}(X)}{{\Lambda_{l,l}(X)}} & \; \text{if} \;\; l\leq \lfloor \rho m\rfloor\\ 
1 & \;\; \text{otherwise}\\
\end{cases}
\label{eq: rho shrinkage whole}
\end{equation}
The estimator $\betainit$ is computed by applying the Lasso to the transformed data $\Q X$ and $\Q Y$:
\begin{equation}
\betainit=\arg\min_{\beta\in \R^{p}}\frac{1}{2n}\|\mathcal{Q}\left(y-X\beta\right)\|_2^2+\lambda\sum_{j=1}^{p}\frac{\|\mathcal{Q} X_{\cdot j}\|_2}{\sqrt{n}} |\beta_j|,
\label{eq: est-beta}
\end{equation}
where $\lambda=A\sigma_{e}\sqrt{\log p/n}$ is a tuning parameter with $A>\sqrt{2}$.

The transformation $\Q$ reduces the effect of the confounding and thus helps for estimation of $\beta$. In Section \ref{sec: estimator rates},  the $\ell_1$ and $\ell_2$-error rates of $\betainit$ are given, thereby extending the results of \citep{cevid2018spectral}.

\subsection{Noise Level Estimator}\label{sec: noise level}
In addition to an initial estimator of $\beta$, we also require a consistent estimator $\widehat{\sigma}_{e}^2$ of the error variance $\sigma_{e}^{2}=\E(e_i^2)$ for construction of confidence intervals. Choosing a noise level estimator which performs well for a wide range of settings is not easy to do in practice \citep{reid2016study}. We propose using the following estimator:
\begin{equation}
\widehat{\sigma}_{e}^2 = \frac{1}{{\rm Tr}(\mathcal{Q}^2)}\|\mathcal{Q}y-\mathcal{Q} X \betainit\|_2^2,
\label{eq: variance est}
\end{equation}
where $\Q$ is the same spectral transformation as in \eqref{eq: est-beta}.

The motivation for this estimator is based on the expression
\begin{equation}
\mathcal{Q}y-\mathcal{Q} X \betainit=\mathcal{Q}\epsilon +\mathcal{Q} X (\beta-\betainit)+\mathcal{Q} X b,
\end{equation}
which follows from the model \eqref{eq: perturb}. The consistency of the proposed noise level estimator, formally shown in Proposition \ref{prop: noise consistency}, follows from the following observations: the initial spectral deconfounding estimator $\betainit$ has a good rate of convergence for estimating $\beta$; the spectral transformation $\Q$ significantly reduces the additional error $X b$ induced by the hidden confounders; $\|\Q \epsilon\|_2^2/{\rm Tr}(\mathcal{Q}^2)$ consistently estimates $\sigma_{\epsilon}^2.$ Additionally, the dense confounding model is shown to lead to a small difference between the noise levels $\sigma_{\epsilon}^2$ and $\sigma_{\err}^{2}$, see Lemma \ref{lem: confounding error} in the supplement. In Section \ref{sec: theory} we show that variance estimator $\widehat{\sigma}_{e}^2$ defined in \eqref{eq: variance est} is a consistent estimator of $\sigma_{\err}^2$.

\subsection{Method Overview and Choice of the Tuning Parameters}
\label{sec: overview}
The Doubly Debiased Lasso needs specification of various tuning parameters. A good and theoretically justified rule of thumb is to use the Trim transform with $\rho = \rho_j = 1/2$, which shrinks the large singular values to the median singular value, see \eqref{eq: rho shrinkage}. Furthermore, similarly to the standard Debiased Lasso \citep{zhang2014confidence}, our proposed method involves the regularization parameters $\lambda$ in the Lasso regression for the initial estimator $\widehat{\beta}^{init}$ (see equation \eqref{eq: est-beta}) and $\lambda_j$ in the Lasso regression for the projection direction $\NT Z_j$ (see equation \eqref{eq: est-gamma}). For choosing $\lambda$ we use 10-fold cross-validation, whereas for $\lambda_j$, we increase slightly the penalty chosen by the 10-fold cross-validation, so that the variance of our estimator, which can be determined from the data up to a proportionality factor $\sigma_e^2$, increases by $25\%$, as proposed in \citep{dezeure2017high}.

The proposed Doubly Debiased Lasso method is summarized in Algorithm \ref{algo: doubly debiased}, which also highlights where each tuning parameter is used.

\begin{algorithm}[htp!]
\caption{Doubly Debiased Lasso}
\begin{flushleft}
\hspace*{\algorithmicindent} \textbf{Input:} Data $X\in \R^{n\times p}, Y\in \R^{n}$; index $j$, tuning parameters $\rho, \rho_j \in (0,1)$ and $\lambda>0$, $\lambda_j>0$ \\
\hspace*{\algorithmicindent} \textbf{Output:} Point estimator $\widehat{\beta}_j$, standard error estimate $\widehat{\sigma}_e^2$ and confidence interval ${\rm CI}(\beta_j)$ \\
\end{flushleft}
\begin{algorithmic}[1]
    \State $\mathcal{Q} \gets$ \textsc{TrimTransform}($X, \rho$) \Comment{{construct $\rho$-trim {as in} 
    \eqref{eq: rho shrinkage whole}}}
    \State $\betainit \gets$ \textsc{Lasso}($\mathcal{Q}X$, $\mathcal{Q}Y$, $\lambda$)
    \Comment{Lasso regression with transformed data, see \eqref{eq: est-beta}}
    
    \vspace*{0.25cm}
    \State $\NT \gets$ \textsc{TrimTransform}($X_{-j}, \rho_j$) \Comment{{construct $\rho_j$-trim as in \eqref{eq: rho shrinkage}}}
    \State $\widehat{\gamma} \gets$ \textsc{Lasso}($\NT X_{-j}$, $\NT X_j$, $\lambda_j$) \Comment{Lasso regression with transformed data, see \eqref{eq: est-gamma}}
    \State $\NT Z_j \gets \NT X_j - \NT X_{-j}\widehat{\gamma}$ \Comment{take the residuals as the projection direction}
    \State $\widehat{\beta}_j \gets$ \textsc{DebiasedLasso}($\betainit, \NT X_{-j}$, $\NT X_j$, $\NT Z_j$) \Comment{{compute Doubly Debiased Lasso as in \eqref{eq: plain estimator}}}     
    \vspace*{0.25cm}
    \State $\widehat{\sigma}_{e}^2 \gets$ \textsc{NoiseLevel}($X, Y, \betainit, \mathcal{Q}$) \Comment{compute noise level as in \eqref{eq: variance est}} 
    \State ${\rm CI}(\beta_j) \gets$ \textsc{ConfidenceInterval}($\widehat{\beta}_j, \NT X_j, \NT Z_j, \widehat{\sigma}_{e}^2, \alpha$) \Comment{compute the $(1-\alpha)$-CI as in \eqref{eq: CI}}
\end{algorithmic}
\label{algo: doubly debiased}
\end{algorithm}

\section{Theoretical Justification}
\label{sec: theory}

This section provides theoretical justifications of the proposed method for the Hidden Confounding Model \eqref{eq: confounder model}. 
The proofs of the main results are presented in 
Sections \ref{sec: key results} and \ref{sec: RE verification}
of the supplementary material
together with several other technical results of independent interest. 
 
\subsection{Model assumptions}
\label{sec: model assumptions}
{In the following, we fix the index $1\leq j\leq p$ and introduce the model assumptions for establishing the asymptotic normality of our proposed estimator $\widehat{\beta}_j$ defined in \eqref{eq: proposed estimator}. }
For the coefficient matrix $\Psi \in \R^{q\times p}$ in \eqref{eq: perturb}, we use $\Psi_{j}\in \R^{q}$ to denote the $j$-th column and $\Psi_{-j} \in \R^{q \times (p-1)}$ denotes the sub-matrix with the remaining $p-1$ columns.
Furthermore, we write $\gamma$ for the best linear approximation of $X_{1, j}\in \R$ by $X_{1, -j}\in \R^{p-1}$, that is $\gamma = \argmin_{\gamma'\in \R^{p-1}} \E(X_{1, j} - X_{1, -j}\gamma')^2$, whose explicit expression is:
$$\gamma = [\E(X_{1,-j}X_{1,-j}^{\intercal})]^{-1}\E(X_{1,-j} X_{1,j}).$$ 
For ease of notation, we do not explicitly express the dependence of $\gamma$ on $j$.
Similarly, define
$$\gamma^{E}= [\E(E_{1,-j}E_{1,-j}^{\intercal})]^{-1}\E(E_{1,-j} E_{1,j}).$$
We denote the corresponding residuals by 
$\eta_{i,j} = X_{i, j} - X^{\intercal}_{i, -j}\gamma$ and $\nu_{i,j}=E_{i,j}-E_{i,-j}^{\intercal}\gamma^{E}$ for $1\leq i\leq n.$  
We use $\sigma_{j}$ to denote the standard deviation of $\nu_{i,j}$. 

The first assumption is on the precision matrix of $E_{i, \cdot}\in \R^{p}$ in \eqref{eq: confounder model}:
\begin{enumerate}
\item[\textbf{(A1)}] The precision matrix $\Omega_{E}=[\E(E_{i,\cdot}E_{i,\cdot}^{\intercal})]^{-1}$ satisfies
$
c_0\leq \lambda_{\min}\left(\Omega_{E}\right)\leq \lambda_{\max}\left(\Omega_{E}\right)\leq C_0 $ {and} $\|(\Omega_{E})_{\cdot,j}\|_0\leq s$ 
where $C_0>0$ and $c_0>0$ are some positive constants and $s$ denotes the sparsity level which can grow with $n$ and $p$.
\end{enumerate}
Such assumptions on well-posedness and sparsity are commonly required for estimation of the precision matrix \citep{meinshausen2006high,lam2009sparsistency,yuan2010high,cai2011constrained} and are also used for confidence interval construction in the standard high-dimensional regression model without unmeasured confounding \citep{van2014asymptotically}.
Here, the conditions are not directly imposed on the covariates $X_{i, \cdot}$, but rather on their unconfounded part $E_{i, \cdot}$.

The second assumption is about the coefficient matrix $\Psi$ in \eqref{eq: perturb}, which describes the effect of the hidden confounding variables $H_{i, \cdot}\in \R^{q}$ on the measured variables $X_{i, \cdot}\in \R^{p}$:
\begin{enumerate}
\item[\textbf{(A2)}] The $q$-th singular value of the coefficient matrix $\Psi_{-j} \in \R^{q \times (p-1)}$ satisfies
\begin{equation}
\lambda_{q}(\Psi_{-j})\gg l(n,p,q) \coloneqq \max\left\{\Mq\sqrt{\frac{qp}{n}}(\log p)^{3/4}, \sqrt{\Mq q}p^{1/4}(\log p)^{{3}/{8}},\sqrt{qn\log p}\right\}
\label{eq: factor condition}
\end{equation}
\normalsize
where $\Mq$ is the sub-Gaussian norm for components of $X_{i,.}$, as defined in Assumption (A3). 
Furthermore, we have
\begin{equation}
\max\left\{\|\Psi (\Omega_{E})_{\cdot, j}\|_{2},\|\Psi_j\|_2,\|\Psi_{-j}(\Omega_{E})_{-j, j}\|_2,\|\phi\|_2\right\}\lesssim \sqrt{q} (\log p)^{c},
\label{eq: upper norm}
\end{equation} 
where $\Psi$ and $\phi$ are defined in \eqref{eq: confounder model} and $0<c\leq 1/4$ is some positive constant.
\end{enumerate}
The condition ${\rm (A2)}$ is crucial for identifying the coefficient $\beta_j$ in the high-dimensional Hidden Confounding Model \eqref{eq: confounder model}. Condition (A2) is referred to as the dense confounding assumption. A few remarks are in order regarding when this identifiability condition holds. 

Since all vectors $\Psi(\Omega_{E})_{\cdot, j}$, $\Psi_j$, $\Psi_{-j}(\Omega_{E})_{-j, j}$ and $\phi$ are $q$-dimensional, the upper bound condition \eqref{eq: upper norm} on their $\ell_2$ norms is mild. If the vector $\phi\in \R^{q}$ has bounded entries and the vectors $\{\Psi_{\cdot,l}\}_{1\leq l\leq p} \in \R^{q}$ are independently generated with zero mean and bounded second moments, then the condition \eqref{eq: upper norm} holds with probability larger than $1-(\log p)^{-2c}$, where $c$ is defined in \eqref{eq: upper norm}.  {
A larger value $c > 1/4$ is possible: the condition then holds with even higher probability, but makes the upper bounds for \eqref{eq: approximation decoup}
 in Lemma \ref{lem: decomposition lemma} and \eqref{eq: var-diff} in Lemma \ref{lem: confounding error} slightly worse, which then requires more stringent conditions on $\lambda_{q}(\Psi_{-j})$ in Theorem \ref{thm: limiting dist}, up to polynomial order of $\log p$.}
 
In the factor model literature  \citep{fan2013large,wang2017asymptotics} the spiked singular value condition $\lambda_{q}(\Psi)\asymp \sqrt{p}$ is quite common and holds under mild conditions. The Hidden Confounding Model is closely related to the factor model, where the hidden confounders $H_{i, \cdot}$ are the factors and the matrix $\Psi$ describes how these factors affect the observed variables $X_{i, \cdot}$. However, for our analysis, our assumption on $\lambda_{q}(\Psi_{-j})$ in \eqref{eq: factor condition} can be much weaker than the classical factor assumption $\lambda_{q}(\Psi_{-j})\asymp \sqrt{p},$ especially for a range of dimensionality where $p \gg n.$ 
In certain dense confounding settings, we can show that condition \eqref{eq: factor condition} holds with 
 high probability. Consider first the special case with a single hidden confounder, that is, $q=1$ and the effect matrix is reduced to a vector $\Psi\in \R^{p}$. In this case, $\lambda_{1}(\Psi_{-j})=\|\Psi_{-j}\|_2$ and the denseness of the effect vector $\Psi_{-j}$ leads to a large $\lambda_{1}(\Psi_{-j})$. The condition \eqref{eq: factor condition} can be satisfied even if only a certain proportion of covariates is affected by hidden confounding. When $q=1$, if we assume that there exists a set $A\subseteq\{1,2,\ldots,p\}$ such that $\{\Psi_l\}_{l\in A}$ are i.i.d. and $|A|\gg l(n,p,q)^2$, where $l(n,p,q)$ is defined in \eqref{eq: factor condition}, then with high probability $\lambda_{q}(\Psi)\gtrsim \sqrt{|A|}\gg l(n,p,q)$. {In the multiple hidden confounders setting, if the vectors $\{\Psi_{l}\}_{l\in A}$ are generated as i.i.d. sub-Gaussian random vectors, which has an interpretation that all covariates are analogously affected by the confounders,} then the spiked singular value condition \eqref{eq: factor condition} is satisfied with  high probability as well. See Lemmas \ref{lem: dense confounding} and \ref{lem: dense confounding general} in Section \ref{sec: verify A2} of the supplementary material for the exact statement. In Section \ref{sec: sim}, we also explore the numerical performance of the method when different proportions of the covariates are affected and observe that the proposed method works well even if the hidden confounders only affect a small percentage of the covariates, say $5\%$.

Under the model \eqref{eq: confounder model}, if the entries of $\Psi$ are assumed to be i.i.d. sub-Gaussian with zero mean and variance $\sigma_{\Psi}^2,$ then we have $\lambda_{q}(\Psi_{-j})\asymp \sqrt{p}\sigma_{\Psi}$ with high probability. 
Together with \eqref{eq: factor condition}, this requires 
$$\sigma_{\Psi}\gg \max\left\{\Mq\sqrt{\frac{q}{n}}(\log p)^{3/4},\sqrt{\frac{qn\log p}{p}}, \frac{\sqrt{q \Mq(\log p)^{3/4}}}{{p}^{1/4}}\right\},$$
So if $p\gg qn\log p$ and $\min\{n,p\}\gg q^3 (\log p)^{3/2}\Mq^2,$ then the required effect size {$\sigma_{\Psi}$} of the hidden confounder $H_{i,\cdot}$ on an individual covariate $X_{i,j}$ can diminish to zero fairly quickly.

The condition \eqref{eq: factor condition} can in fact be empirically checked using the sample covariance matrix $\widehat{\Sigma}_{X}$. Since $\Sigma_{X}=\Psi^{\intercal}\Psi+\Sigma_{E}$, then the condition \eqref{eq: factor condition} implies that $\Sigma_{X}$ has at least $q$ spiked eigenvalues.  If the population covariance matrix $\Sigma_{X}$ has a few spikes, the corresponding sample covariance matrix will also have spiked eigenvalue structure with a high probability \citep{wang2017asymptotics}. Hence, we can inspect the spectrum of the sample covariance matrix $\widehat{\Sigma}_{X}$ and informally check whether it has spiked singular values. See the left panel of Figure \ref{fig: svd} for an illustration.

The third assumption is imposed on the distribution of various terms:
\begin{enumerate}
\item[\textbf{(A3)}] 
The random error $e_{i}$ in \eqref{eq: confounder model}  is assumed to be independent of $(X_{i, \cdot}^{\intercal}, H_{i, \cdot}^{\intercal})^{\intercal}$, {the error vector $E_{i,\cdot}$ is assumed to be independent of the hidden confounder $H_{i,\cdot}$,  and the noise term $\nu_{i,j}=E_{i,j}-E_{i,-j}^{\intercal}\gamma^{E}$ is assumed to be independent of $E_{i,-j}$.} Furthermore, $E_{i, \cdot}$ is a sub-Gaussian random vector and $e_i$ and $\nu_{i,j}$ are sub-Gaussian random variables, whose sub-Gaussian norms satisfy $\max\{\|E_{i, \cdot}\|_{\psi_2},\|e_i\|_{\psi_2},\max_{1\leq l\leq p}\|\nu_{i,l}\|_2\}\leq C$, where $C>0$ is a positive constant independent of $n$ and $p$. For $1\leq l\leq p$, $X_{i,l}$ are sub-Gaussian random variables whose sub-Gaussian norms satisfy $\max_{1\leq l\leq p}\|X_{i,l}\|_{\psi_2}\leq \Mq$, where $1\lesssim \Mq\lesssim \sqrt{n/\log p}.$
\end{enumerate}

The independence assumption between the random error $\err_i$ and $(X_{i, \cdot}^{\intercal}, H_{i, \cdot}^{\intercal})^{\intercal}$ 
is commonly assumed for the SEM {\eqref{eq: hidden SEM} and thus it holds in the induced Hidden Confounding Model \eqref{eq: confounder model} as well}, see for example \citep{pearl2009causality}. {Analogously, when modelling $X_{i,\cdot}$ as a SEM where the hidden variables $H_{i,\cdot}$ are directly influencing $X_{i,\cdot}$, that is, they are parents of the $X_{i,\cdot}$'s, the independence of $E_{i,\cdot}$ from $H_{i,\cdot}$ is a standard assumption.}
The independence assumption between $\nu_{i,j}$ and $E_{i,-j}$ holds automatically if $E_{i, \cdot}$ has a multivariate Gaussian distribution {(but $X_{i,\cdot}$ is still allowed to be non-Gaussian, e.g. due to non-Gaussian confounders)}.  

We emphasize that the individual components $X_{i,j}$ are assumed to be sub-Gaussian, instead of the whole vector $X_{i, \cdot}\in \R^{p}$. The sub-Gaussian norm $\Mq$ is allowed to grow with $q$ and $p$. Particularly, if we assume $H_{i,\cdot}$ to be a sub-Gaussian vector, then condition \eqref{eq: upper norm} implies that $\Mq\lesssim \sqrt{q}(\log p)^{c}\norm{H_{i,\cdot}}_{\psi_2}$. Furthermore, our theoretical analysis also covers the case when the sub-Gaussian norm $\Mq$ is of constant order. This happens, for example, when the entries of $\Psi$ are of order $1/\sqrt{q}$, since $M \asymp \max_{l=1,\ldots,p} \|\Psi_{l}\|_2$.

The final assumption is that the restricted eigenvalue condition \citep{bickel2009simultaneous} for the transformed design matrices $\Q X$ and $\NT X_{-j}$ is satisfied with high probability.
\begin{enumerate}
\item[\textbf{(A4)}] 
With probability at least $1 - \exp(-cn)$, we have
\small
\begin{equation}
{\RE}\left(\tfrac{1}{n}X^{\intercal}\mathcal{Q}^2X\right)=\inf_{\substack{\quad\Sb\subseteq [p]\quad\\ |\Sb|\leq k}}\min_{\substack{\omega\in \R^{p}\\ \|\omega_{\Sb^{c}}\|_1\leq C \Mq \|\omega_{\Sb}\|_1}}\frac{\omega^{\intercal}\left(\tfrac{1}{n}X^{\intercal}\mathcal{Q}^2X\right)\omega}{\|\omega\|_2^2}\geq \tau_*;
\label{eq: RE-a}
\end{equation}
\hspace*{0.1cm}
\begin{equation}
{\RE}\left(\tfrac{1}{n}X_{-j}^{\intercal}(\NT) ^2 X_{-j}\right)=\inf_{\substack{\Sb\subseteq  [p]\backslash \{j\}\\ |\Sb|\leq s}}\min_{\substack{\omega\in \R^{p-1}\\ \|\omega_{\Sb^{c}}\|_1\leq C\Mq \|\omega_{\Sb}\|_1}}\frac{\omega^{\intercal}(\tfrac{1}{n}X_{-j}^{\intercal}(\NT)^2 X_{-j})\omega}{\|\omega\|_2^2}\geq \tau_*
\label{eq: RE-b}
\end{equation} \normalsize
where $c, C, \tau_* > 0$ are positive constants independent of $n$ and $p$ and  {$\Mq$ is the sub-Gaussian norm for components of $X_{i,.}$, as defined in Assumption (A3)}. For ease of notation, the same constants $\tau_*$ and $C$ are used in \eqref{eq: RE-a} and \eqref{eq: RE-b}. 
\end{enumerate}
Such assumptions are common in the high-dimensional statistics literature, see \citep{buhlmann2011statistics}. The restricted eigenvalue condition {\rm (A4)} is similar, but more complicated than the standard restricted eigenvalue condition introduced in \citep{bickel2009simultaneous}. The main complexity is that, rather than for the original design matrix, the restricted eigenvalue condition is imposed on the transformed design matrices $\NT  X_{-j}$ and $\mathcal{Q}X$, after applying the Trim transforms $\NT $ and $\mathcal{Q}$, described in detail in Sections \ref{sec: spectral construction} and \ref{sec: initial betahat}, respectively. In the following, we verify the restricted eigenvalue condition ${\rm (A4)}$ for $\frac{1}{n}X^{\intercal}\mathcal{Q}^2X$ and the argument can be extended to $\frac{1}{n}X_{-j}^{\intercal}(\NT) ^2X_{-j}$.

\begin{Proposition}
Suppose that assumptions (A1) and (A3) hold, $H_{i,\cdot}$ is a sub-Gaussian random vector, $q+\log p\lesssim \sqrt{n}$ and $k=\|\beta\|_0$ satisfies ${\Mq}^2 k q^2\log p\log n/n\rightarrow 0$. Assume further that the loading matrix $\Psi\in \R^{q\times p}$ satisfies 
$\|\Psi\|_{\infty}\lesssim \sqrt{\log (qp)},$ $\lambda_{1}(\Psi)/\lambda_{q}(\Psi)\lesssim 1$ and that
\begin{equation}
\lambda_{q}(\Psi)\gg \frac{\sqrt{Mp}\max\{k^{1/4}q^{5/4},1\}\log (np)}{\min\{n,p\}^{1/4}}.
\label{eq: strong factor}
\end{equation}
{If $\lambda_{\lfloor \rho m\rfloor}(\frac{1}{n}X X^{\intercal}) \geq c  \max\{1,p/n\}$ for $\rho$ defined in \eqref{eq: rho shrinkage whole} and some positive constant $c>0$ independent of $n$ and $p$,} then there exist positive constants $c_1,c_2>0$ such that, with probability larger than $1-p^{-c_2}-\exp(-c_2 n)$, we have ${\rm RE}\left(\frac{1}{n}X^{\intercal}\mathcal{Q}^2X\right)\geq c_1 \lambda_{\min}(\Sigma_{X}).
$
\label{prop: RE verification}
\end{Proposition}

An important condition for  establishing Proposition \ref{prop: RE verification} is the condition \eqref{eq: strong factor}. Under the commonly assumed spiked singular value condition $\lambda_{q}(\Psi)\asymp \sqrt{p}$ \citep{fan2013large,wang2017asymptotics,bai2003inferential,bai2002determining}, the condition \eqref{eq: strong factor} is reduced to $k\ll \min\{n,p\}/({\Mq}^2q^5\log (np)^4).$ As a comparison, for the standard high-dimensional regression model with no hidden confounders, \citep{zhou2009restricted,raskutti2010restricted} verified the restricted eigenvalue condition under the sparsity condition $k\ll n/\log p.$  That is, if $\lambda_{q}(\Psi)\asymp \sqrt{p},$ then the sparsity requirement in Proposition \ref{prop: RE verification} is the same as that for the high-dimensional regression model with no hidden confounders, up to a polynomial order of $q$ and $\log(np),$

In comparison to the condition \eqref{eq: factor condition} in (A2), \eqref{eq: strong factor} can be slightly stronger for a range of dimensionality where $p \gg n^{3/2}.$ However, Proposition \ref{prop: RE verification} does not require the strong spiked singular value condition $\lambda_{q}(\Psi)\asymp \sqrt{p}$.  The proof of Proposition \ref{prop: RE verification} is presented in Appendix \ref{sec: RE verification} in the supplement. The condition $\lambda_{\lfloor \rho m\rfloor}(\frac{1}{n}X X^{\intercal}) \geq c  \max\{1,p/n\}$ can be empirically verified from the data. In Appendix \ref{sec: lower bound}, further theoretical justification for this condition is provided, under mild assumptions.

\subsection{Main Results}
\label{sec: main results}
In this section we present the most important properties of the proposed estimator \eqref{eq: proposed estimator}. 
We always consider asymptotic expressions in the limit where both $n,p \to \infty$ and focus on the high-dimensional regime with $c^{*}=\lim {p}/{n} \in (0,\infty]$. {We mention here that we also give some new results on point estimation of the initial estimator $\betainit$ defined in \eqref{eq: est-beta} in Appendix 
\ref{sec: estimator rates}, as they are established under more general conditions than in \citep{cevid2018spectral}.}

\subsubsection{Asymptotic normality}
\label{sec: normal and efficient}
We first present the limiting distribution of the proposed Doubly Debiased Lasso estimator.
{The proof of Theorem \ref{thm: limiting dist} and important intermediary results for establishing Theorem \ref{thm: limiting dist} are presented in Appendix \ref{sec: key results} in the supplement.}

\begin{Theorem}
{Consider the Hidden Confounding Model \eqref{eq: confounder model}.}
Suppose that conditions {\rm (A1)-(A4)} hold and further assume that $c^{*}=\lim {p}/{n} \in (0,\infty]$, {$k:=\|\beta\|_0 \ll {\sqrt{n}}/(\Mq^3 \log p)$,  $s:=\|(\Omega_{E})_{\cdot,j}\|_0\ll {n}/(\Mq^2\log p)$} 
and  $e_{i} \sim N(0,\sigma_{e}^2)$. 
Let the tuning parameters for $\betainit$ in \eqref{eq: est-beta} and $\widehat{\gamma}$ in \eqref{eq: est-gamma} respectively be $\lambda\asymp \sigma_{e} \sqrt{{\log p}/{n}}+\sqrt{q\log p/\lambda_{q}^2(\Psi)}$ and $\lambda_j\asymp {\sigma_{j}} \sqrt{{\log p}/{n}}+\sqrt{q\log p/\lambda_{q}^2(\Psi_{-j})}$. Furthermore, let $\mathcal{Q}$ and $\NT $ be the Trim transform \eqref{eq: rho shrinkage} with $\min\{\rho,\rho_j\} \geq (q+1)/\min\{n,p-1\}$ and $\max\{\rho,\rho_j\}<1$. Then the Doubly Debiased Lasso estimator \eqref{eq: proposed estimator} satisfies 
\begin{equation}
\frac{1}{\sqrt{V}}\left(\widehat{\beta}_j-\beta_j\right)\cid N(0,1),
\label{eq: limiting}
\end{equation}
where 
\begin{equation}
{\rm V}={\frac{{\sigma}_{e}^2 Z_{j}^{\intercal}(\NT) ^4 Z_{j}}{[Z_{j}^{\intercal}(\NT) ^2 X_j]^2}} \quad \text{and}\quad V^{-1}{\frac{\sigma_{e}^2{\rm Tr}[(\NT) ^4]}{{\sigma_{j}^2{\rm Tr}^2[(\NT) ^2]}}}\cip 1.
\label{eq: var}
\end{equation} 
\label{thm: limiting dist}
\end{Theorem}
\begin{Remark} \rm
The Gaussianity of the random error $e_i$ is mainly imposed to simplify the proof of asymptotic normality. We believe that this assumption is a technical condition and can be removed by applying more refined probability arguments as in \citep{gotze2002asymptotic}, where the asymptotic normality of quadratic forms $(\NT e)^{\intercal}\NT  e$ is established for the general sub-Gaussian case. The argument could be extended to obtain the asymptotic normality for $(\NT \eta_j)^{\intercal}\NT  e$, which is essentially needed for the current result. 
\end{Remark}
\begin{Remark} \rm
For constructing $\mathcal{Q}$ and $\mathcal{P}^{(j)}$, the main requirement is to trim the singular values enough in both cases, that is, $\min\{\rho,\rho_j\} \geq (q+1)/\min\{n,p-1\}$. This condition is mild in the high-dimensional setting with a small number of hidden confounders.
Our results are not limited to the proposed estimator which uses the Trim transform $\NT $ in \eqref{eq: rho shrinkage} and the penalized estimators $\widehat{\gamma}$ and $\betainit$ in \eqref{eq: est-gamma} and \eqref{eq: est-beta}, but hold for any transformation satisfying the conditions given in Section \ref{sec: valid transformations} of the supplementary material and any initial estimator satisfying the error rates presented in Section \ref{sec: estimator rates} of the supplementary material.
\end{Remark}

{
\begin{Remark}\rm
If we further assume the error $\epsilon_i$ in the model \eqref{eq: perturb} to be independent of $X_{i, \cdot}$, then the requirement \eqref{eq: factor condition} of the condition ${\rm (A2)}$ can be relaxed to 
$$\lambda_{q}(\Psi_{-j})\gg \max\left\{M\sqrt{\frac{qp}{n}}(\log p)^{3/4}, \sqrt{qM}p^{1/4}(\log p)^{{3}/{8}}, \sqrt{(s\Mq^2+k\sqrt{n}\Mq^3)q\log p}\right\}.$$
Note that the factor model implies the upper bound $\lambda_{q}(\Psi_{-j})\lesssim \sqrt{p}.$ 
Even if $n\geq p,$ the above condition on $\lambda_{q}(\Psi_{-j})$ can still hold if $p\gg k q\Mq^3 \log p \sqrt{n}$. On the other hand, the condition \eqref{eq: factor condition} together with $\lambda_{q}(\Psi_{-j})\lesssim \sqrt{p}$ imply that $p\gg qn\log p$, which excludes the setting $n\geq p.$ 
\end{Remark}}

There are three conditions on the parameters $s,q,k$ imposed in the Theorem \ref{thm: limiting dist} above. The most stringent one is the sparsity assumption $k\ll \sqrt{n}/[M^{3}\log p]$. In standard high-dimensional sparse linear regression, a related sparsity assumption $k\ll \sqrt{n}/\log p$ has also been used for confidence interval construction \citep{zhang2014confidence,van2014asymptotically,javanmard2014confidence} and has been established in \citep{cai2015regci} as a necessary condition for constructing adaptive confidence intervals. In the high-dimensional Hidden Confounding Model with $M\asymp 1$, the condition on the sparsity of $\beta$ is then of the same asymptotic order as in the standard high-dimensional regression with no hidden confounding. The condition on the sparsity of the precision matrix, $s=\|(\Omega_{E})_{\cdot, j}\|_0\ll {n}/(M^2{\log p})$, is mild in the sense that, for $M\asymp 1$, it is the maximal sparsity level for identifying $(\Omega_{E})_{\cdot, j}$. Implied by \eqref{eq: factor condition}, the condition that the number of hidden confounders $q$ is small is fundamental for all reasonable factor or confounding models. 
\subsubsection{Efficiency}
\label{sec: efficiency}
We investigate now the dependence of the asymptotic variance $V$ in \eqref{eq: var} on the choice of the spectral transformation $\NT$. We further show that the proposed Doubly Debiased Lasso estimator \eqref{eq: proposed estimator} is efficient in the Gauss-Markov sense, with a careful construction of the transformation $\NT$. 

The Gauss-Markov theorem states that the smallest variance of any unbiased linear estimator of $\beta_j$ in the standard low-dimensional regression setting (with no hidden confounding)  is ${\sigma_{e}^2}/{(n \sigma_{j}^2)}$, which we use as a benchmark. The corresponding discussion on efficiency of the standard high-dimensional regression can be found in Section 2.3.3 of \citep{van2014asymptotically}. The expression for the asymptotic variance $V$ of our proposed estimator \eqref{eq: proposed estimator} is given by $\frac{\sigma_{e}^2{\rm Tr}[(\NT) ^4]}{{\sigma_{j}^2{\rm Tr}^2[(\NT) ^2]}}$ (see Theorem \ref{thm: limiting dist}). For the Trim transform defined in \eqref{eq: rho shrinkage}, which trims top $(100\rho_j)\%$ of the singular values, {we have that
\begin{equation*}
\frac{\sigma_{e}^2{\rm Tr}[(\NT) ^4]}{{\sigma_{j}^2{\rm Tr}^2[(\NT) ^2]}}=\frac{\sigma_{e}^2}{{\sigma_{j}^2} } \cdot \frac{\sum_{l=1}^{m}S_{l,l}^4}{(\sum_{l=1}^{m} S_{l,l}^2)^2}, 
\end{equation*}
where we write $m = \min\{n, p-1\}$ and $S_{l,l}=S_{l,l}(X_{-j})\in [0,1]$. Since $S_{l,l}^4\leq S_{l,l}^2$ for every $l$,  $\sum_{l=1}^{m}S_{l,l}^2\geq (1-\rho_j)m$ and $(\sum_{l=1}^{m} S_{l,l}^2)^2\leq m\cdot \sum_{l=1}^{m}S_{l,l}^4,$ we obtain
$$\frac{\sigma_{e}^2 }{{\sigma_{j}^2} m}\leq \frac{\sigma_{e}^2{\rm Tr}[(\NT) ^4]}{{\sigma_{j}^2{\rm Tr}^2[(\NT) ^2]}}\leq \frac{1}{1-\rho_j}\cdot \frac{\sigma_{e}^2 }{{\sigma_{j}^2} m}.$$ In the high-dimensional setting where $p-1\geq n$, we have $m=n$ and then
\begin{equation}    
\frac{\sigma_{e}^2 }{{\sigma_{j}^2} n}
\leq \frac{\sigma_{e}^2{\rm Tr}[(\NT) ^4]}{{\sigma_{j}^2{\rm Tr}^2[(\NT) ^2]}} \leq \frac{1}{1-\rho_j}\cdot\frac{\sigma_{e}^2 }{{\sigma_{j}^2}n}.
\label{eq: power rate}
\end{equation}
}
\begin{Theorem}
Suppose that the assumptions of Theorem \ref{thm: limiting dist} hold.
If $p\geq n+1$ and $\rho_j=\rho_j(n)\rightarrow 0$, then 
the Doubly Debiased Lasso estimator in \eqref{eq: proposed estimator} has asymptotic variance $\frac{\sigma_{e}^2}{\sigma^{2}_{j} n}$, that is, it achieves the Gauss-Markov efficiency bound. 
\label{thm: efficiency}
\end{Theorem}

The above theorem shows that in the $q \ll n$ regime, the Doubly Debiased Lasso achieves the Gauss-Markov efficiency bound if $\rho_j = \rho_j(n) \to 0$ and $\min\{\rho,\rho_j\} \ge (q+1)/n$ (which is also a condition of Theorem \ref{thm: limiting dist}). When using the median Trim transform, i.e. $\rho_j=1/2$, the bound in  \eqref{eq: power rate} implies that the variance of the resulting estimator is at most twice the size of the Gauss-Markov bound. In Section \ref{sec: empirical}, we illustrate the finite-sample performance of the Doubly Debiased Lasso estimator for different values of $\rho_j$; see Figure \ref{fig: trim_level}.

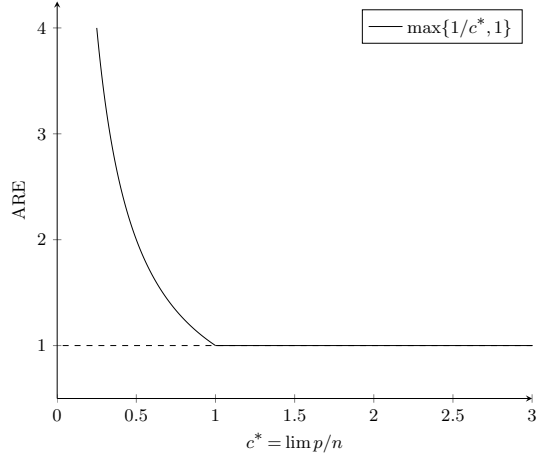
\begin{figure}[!htp]
\centering
\begin{tikzpicture}[scale=0.75]
\begin{axis}[
    axis lines = left,
    xlabel = {$\large c^{*}=\lim p/n$},
    ylabel = {$\large {\rm ARE}$},
    xmin=0, xmax=3,
    ymin=0.5, ymax=4.25, 
]
\addplot [
    domain=0.25:1,
    samples=100, 
]
{1/x};
\addplot [
    domain=1:3, 
    samples=100, 
]{1};
\addplot [dashed] {1};
\addlegendentry{$\large \max\{{1}/{c^{*}},1\}$}
\end{axis}
\end{tikzpicture}
\caption{The plot of ${\rm ARE}$ versus $c^{*} = \lim p/n$, for the setting of $\rho^{*}=0$.}
\label{fig: ARE}
\end{figure}

In general for the high-dimensional setting ${p}/{n}\rightarrow c^{*}\in (0,\infty]$, the Asymptotic Relative Efficiency (ARE) of the proposed Doubly Debiased Lasso estimator with respect to the Gauss-Markov efficiency bound satisfies the following:
\begin{equation}
{\rm ARE} \in \left[\frac{1}{\min\{c^{*},1\}}, \frac{1}{(1-\rho^{*}) \min\{c^{*},1\}}\right],
\label{eq: ARE}
\end{equation}
where $\rho^{*}=\lim_{n\rightarrow \infty}\rho_j(n) \in[0,1)$. The equation \eqref{eq: ARE} reveals how the efficiency of the {Doubly Debiased Lasso} 
is affected by the choice of the percentile $\rho_j=\rho_j(n)$ in transformation $\NT$ and the dimensionality of the problem. Smaller $\rho_j$ leads to a more efficient estimator, as long as the top few singular values are properly shrunk. 
Intuitively, a smaller percentile $\rho_j$ means that less information in $X_{-j}$ is trimmed out and hence the proposed estimator is more efficient. In addition, for the case $\rho^{*}=0$, we have ${\rm ARE}=\max\{1/c^{*},1\}$. With $\rho^{*}=0$, a plot of ${\rm ARE}$ with respect to the ratio $c^* = \lim p/n$ is given in Figure \ref{fig: ARE}. 
We see that for $c^{*}<1$ (that is $p<n$), the relative efficiency of the proposed estimator increases as the dimension $p$ increases and when $c^{*}\geq 1$ (that is $p\geq n$), we have that ${\rm ARE}=1$, saying that the Doubly Debiased Lasso achieves the efficiency bound in the Gauss-Markov sense.

The phenomenon that the efficiency is retained even in presence of hidden confounding is quite remarkable. For comparison, even in the classical low-dimensional setting, the most commonly used approach assumes availability of sufficiently many instrumental variables (IV) satisfying certain stringent conditions under which one can consistently estimate the effects in presence of hidden confounding. In Theorem 5.2 of \citep{wooldridge2010econometric}, the popular IV estimator, two-stage-least-squares (2SLS), is shown to have variance strictly larger than the efficiency bound in the Gauss-Markov setting (with no unmeasured confounding). It has been {also} shown in Theorem 5.3 of \citep{wooldridge2010econometric} that {the 2SLS estimator is efficient in the class of all linear instrumental variable estimators and thus, all linear instrumental variable estimators are strictly less efficient than our Doubly Debiased Lasso.} 
On the other hand, our proposed method not only avoids the difficult step of coming up with a large number of valid instrumental variables, but also achieves the efficiency bound with a careful construction of the spectral transformation $\NT$. This occurs due to a blessing of dimensionality and the assumption of dense confounding, where a large number of covariates are assumed to be affected by a small number of hidden confounders.

\subsubsection{Asymptotic validity of confidence intervals}
The asymptotic normal limiting distribution in Theorem \ref{thm: limiting dist} can be used for construction of confidence intervals for $\beta_j$. Consistently estimating the variance $V$ of our estimator, defined in \eqref{eq: var}, requires a consistent estimator of the error variance $\sigma_{e}^2$.
The following proposition establishes the rate of convergence of the estimator $\widehat{\sigma}_{e}^2$ proposed in \eqref{eq: variance est}:
\begin{Proposition} {Consider the Hidden Confounding Model \eqref{eq: confounder model}.} Suppose that conditions {\rm (A1)-(A4)} hold. Suppose further that $c^{*}=\lim {p}/{n} \in (0,\infty],$ $k\lesssim n/\log p$ and $q\ll \min\{n, p/\log p\}$. Then
with probability larger than $1-\exp(-ct^2)-\frac{1}{t^2}-c(\log p)^{-1/2}-n^{-c}$ for some positive constant $c>0$ and for any $0<t\leq \sqrt{n}$, we have
$$\left|\widehat{\sigma}_{e}^2-\sigma_{e}^2\right|\lesssim \frac{t}{\sqrt{n}}+\Mq^2 k{\frac{\log p}{n}}+\frac{q \log p}{p}+\frac{pq\sqrt{\log p}/n+\Mq^2 kq\log p}{\lambda_{q}^2(\Psi)},$$
where $\Mq$ is the sub-Gaussian norm for components of $X_{i,.}$ defined in Assumption (A3).
\label{prop: noise consistency} 
\end{Proposition}
Together with \eqref{eq: factor condition} of the condition ${\rm (A2)}$, we apply the above proposition and establish $\widehat{\sigma}_{e}^2-\sigma_{e}^2\cip 0.$
{As a remark, the estimation error $\left|\widehat{\sigma}_e^2-\sigma_{\epsilon}^2\right|$ is of the same order of magnitude as $\left|\widehat{\sigma}_e^2-\sigma_{e}^2\right|$ since the difference $\sigma_{\epsilon}^2-\sigma_e^2$ is small in the dense confounding model, see Lemma \ref{lem: confounding error} in the supplement.} 

Proposition \ref{prop: noise consistency}, together with Theorem \ref{thm: limiting dist}, imply the asymptotic coverage and precision properties of the proposed confidence interval ${\rm CI}(\beta_j)$, described in \eqref{eq: CI}:
\begin{Corollary}
Suppose that the conditions of Theorem \ref{thm: limiting dist} hold, then the confidence interval defined in \eqref{eq: CI} satisfies the following properties:
\begin{equation}
\liminf_{n,p\rightarrow \infty} \P\left(\beta_j\in {\rm CI}(\beta_j)\right)\geq 1-\alpha, 
\label{eq: coverage}
\end{equation}
\begin{equation}
\limsup_{n,p\rightarrow \infty}\P\left({\bf L}\left({\rm CI}(\beta_j)\right)\geq(2+c)z_{1-\frac{\alpha}{2}}\sqrt{\frac{{\sigma}_{e}^2{\rm Tr}[(\NT) ^4]}{{{\sigma}_{j}^2{\rm Tr}^2[(\NT) ^2]}}}\right)=0,
\label{eq: precision}
\end{equation}
for any positive constant $c>0$, where ${\bf L}\left({\rm CI}(\beta_j)\right)$ denotes the length of the proposed confidence interval.
\end{Corollary}
Similarly to the efficiency results {in Section \ref{sec: efficiency}}, the exact length depends on the construction of the spectral transformation $\NT $. Together with \eqref{eq: power rate}, the above proposition shows that the length of constructed confidence interval is shrinking at the rate of $n^{-1/2}$ for the Trim transform in the high-dimensional setting. Specifically, for the setting $p\geq n+1$, if we choose $\rho_j=\rho_j(n)\geq (q+1)/n$ and $\rho_j(n)\rightarrow 0$, the constructed confidence interval has asymptotically optimal length.

\section{Empirical results}
\label{sec: empirical}
In this section we consider the practical aspects of Doubly Debiased Lasso methodology and illustrate its empirical performance on both real and simulated data. The overview of the method and the tuning parameters selection can be found in Section \ref{sec: overview}.

\begin{figure}[htp]
    \hspace{-1cm}
    \centering
    \includegraphics[scale=0.8]{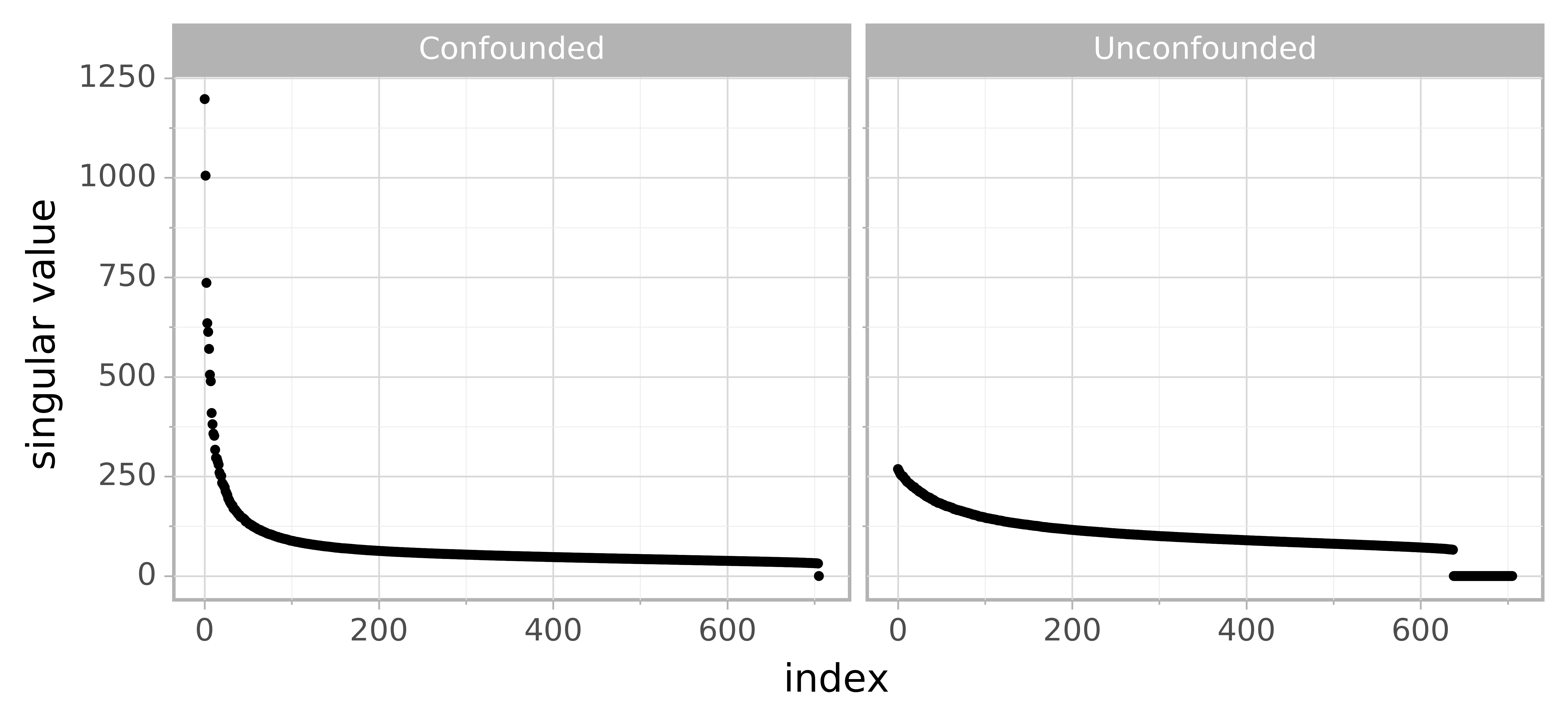}
    \caption{Left: Spiked singular values of the standardized gene expression matrix (see Section \ref{sec: real_data}) indicate possible confounding. Right: Singular values after regressing out the $q=65$ confounding proxies given in the dataset (thus labeled as ``unconfounded''). The singular values in both plots are sorted decreasingly.
    }
    \label{fig: svd}
\end{figure}

In order to investigate whether the given data set is potentially confounded, one can inspect the principal components of the design matrix $X$, or equivalently consider its SVD. Spiked singular value structure (see Figure \ref{fig: svd}) indicates the existence of hidden confounding, as much of the variance of our data can be explained by a small number of latent factors. {This also serves as an informal check of the spiked singular value condition in the assumption (A2).}

The scree plot can also be used for choosing the trimming thresholds, if one wants to depart from the default median rule (see Section \ref{sec: overview}). We have seen from the theoretical considerations in Section \ref{sec: theory} that we can reduce the estimator variance by decreasing the trimming thresholds for the spectral transformation $\NT$. On the other hand, it is crucial to choose them so that the number of shrunk singular values is still sufficiently large compared to the number of confounders. However, exactly estimating the number of confounders, e.g. by detecting the elbow in the scree plot \citep{wang2017asymptotics}, is not necessary with our method, since the efficiency of our estimator decreases relatively slowly as we decrease the trimming threshold.

In what follows, we illustrate the empirical performance of the Doubly Debiased Lasso in practice. We compare the performance with the standard Debiased Lasso \citep{zhang2014confidence}, even though it is not really a competitor for dealing with hidden confounding. Our goal is to illustrate and quantify the error and bias when using the naive and popular approach which ignores potential hidden confounding.
We first investigate the performance of our method on simulated data for a range of data generating mechanisms and then investigate its behaviour on a gene expression dataset from the GTEx project \citep{lonsdale2013genotype}.

\subsection{Simulations}
\label{sec: sim}
In this section, we compare the Doubly Debiased Lasso with the standard Debiased Lasso in several different simulation settings for estimation of $\beta_j$ and construction of the corresponding confidence intervals. 

In order to make comparisons with the standard Debiased Lasso as fair as possible, we use the same procedure for constructing the standard Debiased Lasso, but with $\mathcal{Q} = {\rm I}_p$, $\NT  = {\rm I}_{p-1}$, whereas for the Doubly Debiased Lasso, $\NT $, $\mathcal{Q}$ are taken to be median Trim transform matrices, unless specified otherwise. Finally, to investigate the usefulness of double debiasing, we additionally include the standard Debiased Lasso estimator with the same initial estimator $\betainit$ as our proposed method, see Section \ref{sec: initial betahat}. Therefore, this corresponds to the case where $\mathcal{Q}$ is the median Trim transform, whereas $\NT  = {\rm I}_{p-1}$.

We will compare the (scaled) bias and variance of the corresponding estimators. For a fixed index $j$, from the equation (\ref{eq: error decomp}) we have
\begin{equation*}
    V^{-1/2}(\widehat{\beta}_j - \beta_j) = N(0,1) + B_{\beta} + B_b, 
\end{equation*}
where the estimator variance ${\rm V}$ is defined in \eqref{eq: var} and the bias terms $B_\beta$ and $B_b$ are given by
\begin{gather*}
B_\beta =  V^{-1/2} \frac{Z_j^{\intercal} (\NT) ^2 X_{-j}(\betainit_{-j} - \beta_{-j})}{Z_j^{\intercal}(\NT) ^2X_j}, \qquad
B_b = V^{-1/2} \frac{Z_j^{\intercal}(\NT) ^2Xb}{Z_j^{\intercal}(\NT) ^2X_j}.
\end{gather*}
Larger estimator variance makes the confidence intervals wider. However, large bias makes the confidence intervals inaccurate. We quantify this with the scaled bias terms $B_\beta$, which is due to the error in estimation of $\beta$, and $B_b$, which is due to the perturbation $b$ arising from the hidden confounding. Having small $|B_\beta|$ and $|B_b|$ is essential for having a correct coverage, since the construction of confidence intervals is based on the approximation $V^{-1/2}(\widehat{\beta}_j - \beta_j) \approx N(0,1)$. We investigate the validity of the confidence interval construction by measuring the coverage of the nominal $95\%$ confidence interval. We present here a wide range of simulations settings and further simulations can be found in the Appendix \ref{sec: additional simulations}.

\paragraph*{Simulation parameters} Unless specified otherwise, in all simulations we fix $q = 3$, $s = 5$ and $\beta = (1, 1, 1, 1, 1, 0, \ldots 0)^{\intercal}$ and we target the coefficient $\beta_1=1$. The rows of the unconfounded design matrix $E$ are generated from $N(0, \Sigma_E)$ distribution, where $\Sigma_E = {\rm I}_p$, as a default. The matrix of confounding variables $H$, the additive error $e$ and the coefficient matrices $\Psi$ and $\phi$ all have i.i.d. $N(0, 1)$ entries, unless stated otherwise.
Each simulation is averaged over $5,000$ independent repetitions.

\begin{figure}[htp]
    \hspace{-0.65cm}
    \includegraphics[width=1.04\linewidth]{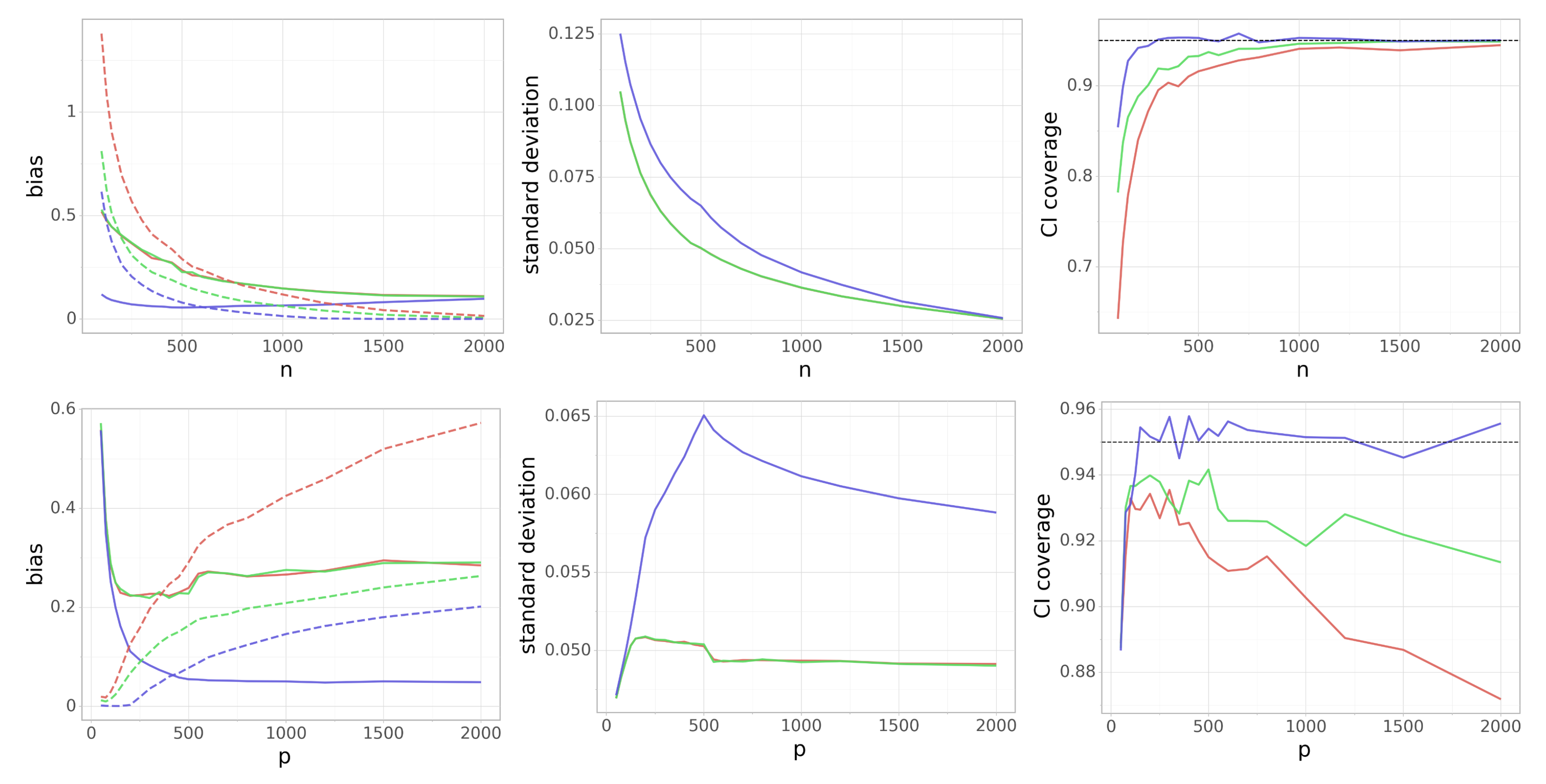}
    \caption{\textit{(Varying dimensions)} Dependence of the (scaled) absolute bias terms $|B_\beta|$ and $|B_b|$ (left), standard deviation $V^{1/2}$ (middle) and the coverage of the $95\%$ confidence interval (right) on the number of data points $n$ (top row) and the number of covariates $p$ (bottom row). On the left side, $|B_\beta|$ and $|B_b|$ are denoted by a dashed and a solid line, respectively. In the top row we fix $p=500$, whereas in the bottom row we have $n=500$. Blue color corresponds to the Doubly Debiased Lasso, red color represents the standard Debiased Lasso and green color corresponds also to the Debiased Lasso estimator, but with the same $\betainit$ as our proposed method. Note that the last two methods have almost indistinguishable $|B_b|$ and $V$.} 
        \label{fig: n_p_growing}
\end{figure}

\paragraph*{Varying dimensions $n$ and $p$}
In this simulation setting we investigate how the performance of our estimator depends on the dimensionality of the problem. The results can be seen in Figure \ref{fig: n_p_growing}. In the first scenario, shown in the top row, we have $p=500$ and $n$ varying from $50$ to $2,000$, thus covering both low-dimensional and high-dimensional cases. In the second scenario, shown in the bottom row, the sample size is fixed at $n=500$ and the number of covariates $p$ varies from $100$ to $2,000$. We provide analogous simulations in Appendix \ref{sec: additional simulations}, where both the random variables and the model parameters are generated from non-Gaussian distributions.

We see that the absolute bias term $|B_b|$ due to confounding is substantially smaller for Doubly Debiased Lasso compared to the standard Debiased Lasso, regardless of which initial estimator is used. This is because $\NT$ additionally removes bias by shrinking large principal components of $X_{-j}$. This spectral transformation helps also to make the absolute bias term $|B_\beta|$ smaller for the Doubly Debiased Lasso compared to the Debiased Lasso, even when using the same initial estimator $\betainit$. This comes however at the expense of slightly larger variance, but we can see that the decrease in bias reflects positively on the validity of the constructed confidence intervals. Their coverage is significantly more accurate for Doubly Debiased Lasso, over a large range of $n$ and $p$. 

There are two challenging regimes for estimation under confounding. Firstly, when the dimension $p$ is much larger than the sample size $n$, the coverage can be lower than $95\%$, since in this regime it is difficult to estimate $\beta$ accurately and thus the term $|B_\beta|$ is fairly large, even after the bias correction step. We see that the absolute bias $|B_\beta|$ grows with $p$, but it is much smaller for the Doubly Debiased Lasso which positively impacts the coverage. Secondly, in the regime where $p$ is relatively small compared to $n$, $|B_b|$ begins to dominate and leads to undercoverage of confidence intervals. $B_b$ is caused by the hidden confounding and does not disappear when $n \to \infty$, while keeping $p$ constant. The simulation results agree with the asymptotic analysis of the bias term in \eqref{eq: bias control} in the Supplementary material, where the term $|B_{b}|$ vanishes as $\lambda_{q}(\Psi)$ increases, in addition to increasing the sample size $n$. In the regime considered in this simulation, $|B_b|$ can even grow, since the bias becomes increasingly large compared to the estimator's variance. However, it is important to note that even in these difficult regimes, Doubly Debiased Lasso performs significantly better than the standard Debiased Lasso (irrespective of the initial estimator) as it manages to additionally decrease the estimator's bias. 

\begin{figure}[htp]
    \hspace{-0.65cm}
    \includegraphics[width=1.04\linewidth]{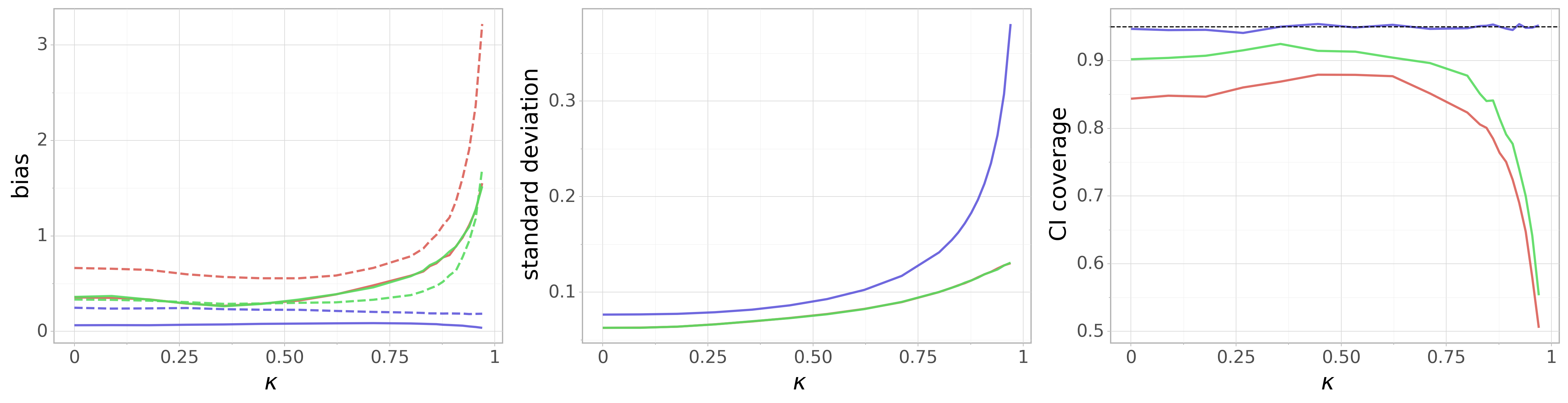}
    \caption{\textit{(Toeplitz covariance for $\Sigma_E$)} Dependence of the (scaled) absolute bias terms $|B_\beta|$ and $|B_b|$ (left), standard deviation $V^{1/2}$ (middle) and the coverage of the $95\%$ confidence interval (right) on the parameter $\kappa$ of the Toeplitz covariance structure. $n=300$ and $p=1,000$ are fixed. On the leftmost plot, $|B_\beta|$ and $|B_b|$ are denoted by a dashed and a solid line, respectively. Blue color corresponds to the Doubly Debiased Lasso, red color represents the standard Debiased Lasso and green color corresponds also to the Debiased Lasso estimator, but with the same $\betainit$ as our proposed method. Note that the last two methods have almost indistinguishable $|B_b|$ and $V$.}
    \label{fig: toeplitz}
\end{figure}

\paragraph*{Toeplitz covariance structure for $\Sigma_E$} 
Now we fix $n=300, p=1,000$, but we generate the covariance matrix $\Sigma_E$ of the unconfounded part of the design matrix $X$ to have Toeplitz covariance structure: $(\Sigma_E)_{i,j} = \kappa^{|i-j|}$, where we vary $\kappa$ across the interval $[0, 0.97]$. As we increase $\kappa$, the covariates $X_1, \ldots, X_5$ in the active set get more correlated, so it gets harder to distinguish their effects on the response and therefore to estimate $\beta$. Similarly, it gets as well harder to estimate $\gamma$ in the regression of $X_j$ on $X_{-j}$, since $X_j$ can be explained well by many linear combinations of the other covariates that are correlated with $X_j$. In Figure \ref{fig: toeplitz} we can see that Doubly Debiased Lasso is much less affected by correlated covariates. The (scaled) absolute bias terms $|B_b|$ and $|B_\beta|$ are much larger for standard Debiased Lasso, which causes the coverage to worsen significantly for values of $\kappa$ that are closer to $1$.

\begin{figure}[htp]
    \hspace{-0.65cm}
    \includegraphics[width=1.04\linewidth]{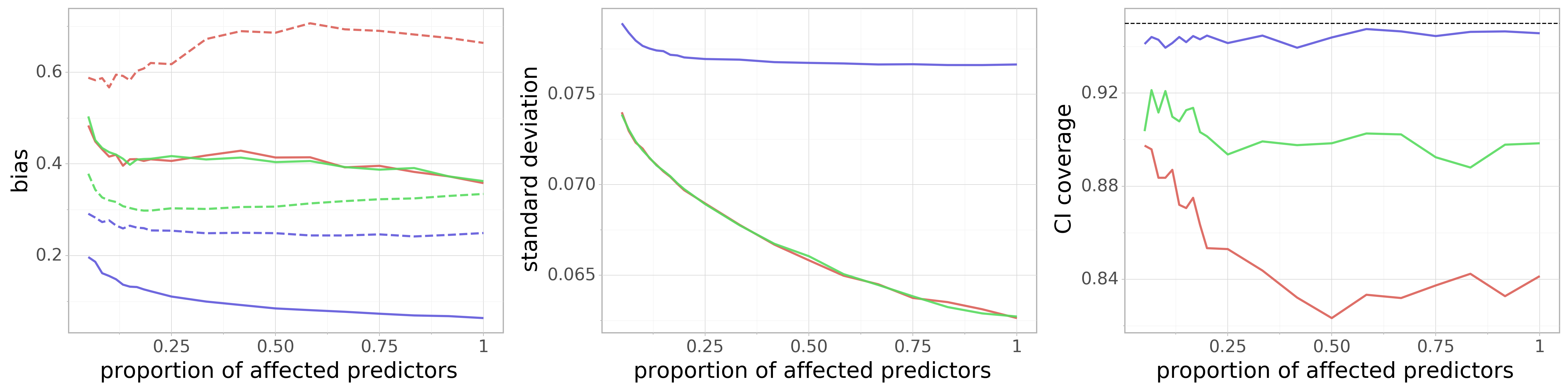}
    \caption{\textit{(Proportion confounded)} Dependence of the (scaled) absolute bias terms $|B_\beta|$ and $|B_b|$ (left), standard deviation $V^{1/2}$ (middle) and the coverage of the $95\%$ confidence interval (right) on proportion of confounded covariates. $n=300$ and $p=1,000$ are fixed. On the leftmost plot, $|B_\beta|$ and $|B_b|$ are denoted by a dashed and a solid line, respectively. Blue color corresponds to the Doubly Debiased Lasso, red color represents the standard Debiased Lasso and green color corresponds also to the Debiased Lasso estimator, but with the same $\betainit$ as our proposed method. Note that the last two methods have almost indistinguishable $|B_b|$ and $V$.}
    \label{fig: proportion}
\end{figure}

\paragraph*{Proportion of confounded covariates}
In order to investigate how the confounding denseness affects the performance of our method, we now again fix $n=300$ and $p=1,000$, but we change the proportion of covariates $X_i$ that are affected by each confounding variable. We do this by setting to zero a desired proportion of entries in each row of the matrix $\Psi \in \R^{q \times p}$, which describes the effect of the confounding variables on each predictor. Its non-zero entries are still generated as $N(0,1)$. We set once again $\Sigma_E = {\rm I}_p$ and we vary the proportion of nonzero entries of $\Psi$ from $5\%$ to $100\%$. The results can be seen in Figure \ref{fig: proportion}. We can see that Doubly Debiased Lasso performs well even when only a very small number ($5\%$) of the covariates are affected by the confounding variables, which agrees with our theoretical discussion for assumption \textbf{(A2)}. We can also see that the coverage of the standard Debiased Lasso is poor even for a small number of affected variables and it worsens as the confounding variables affect more and more covariates. The coverage improves to some extent when we use a better initial estimator, but is still worse than our proposed method.

In Appendix \ref{sec: additional simulations} we also show how the performance changes with the strength of confounding, by gradually decreasing the size of the entries of the loading matrix $\Psi$.

\begin{figure}[htp]
    \hspace{-0.65cm}
    \includegraphics[width=1.04\linewidth]{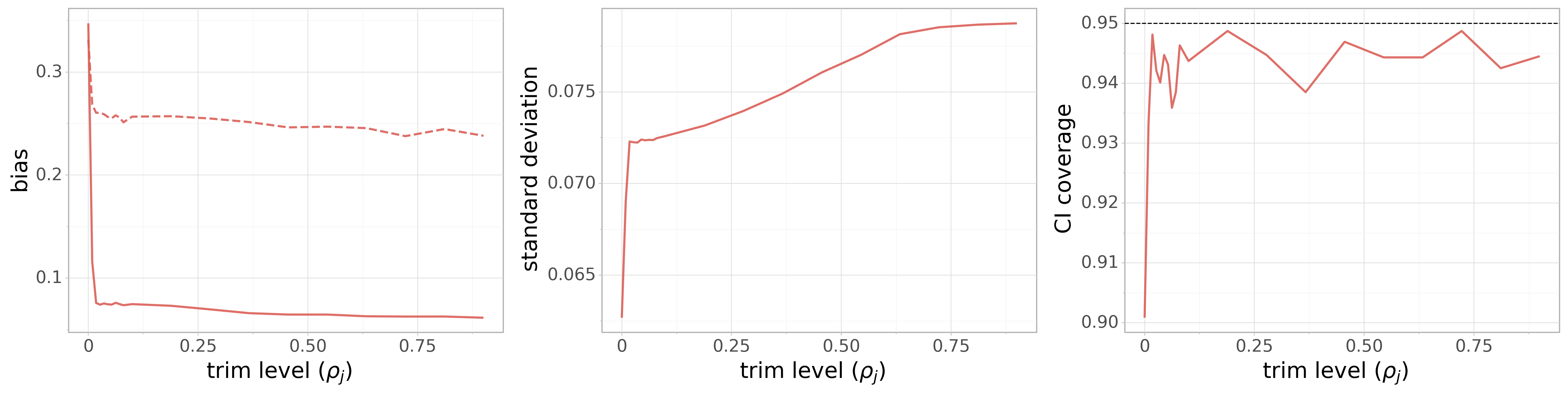}
    \caption{\textit{(Trimming level)} Dependence of the (scaled) absolyte bias terms $|B_\beta|$ and $|B_b|$ (left), standard deviation $V^{1/2}$ (middle) and the coverage of the $95\%$ confidence interval (right) on the trimming level $\rho_j$ of the Trim transform (see Equation (\ref{eq: rho shrinkage})). The sample size is fixed at $n=300$ and the dimension at $p=1,000$. On the leftmost plot, $|B_\beta|$ and $|B_b|$ are denoted by a dashed and a solid line, respectively. The case $\rho_j=0$ corresponds to Debiased Lasso with the spectral deconfounding initial estimator $\betainit$, described in \eqref{eq: est-beta}.}
    \label{fig: trim_level}
\end{figure}

\paragraph*{Trimming level}
We investigate here the dependence of the performance on the choice of the trimming threshold for the Trim transform (\ref{eq: rho shrinkage}), parametrized by the proportion of singular values $\rho_j$ which we shrink. The spectral transformation $\mathcal{Q}$ used for the initial estimator $\betainit$ is fixed to be the default choice of Trim transform with median rule. We fix $n=300$ and $p=1,000$ and consider the same setup as in Figure \ref{fig: n_p_growing}. We take $\tau = \Lambda_{\floor{\rho_j m},\floor{\rho_j m}}$ to be the $\rho_j$-quantile of the set of singular values of the design matrix $X$, where we vary $\rho_j$ across the interval $[0, 0.9]$. When $\rho_j=0$, $\tau$ is the maximal singular value, so there is no shrinkage and our estimator reduces to the standard Debiased Lasso (with the initial estimator $\betainit$). The results are displayed in Figure \ref{fig: trim_level}. We can see that Doubly Debiased Lasso is quite insensitive to the trimming level, as long as the number of shrunken singular values is large enough compared to the number of confounding variables $q$. In the simulation $q=3$ and the (scaled) absolute bias terms $|B_b|$ and $|B_\beta|$ are still small when $\rho_j \approx 0.02$, corresponding to shrinking $6$ largest singular values. We see that the standard deviation decreases as $\rho_j$ decreases, i.e. as the trimming level $\tau$ increases, which matches our efficiency analysis in Section \ref{sec: normal and efficient}. However, we see that the default choice $\tau = \Lambda_{\floor{m/2}, \floor{m/2}}$ has decent performance as well. 
In Appendix \ref{sec: additional simulations} we also explore whether the choice of spectral transformation significantly affects the performance, with a focus on the PCA adjustment, which maps first several singular values to $0$, while keeping the others intact.

\begin{figure}[htp]
    \hspace{-0.65cm}
    \includegraphics[width=1.04\linewidth]{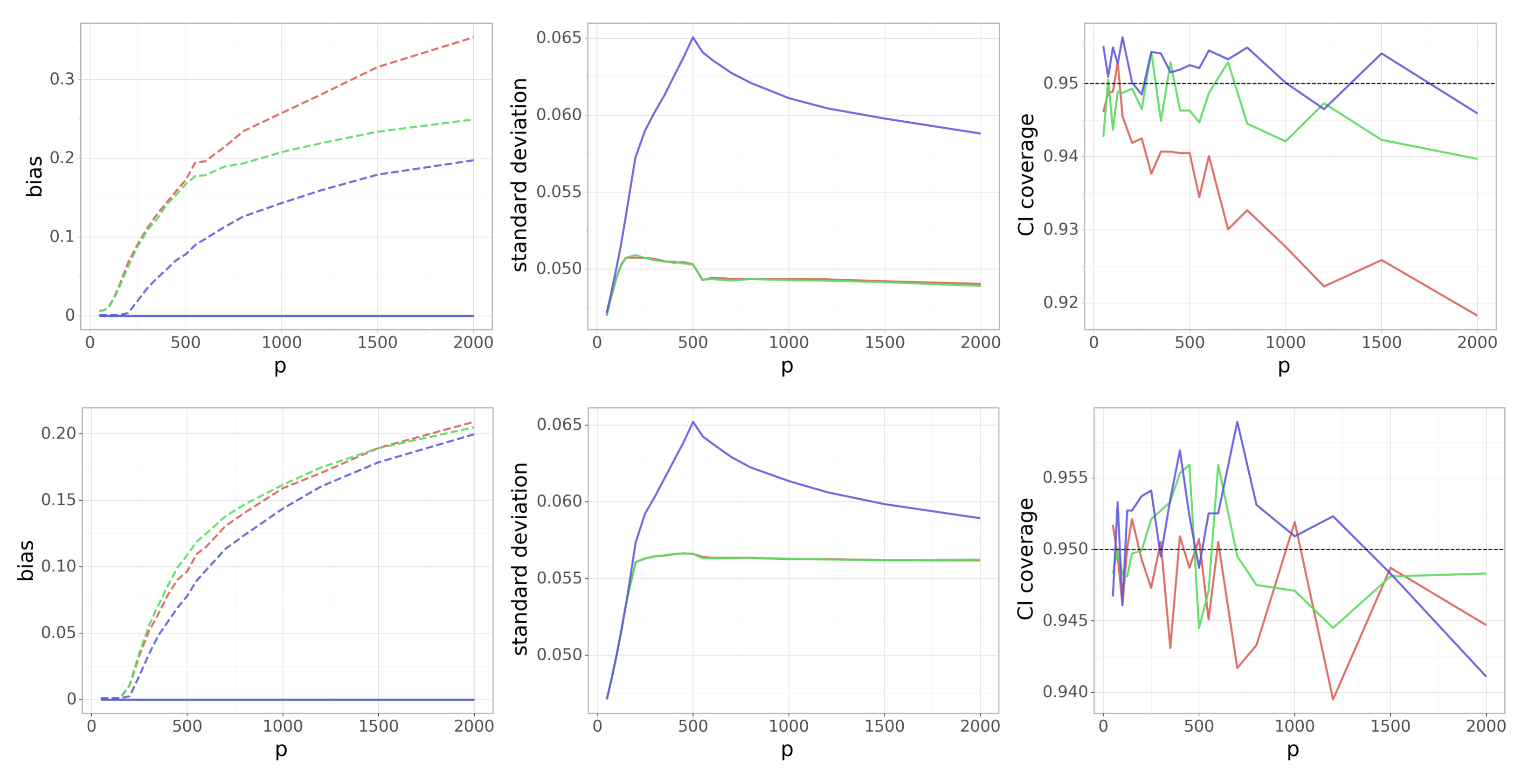}
    \caption{\textit{(No confounding bias)} Dependence of the (scaled) absolute bias terms $|B_\beta|$ and $|B_b|$ (left), standard deviation $V^{1/2}$ (middle) and the coverage of the $95\%$ confidence interval (right) on the number of covariates $p$, while keeping $n=500$ fixed. In the plots on the left, $|B_\beta|$ and $|B_b|$ are denoted by a dashed and a solid line, respectively, but $B_b=0$ since we have enforced $b=0$. Top row corresponds to the spiked covariance case $\Sigma_X = \Psi^T\Psi + I$, whereas for the bottom row we set $\Sigma_X = I$. Blue color corresponds to the Doubly Debiased Lasso, red color represents the standard Debiased Lasso and green color corresponds also to the Debiased Lasso estimator, but with the same $\betainit$ as our proposed method. Note that the last two methods have almost indistinguishable $V$.}
    \label{fig: no_bias}
\end{figure}

\paragraph*{No confounding bias}
We consider now the same simulation setting as in Figure \ref{fig: n_p_growing}, where we fix $n=500$ and vary $p$, but where in addition we remove the effect of the perturbation $b$ that arises due to the confounding. We generate from the model \eqref{eq: confounder model}, but then adjust for the confounding bias: $Y \leftarrow (Y - Xb)$, where $b$ is the induced coefficient perturbation, as in Equation \eqref{eq: perturb}.
In this way we still have a perturbed linear model, but where we have enforced $b=0$ while keeping the same spiked covariance structure of $X$: $\Sigma_X = \Sigma_E + \Psi^{\intercal}\Psi$ as in \eqref{eq: confounder model}. The results can be seen in the top row of Figure \ref{fig: no_bias}. We see that Doubly Debiased Lasso still has smaller absolute bias $|B_\beta|$, slightly higher variance and better coverage than the standard Debiased Lasso, even in absence of confounding. The bias term $B_b$ equals $0$, since we have put $b=0$. We can even observe a decrease in estimation bias for large $p$, and thus an improvement in the confidence interval coverage. This is due to the fact that $X$ has a spiked covariance structure and trimming the large singular values reduces the correlations between the predictors. This phenomenon is also illustrated in the additional simulations in the Appendix \ref{sec: additional simulations}, where we set $q=0$ and put $E$ to have either Toeplitz or equicorrelation covariance structure with varying degree of spikiness (by varying the correlation parameters). 

In the bottom row of Figure \ref{fig: no_bias} we repeat the same simulation, but where we set $q=0$ and take $\Sigma_X = \Sigma_E = I$ in order to investigate the performance of the method in the setting without confounding, but where the covariance matrix of the predictors is not spiked. We see that there is not much difference in the bias and only a slight increase in the variance of our estimator and thus also there is not much difference in the coverage of the confidence intervals. We conclude that our method can provide certain robustness against dense confounding: if there is such confounding, our proposed method is able to significantly reduce the bias caused by it; on the other hand, if there is no confounding, in comparison to the standard Debiased Lasso, our proposed method still has essentially as good performance, with a small increase in variance.

\begin{figure}[htp]
    \hspace{-0.65cm}
    \includegraphics[width=1.04\linewidth]{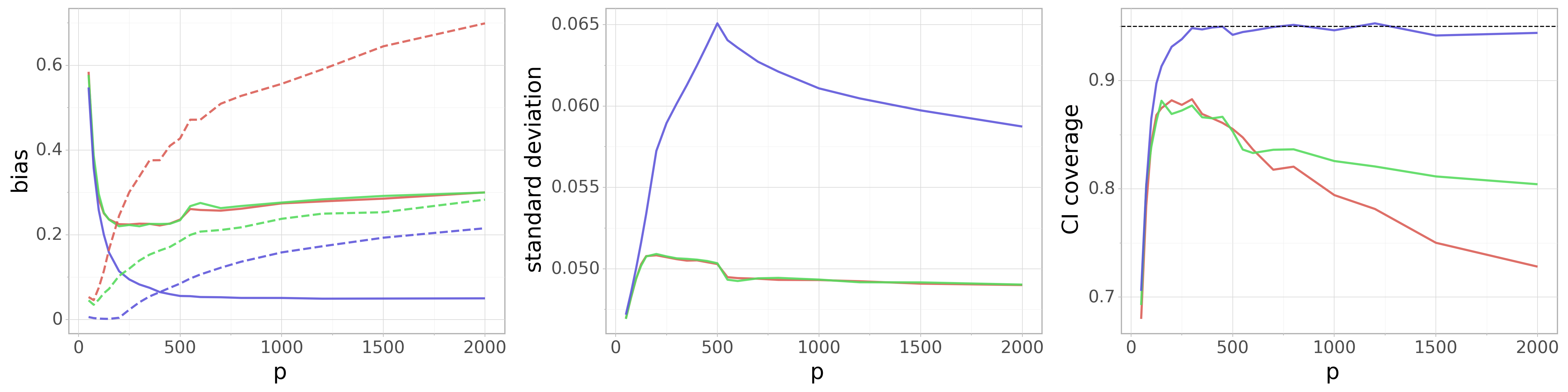}
    \caption{\textit{(Measurement error)} Dependence of the (scaled) absolute bias terms $|B_\beta|$ and $|B_b|$ (left), standard deviation $V^{1/2}$ (middle) and the coverage of the $95\%$ confidence interval (right) on the number of covariates $p$ in the measurement error model (\ref{eq: measurement-error}). The sample size is fixed at $n=500$. On the leftmost plot, $|B_\beta|$ and $|B_b|$ are denoted by a dashed and a solid line, respectively. Blue color corresponds to the Doubly Debiased Lasso, red color represents the standard Debiased Lasso and green color corresponds also to the Debiased Lasso estimator, but with the same $\betainit$ as our proposed method. Note that the last two methods have almost indistinguishable $|B_b|$ and $V$.}
    \label{fig: measurement_error}
\end{figure}

\paragraph*{Measurement error}
We now generate from the measurement error model (\ref{eq: measurement-error}), which can be viewed as a special case of our model \eqref{eq: confounder model}. The measurement error $W = \Psi^{\intercal}
H$ is generated by $q=3$ latent variables $H_{i, \cdot}\in \R^{q}$ for $1\leq i\leq n$. We fix the number of data points to be $n=500$ and vary the number of covariates $p$ from $50$ to $1,000$, as in Figure \ref{fig: n_p_growing}. The results are displayed in Figure \ref{fig: measurement_error}, where we can see a similar pattern as before: Doubly Debiased Lasso decreases the bias at the expense of a slightly inflated variance, which in turn makes the inference much more accurate and the confidence intervals have significantly better coverage.

\subsection{Real data}
\label{sec: real_data}
We investigate here the performance of Doubly Debiased Lasso on a genomic dataset. The data are obtained from the GTEx project \citep{lonsdale2013genotype}, where the gene expression has been measured postmortem on samples coming from various tissue types. 
For our purposes, we use fully processed and normalized gene expression data for the skeletal muscle tissue. The gene expression matrix $X$ consists of measurements of expressions of $p = 12,646$ protein-coding genes for $n = 706$ individuals. Genomic datasets are particularly prone to confounding \citep{leek2007capturing, gagnon2012using, gerard2020empirical}, and for our analysis we are provided with $q=65$ proxies for hidden confounding, 
computed with genotyping principal components and PEER factors. 

We investigate the associations between the expressions of different genes by regressing one target gene expression $X_i$ on the expression of other genes $X_{-i}$. Since the expression of many genes is very correlated, researchers often use just 
$\sim 1,000$ carefully chosen landmark genes as representatives of the whole gene expression \citep{subramanian2017next}. We will use several such landmark genes as the responses in our analysis.

\begin{figure}[htp]
    \hspace{-0.7cm}
    \vspace{-0.35cm}
    \centering
    \includegraphics[scale=0.48]{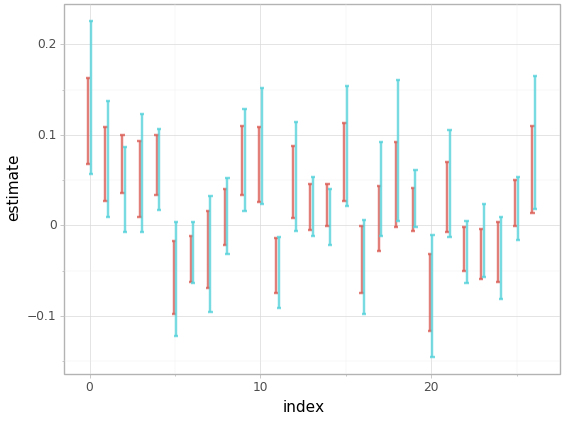}
    \caption{Comparison of $95\%$ confidence intervals obtained by Doubly Debiased Lasso (blue) and Doubly Debiased Lasso (red) for regression of the expression of one target landmark gene on the other gene expressions.}
    \label{fig: confidenceintervals}
\end{figure}

In Figure \ref{fig: confidenceintervals} we can see a comparison of $95\%$-confidence intervals that are obtained from Doubly Debiased Lasso and standard Debiased Lasso. For a fixed response landmark gene $X_i$, we choose $25$ predictor genes $X_j$ where $j\neq i$ such that their corresponding coefficients of the Lasso estimator for regressing $X_i$ on $X_{-i}$ are non-zero. The covariates are ordered according to decreasing absolute values of their estimated Lasso coefficients. We notice that the confidence intervals follow a similar pattern, but that the Doubly Debiased Lasso, besides removing bias due to confounding, is more conservative as the resulting confidence intervals are wider.

\begin{figure}[htp]
    \hspace{-0.6cm}
    \centering
    \includegraphics[scale=0.44]{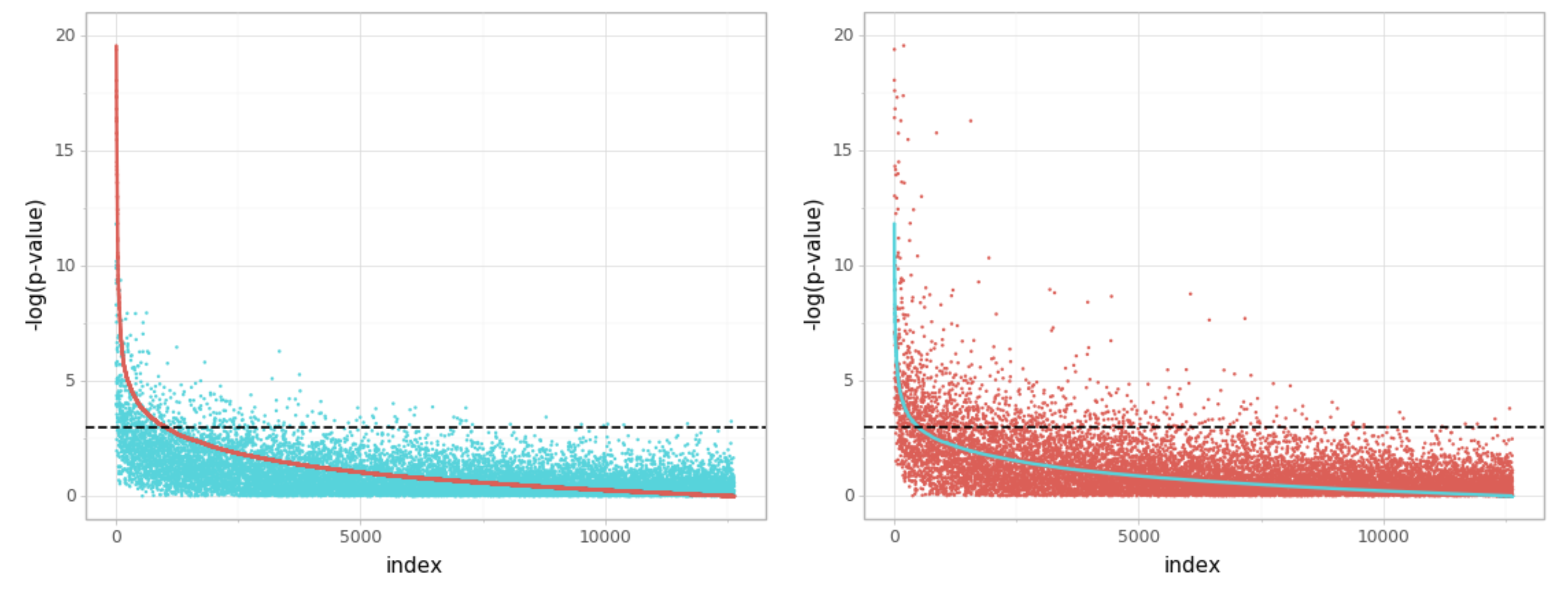}
    \caption{Comparison of p-values for two-sided test of the hypothesis $\beta_j = 0$, obtained by Doubly Debiased Lasso (red) and Doubly Debiased Lasso (blue) for regression of the expression of one target gene on the other gene expressions. The covariates are ordered by decreasing significance, either estimated by the Debiased Lasso (left) or by the Doubly Debiased Lasso (right). Black dotted line indicates the $5\%$ significance level.}
    \label{fig: ordering}
\end{figure}

\begin{figure}[htp]
    \centering
    \includegraphics[scale=0.38]{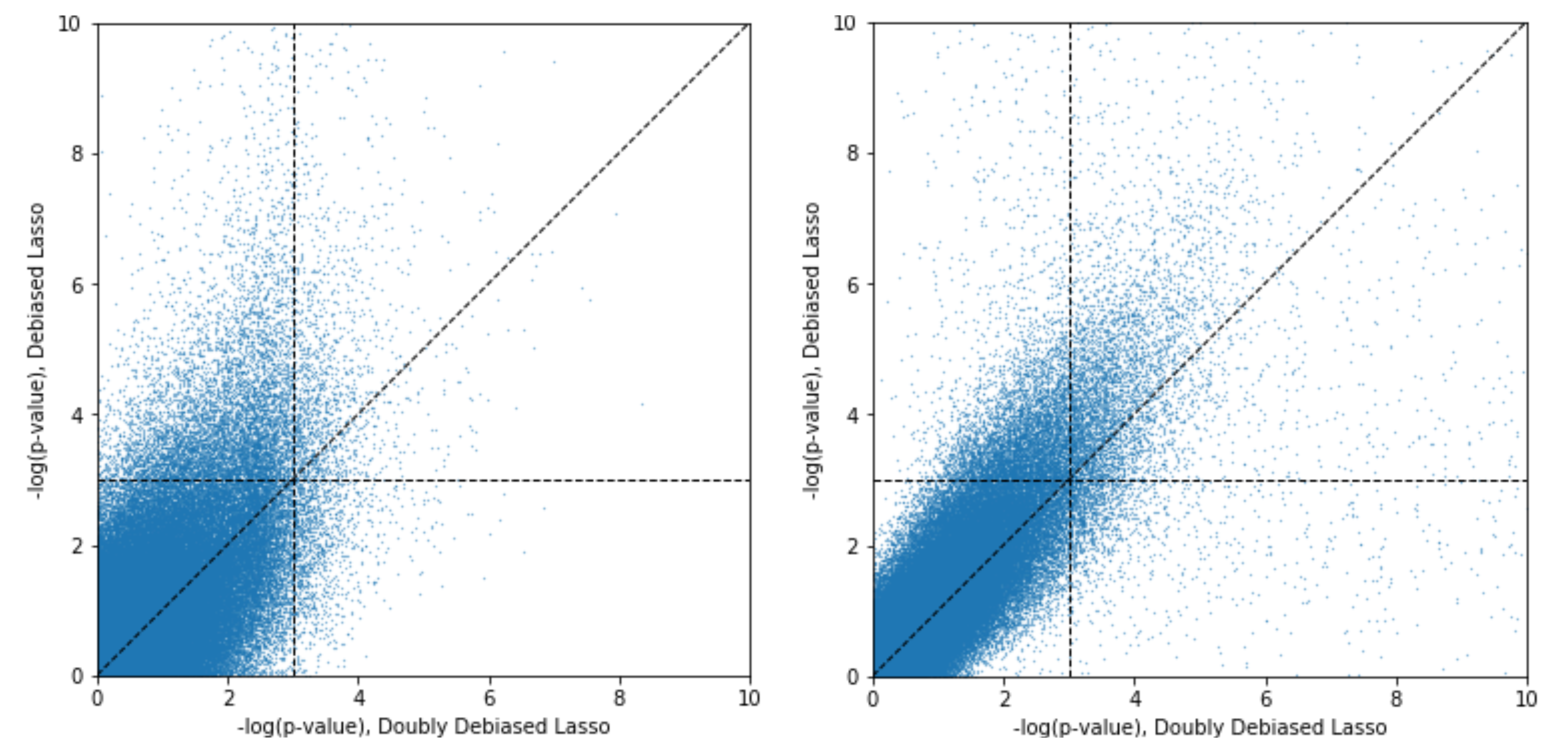}
    \caption{Comparison of p-values for two-sided test of the hypothesis $\beta_j = 0$, obtained by Doubly Debiased Lasso and standard Debiased Lasso for regression of the expression of one target gene on the other gene expressions. The points are aggregated over $10$ landmark response genes. The p-values are either determined using the original gene expression matrix (left) or the matrix where we have regressed out the given $q=65$ confounding proxies (right). Horizontal and vertical black dashed lines indicate the $5\%$ significance level.}
    \label{fig: scatter}
\end{figure}

This behavior becomes even more apparent in Figure \ref{fig: ordering}, where we compare all p-values for a fixed response landmark gene. We see that Doubly Debiased Lasso is more conservative and it declares significantly less covariates significant than the standard Debiased Lasso. Even though the p-values of the two methods are correlated (see also Figure \ref{fig: scatter}), we see that it can happen that one method declares a predictor significant, whereas the other does not.  

\paragraph*{Robustness against hidden confounding}
We now adjust the data matrix $X$ by regressing out the $q=65$ provided hidden confounding proxies. By regressing out these covariates, we obtain an estimate of the unconfounded gene expression matrix $\tilde{X}$. We compare the estimates for the original gene expression matrix with the estimates obtained from the adjusted matrix. 

\begin{figure}[htp]
    \hspace{-0.6cm}
    \centering
    \includegraphics[scale=0.6]{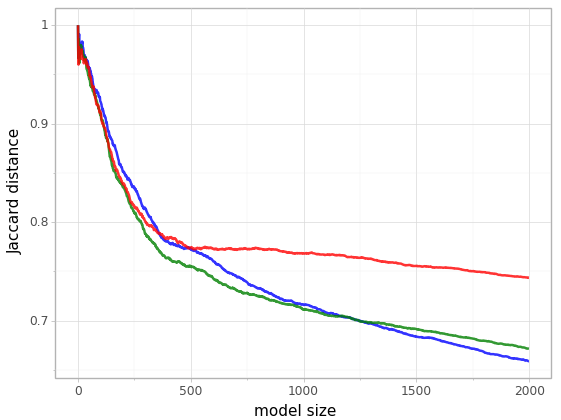}
    \caption{Comparison of the sets of the most significant covariates chosen based on the original expression matrix $X$ and the deconfounded gene expression matrix $\tilde{X}$, for different cardinalities of the sets (model size). The set differences are measured by Jaccard distance. Red line represents the standard Debiased Lasso method, whereas the blue and green lines denote the Doubly Debiased Lasso that uses $\rho=0.5$ and $\rho=0.1$ for obtaining the trimming threshold, respectively; see Equation (\ref{eq: rho shrinkage}).}
    \label{fig: jaccard}
\end{figure}

For a fixed response landmark gene expression $X_i$, we can determine significance of the predictor genes by considering the p-values. One can perform variable screening by considering the set of most significant genes. For Doubly Debiased Lasso and the standard Lasso we compare the sets of most significant variables determined from the gene expression matrix $X$ and the deconfounded matrix $\tilde{X}$. The difference of the chosen sets is measured by the Jaccard distance. A larger Jaccard distance indicates a larger difference between the chosen sets. The results can be seen in Figure \ref{fig: jaccard}. The results are averaged over $10$ different response landmark genes. We see that the Doubly Debiased Lasso gives more similar sets for the large model size, indicating that the analysis conclusions obtained by using Doubly Debiased Lasso are more robust in presence of confounding variables. However, for small model size we do not see large gains. In this case the sets produced by any method are quite different, i.e. the Jaccard distance is very large. This indicates that the problem of determining the most significant covariates is quite difficult, since $X$ and $\tilde{X}$ differ a lot.

In Figure \ref{fig: scatter} we can see the relationship between the p-values obtained by Doubly Debiased Lasso and the standard Debiased Lasso for the original gene expression matrix $X$ and the deconfounded matrix $\tilde{X}$. The p-values are aggregated over $10$ response landmark genes and are computed for all possible predictor genes. We can see from the left plot that the Doubly Debiased Lasso is much more conservative for the confounded data. The cloud of points is skewed upwards showing that the standard Debiased Lasso declares many more covariates significant in presence of the hidden confounding. On the other hand, in the right plot we can see that the p-values obtained by the two methods are much more similar for the unconfounded data and the point cloud is significantly less skewed upwards. The remaining deviation from the $y=x$ line might be due to the remaining confounding, not accounted for by regressing out the given confounder proxies.

\section{Discussion}
We propose the Doubly Debiased Lasso estimator for hypothesis testing and confidence interval construction for single regression coefficients in high-dimensional settings with ``dense'' confounding. We present theoretical and empirical justifications and argue that our double debiasing leads to robustness against hidden confounding. In case of no confounding, the price to be paid is (typically) small, with a small increase in variance but even a decrease in estimation bias, in comparison to the standard Debiased Lasso \citep{zhang2014confidence}; but there can be substantial gain when ``dense" confounding is present. 

It is ambitious to claim significance based on observational data. One always needs to make additional assumptions to guard against confounding. We believe that our robust Doubly Debiased Lasso is a clear improvement over the use of standard inferential high-dimensional techniques, yet it is simple and easy to implement, requiring two additional SVDs only, with no additional tuning parameters {when using} {our default choice of trimming $\rho=\rho_j = 50\%$ of the singular values in Equations \eqref{eq: rho shrinkage} and \eqref{eq: rho shrinkage whole}.}

\section*{Acknowledgements} 
We thank Yuansi Chen for providing the code to preprocess the raw data from the GTEx project. We also thank Matthias L\"{o}ffler for his help and useful discussions about random matrix theory.

\bibliographystyle{imsart-number} 
\bibliography{bibfile}

\newpage
\appendix
\renewcommand{\thefigure}{A\arabic{figure}}
\setcounter{figure}{0}

\begin{center}
\Large Supplementary material for ``Doubly Debiased Lasso:
High-Dimensional Inference under Hidden Confounding''
\end{center}

In Appendix \ref{sec: key results}, we present the proof of Theorem \ref{thm: limiting dist} and important intermediary results for establishing Theorem \ref{thm: limiting dist}. In Appendix \ref{sec: RE verification}, we present the proof of Proposition \ref{prop: RE verification}, which relies on a finite-sample analysis of the factor model. Additional Proofs are presented in Appendix \ref{sec:proofs}.

\section{Intermediary Results and Proof of Theorem \ref{thm: limiting dist}}
\label{sec: key results}
In the following, we list three intermediary  results in Sections \ref{sec: valid transformations} to \ref{sec: estimator rates} as the key components of proving our main result Theorem \ref{thm: limiting dist} and then provide  the proof of Theorem \ref{thm: limiting dist} in Section \ref{sec: proof sketch}. We verify the condition {\rm (A2)} in Section \ref{sec: verify A2}. 
{All our theoretical derivations are done for the Hidden Confounding Model \eqref{eq: confounder model}, but they additionally hold more generally for the perturbed linear model \eqref{eq: perturb}.}
\subsection{Valid spectral transformations}
\label{sec: valid transformations}
The first intermediary result is on the properties of the spectral transformation we use. We will show that 
the limiting distribution in Theorem \ref{thm: limiting dist} holds generally for the estimator \eqref{eq: proposed estimator} using any spectral transformations $\NT $ and $\mathcal{Q}$ that satisfy the following:
\begin{enumerate}
\item[(P1)] {\bf Spectral Transformation Property.}  $\NT =U(X_{-j}) S(X_{-j}) U(X_{-j})^{\intercal}$ and $\mathcal{Q}=U(X) S(X) U(X)^{\intercal}$ satisfy
\begin{align}
\frac{1}{n}\|\NT X_{-j}\|_2^2\lesssim \max\left\{1,\frac{p}{n}\right\} \quad &\text{and} \quad \frac{1}{n}\|\mathcal{Q}X\|_2^2\lesssim \max\left\{1,\frac{p}{n}\right\}
\label{eq: deconfounding}\\
{\rm Tr}[(\NT) ^4]=\sum_{l=1}^{n}[S_{l,l}(X_{-j})]^{4}\gtrsim m \quad &\text{and} \quad {\rm Tr}(\mathcal{Q}^4)=\sum_{l=1}^{n}[S_{l,l}(X)]^{4}\gtrsim m.
\label{eq: lower bound}
\end{align}
with $m=\min\{n,p-1\}$.
\end{enumerate}
The first requirement means that $\NT $ and $\mathcal{Q}$ need to shrink the leading singular values of $X_{-j}$ and $X$  to a sufficiently small level, respectively. 
On the other hand, the second requirement says that the overall shrinkage of all singular values together is not too big.

For the proof of Theorem \ref{thm: limiting dist} and its intermediate results, we extensively use that our spectral transformations satisfy the property (P1). Therefore, we first need to show that the Trim transform $\NT$ defined in \eqref{eq: rho shrinkage} and $\mathcal{Q}$ defined in \eqref{eq: rho shrinkage whole} satisfy the property (P1).
Since $S_{l,l} = 1$ for $l > \floor{\rho m}$, we have that at least $\floor{(1-\rho)m}$ diagonal elements of $S$ are equal to $1$, which immediately gives us \eqref{eq: lower bound} for $\Q$ whenever $\rho < 1$. Similarly, \eqref{eq: lower bound} for $\NT$ holds for any $\rho_j\in (0,1).$ However, in order to show the condition \eqref{eq: deconfounding}, we need to better understand the behaviour of the singular values of the random matrix $X$.

\begin{Proposition} Suppose $E_{i, \cdot}\in \R^{p}$ is a sub-Gaussian random vector and $\lambda_{\max}(\Sigma_{E})\leq C$, for some positive constant $C>0$, then with probability larger than $1-\exp(-cn)$, 
\begin{equation*}
\lambda_{q+1}\left(\tfrac{1}{n}X^\intercal X\right) \lesssim \max\{1,{p}/{n}\},
\end{equation*}
for some positive constant $c>0$.
\label{prop: fund-spectral}
\end{Proposition}
The above proposition is proved in the Section \ref{sec: proof spectral} by applying the Weyl's inequality. This now allows us to conclude that the Trim transform  satisfies the property (P1):

\begin{Corollary}
Let $\NT $ and $\mathcal{Q}$ be the spectral transformation matrices obtained by applying the Trim transformation \eqref{eq: rho shrinkage} and \eqref{eq: rho shrinkage whole}, respectively. Suppose that the conditions of Proposition \ref{prop: fund-spectral} hold and that $\min\{\rho,\rho_j\} \geq (q+1)/\min\{n,p-1\}$ and $\max\{\rho,\rho_j\}<1$. Then the Trim transformations $\NT $ and $\mathcal{Q}$ satisfy {\rm (P1)}.
\label{cor: trim valid}
\end{Corollary}

\subsection{Approximate sparsity and perturbation size}
\label{sec: approximate sparsity}
The essential step of bias correction is to decouple the correlation between the variable of interest $X_{1,j}$ and other covariates $X_{1,-j}\in \R^{p-1}$. In order to get an informative projection direction $\NT Z_j$, one needs to estimate the best linear approximation vector $\gamma = [\E(X_{1,-j}X_{1,-j}^{\intercal})]^{-1}\E(X_{1,-j} X_{1,j})\in \R^{p-1}$ well. Recall that the results for the standard Debiased Lasso \citep{van2014asymptotically} are based on the fact that the sparsity of the precision matrix $\Sigma_X^{-1}$ gives sparsity of $\gamma$, thus justifying the estimation accuracy of the Lasso regression of $X_{1,j}$ on $X_{1, -j}$. However, even though the assumption (A1) ensures the sparsity of the precision matrix of the unconfounded part $E$, $\gamma$ will not be sparse, since the confounding variables $H$ introduce additional correlations between the covariates $X$.

Recall the definitions  
$$\eta_{i,j}=X_{i, j} - X^{\intercal}_{i, -j}\gamma \quad \text{and}\quad \nu_{i,j}=E_{i,j}-E_{i,-j}^{\intercal}\gamma^{E},$$
where $\gamma^{E}= [\E(E_{1,-j}E_{1,-j}^{\intercal})]^{-1}\E(E_{1,-j} E_{1,j}).$

The following Lemma \ref{lem: decomposition lemma} shows that in the  presence of confounding variables, the vector $\gamma$ can be decomposed into a main sparse component $\gamma^{E}$ and an additional small perturbation vector $\gamma^A$. The proof of the following Lemma is presented in Section \ref{sec: proof decomp}.

\begin{Lemma} Suppose that the conditions {\rm (A1)} and {\rm (A2)} hold, then 
the vector $\gamma=[\E(X_{1,-j}X_{1,-j}^{\intercal})]^{-1}\E(X_{1,-j} X_{1,j})$ defined as the minimizer of $\E (X_{1,j}-X_{1,-j}^{\intercal}\gamma')^2$, can be decomposed as $\gamma=\gamma^{E}+\gamma^{A}$, where $\gamma^{E}=[\E (E_{1,-j}E_{1,-j}^{\intercal})]^{-1}\E E_{1,j}E_{1,-j}$ is a sparse vector with at most $s$ non-zero components and the approximation error $\gamma^{A}$ satisfies
\begin{equation}
\|\gamma^{A}\|_2 \leq \max_{1\leq l\leq q} \frac{C_0|\lambda_{l}(\Psi_{-j})|}{c_0\lambda_{l}^2(\Psi_{-j})+1}\|\Psi_j+\Psi_{-j}\gamma^{E}\|_2{\lesssim \frac{\sqrt{q}(\log p)^{1/4}}{\lambda_{q}(\Psi_{-j})}}.
\label{eq: approximation decoup}
\end{equation}
Furthermore, the difference $\delta_{i,j}=\eta_{i,j}-\nu_{i,j}$ satisfies
{
\begin{equation}
{\rm Var}(\delta_{i,j})\lesssim \frac{\|\Psi_{j}-\Psi_{-j}\gamma^{E}\|_2^2}{1+\lambda_{q}^2(\Psi_{-j})}\lesssim \frac{q(\log p)^{1/2}}{1+\lambda_{q}^2(\Psi_{-j})}.
\label{eq: residue upper}
\end{equation}}
\label{lem: decomposition lemma} 
\end{Lemma}

The main component $\gamma^{E}$ is fully determined by the covariance structure of $E_{i, \cdot}$. From the block matrix inverse formula, we get that $\gamma^{E}$ is proportional to $(\Omega_{E})_{j,-j}\in \R^{p-1}$ and therefore sparse with at most $s$ non-zero components. Since the additional component $\gamma^{A}$ converges to zero as in \eqref{eq: approximation decoup}, the regression vector $\gamma$ is approximately sparse.

In a similar fashion, we will show that 
the perturbation $b$ in  \eqref{eq: perturb}, which is induced by the confounding variables, is of a small order of magnitude as well. 
{
\begin{Lemma} Suppose that the conditions {\rm (A1)} and {\rm (A2)} hold, 
then 
\begin{equation}
|b_j| \lesssim \frac{q (\log p)^{1/2}}{1+\lambda_{q}^2(\Psi)}, \quad  \|b\|_2 \lesssim \frac{\sqrt{q}(\log p)^{1/4}}{\lambda_{q}(\Psi)},
\label{eq: b approximation upper}
\end{equation}
and 
\begin{equation}
\left|\sigma_{\epsilon}^2-\sigma_e^2\right|=\left|\phi^{\intercal}\left({\rm I}_q-\Psi\Sigma_{X}^{-1}\Psi^{\intercal}\right)\phi\right|\lesssim \frac{q (\log p)^{1/2}}{1+\lambda_{q}^2(\Psi)}. 
\label{eq: var-diff}
\end{equation}
\label{lem: confounding error}
\end{Lemma}
}
The above lemma also shows that the variance of the error $\epsilon_i$ in \eqref{eq: perturb} is close to that of the random error $e_i$.  
The proof of the above lemma is presented in Section \ref{sec: proof confounding}.

\subsection{Error rates of $\betainit$ and $\widehat{\gamma}$}
\label{sec: estimator rates}
In order to show the asymptotic normality of the proposed Doubly Debiased Lasso estimator \eqref{eq: proposed estimator}, we need that the estimators $\betainit$ and $\widehat{\gamma}$ estimate the target values $\beta$ and $\gamma$ well. 
In the following proposition, we show that the estimator $\widehat{\gamma}$ described in \eqref{eq: est-gamma} accurately estimates $\gamma$ with a high probability. 
The proof of Proposition \ref{prop: estimation-gamma} is presented in Section \ref{sec: proof estimation}.

\begin{Proposition}
Suppose that the conditions ${\rm (A1)-(A4)}$ hold. If the spectral transformation $\NT $ satisfies  ${\rm (P1)}$ and the tuning parameter $\lambda_j$ in  \eqref{eq: est-gamma} is chosen as 
$
\lambda_j\geq A\sigma_{j} \sqrt{\frac{\log p}{n}}+ \sqrt{\frac{q \log p}{1+\lambda_{q}^2(\Psi_{-j})}},
$ 
for some positive constant $A>0$, then with probability larger than $1-e\cdot p^{1-c(A/C_1)^2}-\exp(-c n)-(\log p)^{-1/2}$ for some positive constant $c>0$, the estimator $\widehat{\gamma}$ proposed in \eqref{eq: est-gamma} satisfies
\begin{equation}
\|\widehat{\gamma}-\gamma^{E}\|_1\lesssim \|W_{-j,-j}(\widehat{\gamma}-\gamma^{E})\|_1 \lesssim \frac{\Mq^2}{\tau_*}s\lambda_j+\frac{1}{\lambda_j}\frac{\|\NT  X_{-j}\gamma^{A}\|_2^2}{{n}}, 
\label{eq: accuracy-gamma-1}
\end{equation}
\begin{equation}
\|\widehat{\gamma}-\gamma^{E}\|_2 \lesssim \frac{{\Mq}}{\tau_*}\sqrt{s}\lambda_j+\frac{1}{\lambda_j}\frac{\|\NT  X_{-j}\gamma^{A}\|_2^2}{{n}},
\label{eq: accuracy-gamma-2}
\end{equation}
\begin{equation}
\frac{1}{\sqrt{n}}\|\NT X_{-j}(\widehat{\gamma}-\gamma^{E})\|_2 \lesssim \frac{{\Mq}}{\tau_*} \sqrt{s}\lambda_j+\frac{\|\NT  X_{-j}\gamma^{A}\|_2}{\sqrt{n}},  
\label{eq: prediction-accuracy-gamma}
\end{equation}
where $W\in \R^{p\times p}$ as a diagonal matrix with  diagonal entries as $W_{l,l}={\|\NT X_{\cdot,l}\|_2}/{\sqrt{n}}$ for $1\leq l \leq p$, $\tau_*>0$ is the lower bound for the restricted eigenvalue defined in \eqref{eq: RE-b} and {$\Mq$ is the sub-Gaussian norm for components of $X_{i,.}$, as defined in Assumption (A3).}
\label{prop: estimation-gamma}
\end{Proposition}

Throughout our analysis, we shall choose $\lambda_j$ as
\begin{equation}
\lambda_j\asymp A\sigma_{j} \sqrt{\frac{\log p}{n}}+ \sqrt{\frac{q \log p}{1+\lambda_{q}^2(\Psi_{-j})}},
\label{eq: tuning gamma}
\end{equation}
though Proposition \ref{prop: estimation-gamma} shows that the results also hold for a larger $\lambda_j$.  Furthermore, we combine \eqref{eq: deconfounding} and \eqref{eq: approximation decoup} and establish 
\begin{equation}
\frac{1}{n}\|\NT  X_{-j}\gamma^{A}\|_2^2\lesssim \max\left\{1,\frac{p}{n}\right\}\cdot \frac{q \sqrt{\log p}}{\lambda^2_{q}(\Psi_{-j})}.
\label{eq: approx proj error}
\end{equation}

In addition, we show an analogous result that the initial spectral deconfounding estimator $\betainit$ proposed in \eqref{eq: est-beta} accurately estimates $\beta$ with a high probability:
\begin{Proposition}
Suppose that the conditions ${\rm (A1)-(A4)}$ hold. If the spectral transformation $\mathcal{Q}$ satisfies  ${\rm (P1)}$ and the tuning parameter $\lambda$ in  \eqref{eq: est-beta} is chosen as 
$
\lambda\geq A\sigma_{e} \sqrt{\frac{\log p}{n}}+ \sqrt{\frac{q \log p}{1+\lambda_{q}^2(\Psi)}},
$
for some positive constant $A>0$, then with probability larger than $1-e\cdot p^{1-c(A/C_1)^2}-\exp(-c n)-(\log p)^{-1/2}$ for some positive constant $c>0$, the estimator $\betainit$ proposed in \eqref{eq: est-beta} satisfies
\begin{equation}
\|\betainit-\beta\|_1 \lesssim \|\widetilde{W}(\betainit-\beta)\|_1  \leq \frac{\Mq^2}{\tau_*} k \lambda+\frac{1}{\lambda}\frac{\|\mathcal{Q} X b\|^2_2}{{n}}, 
\label{eq: accuracy-beta-1}
\end{equation}
\begin{equation}
\|\betainit-\beta\|_2 \leq \frac{\Mq}{\tau_*} \sqrt{k}\lambda+\frac{1}{\lambda}\frac{\|\mathcal{Q} X b\|^2_2}{{n}}, \label{eq: accuracy-beta-2}
\end{equation}
\begin{equation}
\frac{1}{\sqrt{n}}\|\mathcal{Q}X(\betainit-\beta)\|_2 \leq \frac{\Mq}{\tau_*} \sqrt{k}\lambda+\frac{\|\mathcal{Q} X b\|_2}{\sqrt{n}}, 
\label{eq: prediction-accuracy-beta}
\end{equation} 
where $\widetilde{W}\in \R^{p\times p}$ as a diagonal matrix with  diagonal entries as $\widetilde{W}_{l,l}={\|\mathcal{Q} X_{\cdot,l}\|_2}/{\sqrt{n}}$ for $1\leq l \leq p$, $\tau_*>0$ is the lower bound for the restricted eigenvalue defined in \eqref{eq: RE-a} and {$\Mq$ is the sub-Gaussian norm for components of $X_{i,.}$, as defined in Assumption (A3).}.
\label{prop: estimation-beta-refined}
\end{Proposition}
This extends the results in \citep{cevid2018spectral}, where only the rate of convergence of $\|\betainit-\beta\|_1$ has been established, but not of $\|\betainit-\beta\|_2$ and $\frac{1}{\sqrt{n}}\|\mathcal{Q}X(\betainit-\beta)\|_2$ {and furthermore, the assumption (A2) is weaker than the assumption $\lambda_{q}(\Psi)\gtrsim \sqrt{p}$ required in Theorem 1 of \citep{cevid2018spectral}}. 
The proof of Proposition \ref{prop: estimation-beta-refined} is presented in Section \ref{sec: proof estimation refined}. We shall choose $$\lambda\asymp A\sigma_{e} \sqrt{\frac{\log p}{n}}+ \sqrt{\frac{q \log p}{1+\lambda_{q}^2(\Psi)}},$$ though Proposition \ref{prop: estimation-beta-refined} shows that the results also hold for a larger $\lambda$.  Furthermore, similar to \eqref{eq: approx proj error}, we combine \eqref{eq: deconfounding} and \eqref{eq: b approximation upper} and establish 
\begin{equation}
\frac{1}{n}\|\mathcal{Q} Xb\|_2^2\lesssim \max\left\{1,\frac{p}{n}\right\}\cdot \frac{q \sqrt{\log p}}{\lambda^2_{q}(\Psi)}.
\label{eq: approx proj error Q}
\end{equation}
As a remark, if we further assume the error $\epsilon_i$ in the model \eqref{eq: perturb} to be independent of $X_{i, \cdot},$ then we can take $\lambda= A \sigma_{\epsilon} \sqrt{{\log p}/{n}}$ and establish a slightly better rate of convergence.

\subsection{Proof of Theorem \ref{thm: limiting dist}}
\label{sec: proof sketch}
We write $$V = \frac{Z_j^{\intercal}(\NT) ^4Z_j \cdot \sigma_e^2}{(Z_j^{\intercal} (\NT) ^2 X_j)^2} \quad \text{and}\quad \sqrt{V} = \frac{\sqrt{Z_j^{\intercal}(\NT) ^4Z_j \cdot \sigma_e^2}}{Z_j^{\intercal} (\NT) ^2 X_j}.$$
Note that the following limiting result \eqref{eq: variance limit 1} shows that $Z_j^{\intercal} (\NT) ^2 X_j$ converges to a positive value in probability.  From the equation (\ref{eq: error decomp}), we have the following expression 
\begin{equation}
\frac{1}{\sqrt{V}}(\widehat{\beta}_j - \beta_j)=\frac{1}{\sqrt{V}}\frac{(\NT Z_{j})^{\intercal}\NT  \epsilon}{(\NT Z_{j})^{\intercal}\NT X_j}+ B_{\beta} + B_b,
\label{eq: scaled decomp}
\end{equation}
where $B_\beta$ and $B_b$ are the (scaled) bias terms defined as 
$$B_\beta = \frac{Z_j^{\intercal} (\NT) ^2 X_{-j}(\betainit_{-j} - \beta_{-j})}{\sqrt{Z_j^{\intercal}(\NT) ^4Z_j \cdot \sigma_e^2}} \quad\text{and} \quad B_b = \frac{Z_j^{\intercal}(\NT) ^2Xb}{\sqrt{Z_j^{\intercal}(\NT) ^4Z_j \cdot \sigma_e^2}}.$$ 

We decompose 
\begin{equation*}
\frac{1}{\sqrt{V}}\frac{(\NT Z_{j})^{\intercal}\NT  \epsilon}{(\NT Z_{j})^{\intercal}\NT X_j}=\frac{1}{\sqrt{V}}\frac{(\NT Z_{j})^{\intercal}\NT  \err}{(\NT Z_{j})^{\intercal}\NT X_j}+\frac{1}{\sqrt{V}}\frac{(\NT Z_{j})^{\intercal}\NT  \Delta}{(\NT Z_{j})^{\intercal}\NT X_j}
\end{equation*}
with $\Delta_i=\psi^{\intercal}H_{i, \cdot}-b^{\intercal}X_{i, \cdot}$ for $1\leq i\leq n.$
Since $e_i$ is Gaussian and independent of $X_{i, \cdot}$ and $Z_j$ is a function of $X$, we establish
\begin{equation}
\frac{1}{\sqrt{V}}\frac{(\NT Z_{j})^{\intercal}\NT  \err}{(\NT Z_{j})^{\intercal}\NT X_j} \mid X \sim N(0,1).
\label{eq: var normal}
\end{equation}
It follows from Lemma \ref{lem: confounding error} that 
\begin{equation}
\frac{1}{n}\E \|\Delta\|_2^2=\E |\Delta_i|^2=\phi^{\intercal}\left({\rm I}_q-\Psi\Sigma_{X}^{-1}\Psi^{\intercal}\right)\phi\lesssim \frac{q\sqrt{\log p}}{1+\lambda_{q}^2(\Psi)}.
\label{eq: approximation expectation}
\end{equation}
By Cauchy inequality, we have 
$$\left|\frac{1}{\sqrt{V}}\frac{(\NT Z_{j})^{\intercal}\NT  \Delta}{(\NT Z_{j})^{\intercal}\NT X_j}\right|\leq \frac{1}{\sigma_e^2}{\|\Delta\|_2}.$$ 
Combined with \eqref{eq: approximation expectation}, we establish that, with probability larger than $1-(\log p)^{-1/2},$
\begin{equation}
\left|\frac{1}{\sqrt{V}}\frac{(\NT Z_{j})^{\intercal}\NT  \Delta}{(\NT Z_{j})^{\intercal}\NT X_j}\right|\lesssim \sqrt{\frac{n q \log p}{1+\lambda_{q}^2(\Psi)}}.
\label{eq: approx error in limiting}
\end{equation}
 If $\lambda_{q}^2(\Psi)\gg \max\{1,qn \log p\},$ we combine \eqref{eq: var normal} and \eqref{eq: approx error in limiting} and establish
\begin{equation}
\frac{1}{\sqrt{V}}\frac{(\NT Z_{j})^{\intercal}\NT  \epsilon}{(\NT Z_{j})^{\intercal}\NT X_j} \cid N(0,1).
\label{eq: limiting normal}
\end{equation}

We establish in the following lemma that $B_b$ and $B_{\beta}$ converges to $0$ in probability under certain model conditions. The proof of this lemma is presented in Section \ref{sec: proof limiting}. The proof relies on our established intermediary results: Corollary \ref{cor: trim valid}, Lemmas \ref{lem: decomposition lemma} and \ref{lem: confounding error}, and Propositions \ref{prop: estimation-gamma} and \ref{prop: estimation-beta-refined}.

\begin{Lemma}
Suppose that the conditions of Theorem \ref{thm: limiting dist} hold.  Then we have
\begin{equation}
\frac{(\NT Z_{j})^{\intercal}\NT X_j}{{{\rm Tr}[(\NT) ^2]}\sigma_{j}^2}\cip 1 
\label{eq: variance limit 1}
\end{equation}
\begin{equation}
\frac{Z_{j}^{\intercal}(\NT) ^4 Z_{j}}{{{\rm Tr}[(\NT) ^4]}\sigma_{j}^2} \cip 1 
\label{eq: variance limit 2}
\end{equation}
\begin{equation}
B_{\beta} \cip 0 \qquad
B_{b} \cip 0.
\label{eq: bias control}
\end{equation}
\label{lem: limiting dist} 
\end{Lemma}

By the decomposition \eqref{eq: scaled decomp} together with \eqref{eq: limiting normal} and \eqref{eq: bias control}, we establish the limiting distribution in \eqref{eq: limiting}. The asymptotic expression of the variance ${\rm V}$ in \eqref{eq: var} follows from \eqref{eq: variance limit 1} and \eqref{eq: variance limit 2}.

\subsection{Verification of Assumption {\rm (A2)}}
\label{sec: verify A2}
In the following, we verify the condition {\rm (A2)} for a general class of models, whose proof can be found in Section \ref{sec: proof dense confounding}. 
\begin{Lemma}
Suppose that $\{\Psi_{\cdot,l}\}_{1\leq l\leq p}$ are generated as i.i.d. $q$-dimensional sub-Gaussian random vectors with mean zero and covariance $\Sigma_{\Psi}\in \R^{q\times q}$. If $q\ll p$,  $\lambda_{\max}(\Sigma_{\Psi})/\lambda_{\min}(\Sigma_{\Psi})\leq C$  and $\|\phi\|_{\infty}/\lambda_{\min}(\Sigma_{\Psi})\leq C$ for some positive constant $C>0$, then with probability larger than $1-(\log p)^{2c}$, we have  
\begin{equation}
\lambda_{q}(\Psi)\geq \lambda_{q}(\Psi_{-j})\gtrsim \sqrt{p} \sqrt{\lambda_{\min}(\Sigma_{\Psi})}
\label{eq: result 1}
\end{equation}
\begin{equation}\max\left\{\|\Psi (\Omega_{E})_{\cdot, j}\|_{2},\|\Psi_j\|_2,\|\Psi_{-j}(\Omega_{E})_{-j, j}\|_2,\|\phi\|_2\right\}\lesssim \sqrt{\lambda_{\max}(\Sigma_{\Psi})}\cdot \sqrt{q} (\log p)^{c},
\label{eq: result 2}
\end{equation}
where $c>0$ is a positive constant. 
\label{lem: dense confounding}
\end{Lemma}

The conclusion of Lemma \ref{lem: dense confounding} can be generalized to hold if a fixed proportion of the $p$ columns of $\Psi$ are i.i.d. sub-Gaussian in $\R^{q}$. This generalized result is stated in the following lemma, whose proof is presented in Section \ref{sec: general dense confounding}: 
\begin{Lemma}
Suppose that there exists a set $A\subseteq \{1,2,\ldots,p\}$ such that $\{\Psi_{\cdot,l}\}_{l\in A}$ are generated as i.i.d sub-Gaussian random vector with mean zero and covariance $\Sigma_{\Psi}\in \R^{q\times q}$ and $\{\Psi_{\cdot,l}\}_{l\in A^{c}}$ are generated as independent $q$-dimensional sub-Gaussian random vectors with sub-Gaussian norm $C_1$. If $\max\{C_1,\lambda_{\max}(\Sigma_{\Psi})\}/\lambda_{\min}(\Sigma_{\Psi})\leq C,$ $\|\psi\|_{\infty}/\lambda_{\min}(\Sigma_{\Psi})\leq C$ and $\max\{C_1,\lambda_{\max}(\Sigma_{\Psi})\}\leq C$ for some positive constant $C>0$ and $|A|$ satisfies \begin{equation}
|A|\gg q\quad \text{and}\quad
|A|\gg \max\left\{\sqrt{\frac{qp}{n}}(\log p)^{3/4},\sqrt{qn\log p},  q^{3/2}(\log p)^{3/4}\right\},
\label{eq: confounding number condition}
\end{equation}
then the assumption ${\rm (A2)}$ holds with probability larger than $1-(\log p)^{2c}$. 
\label{lem: dense confounding general}
\end{Lemma}

\section{Proof of Proposition \ref{prop: RE verification}}
\label{sec: RE verification}

We express the hidden confounding model as 
\begin{equation}
X_{n\times p}=D_{n\times p}+E_{n\times p} \quad \text{with}\quad D_{n\times p}=H_{n\times q} \Psi_{q\times p}.
\label{eq: matrix factor}
\end{equation}
For a given $q$, a natural way to estimate $\Psi$ and $H$ is to solve the optimization problem 
$
\argmin_{H\in \R^{n\times q}, \Psi\in \R^{q\times p}}\|X-H\Psi\|_{F}^2,
$ where $\|\cdot\|_{F}$ denotes the matrix Frobenius norm. 
Since the solution of this optimization problem is not unique, we introduce an additional constraint $H^{\intercal} H/n={\rm I}_{q}$ for the parameter identification. Then the minimizer is defined as
\begin{equation*}
\begin{aligned}
(\widetilde{H},\widetilde{\Psi})&=\argmin_{H\in \R^{n\times q}, \Psi\in \R^{q\times p},H^{\intercal} H/n={\rm I}_{q}}\|X-H\Psi\|_{F}^2\\
&=\argmin_{H\in \R^{n\times q}, \Psi\in \R^{q\times p},H^{\intercal} H/n={\rm I}_{q}}-2 {\rm Tr}(\Psi^{\intercal} H^{\intercal} X)+n{\rm Tr}(\Psi^{\intercal}\Psi).
\end{aligned}
\end{equation*}
We compute the derivative of $-2 {\rm Tr}(\Psi^{\intercal} H^{\intercal} X)+n{\rm Tr}(\Psi^{\intercal}\Psi)$ with respect to $\Psi$ and set it to be zero. Then we obtain the solution 
\begin{equation}
\frac{1}{n}\widetilde{H}^{\intercal} X=\widetilde{\Psi}\quad \text{with}\quad
\widetilde{H}=\argmax_{H\in \R^{n\times q},H^{\intercal} H/n={\rm I}_{q}} {\rm Tr}(H^{\intercal} X X^{\intercal} H).
\label{eq: factor def}
\end{equation}
That is, the columns of $\widetilde{H}\in \R^{n\times q}$ are $\sqrt{n}$ times the first $q$ eigenvectors, corresponding to the top $q$ eigenvalues of $X X^{\intercal}\in \R^{n\times n}.$ 
Then the PCA adjusted covariates are defined as 
\begin{equation*}
\widetilde{X}^{\rm PCA}=X-\widetilde{D} \quad \text{with}\quad \widetilde{D}=\widetilde{H}\widetilde{\Psi}.
\end{equation*}
That is, we remove from $X$ the eigen-decomposition corresponding to the top $q$ eigenvalues, which is denoted as $\widetilde{D}.$ Define $R=D-\widetilde{D}.$ Then we have $\widetilde{X}^{\rm PCA}=R+E$ and  
\begin{equation*}
\frac{1}{n}(\widetilde{X}^{\rm PCA})^{\intercal}\widetilde{X}^{\rm PCA}-\Sigma_{E}=\left(\frac{1}{n} E^{\intercal}E-\Sigma_{E}\right)+\frac{1}{n}R^{\intercal} E+\frac{1}{n}E^{\intercal} R+\frac{1}{n}R^{\intercal}R.
\end{equation*}
We further have 
\begin{equation}
\begin{aligned}
&\min_{\|\omega_{\Sb^c}\|_1\leq \Cq\cdot \|\omega_{\Sb}\|_1, \|\omega\|_2=1} \omega^{\intercal}\left(\frac{1}{n}(\widetilde{X}^{\rm PCA})^{\intercal}\widetilde{X}^{\rm PCA}-\Sigma_{E}\right)\omega\\
&\geq \min_{\|\omega_{\Sb^c}\|_1\leq \Cq\cdot  \|\omega_{\Sb}\|_1, \|\omega\|_2=1} \omega^{\intercal}\left(\frac{1}{n} E^{\intercal}E-\Sigma_{E}\right)\omega\\
&-\max_{\|\omega_{\Sb^c}\|_1\leq \Cq \cdot \|\omega_{\Sb}\|_1, \|\omega\|_2=1} \omega^{\intercal}\frac{2}{n}R^{\intercal} E\omega-\max_{\|\omega_{\Sb^c}\|_1\leq \Cq \cdot \|\omega_{\Sb}\|_1, \|\omega\|_2=1} \omega^{\intercal}\frac{1}{n}R^{\intercal}R \omega.
\end{aligned}
\label{eq: RE decomposition}
\end{equation}
In the following, we shall control the three terms on the right-hand-side of \eqref{eq: RE decomposition}.

Note that Theorem 1.6 in \citep{zhou2009restricted} (with $k_0$ in this theorem taken as $\Cq$) implies that, if $$n\gtrsim \Mq^2 \frac{k \log p}{n},$$
then with probability larger than $1-p^{-c}$ for some positive constant $c>0,$
$$\max_{\|\omega_{\Sb^c}\|_1\leq \Cq \cdot \|\omega_{\Sb}\|_1, \|\omega\|_2=1}\left|\sqrt{\frac{\omega^{\intercal}\frac{1}{n}E^{\intercal}E\omega}{\omega^{\intercal}\Sigma_{E}\omega}}-1\right|\leq 0.1.$$
That is, there exists a positive constant $C'>0$ such that 
\begin{equation}
 \min_{\|\omega_{\Sb^c}\|_1\leq \Cq\cdot  \|\omega_{\Sb}\|_1, \|\omega\|_2=1}\omega^{\intercal}\frac{1}{n}E^{\intercal}E\omega\geq 0.9\cdot \lambda_{\min}(\Sigma_{E}),
 \label{eq: RE existing lower}
 \end{equation}
 and 
\begin{equation}
\max_{\|\omega_{\Sb^c}\|_1\leq \Cq \cdot \|\omega_{\Sb}\|_1, \|\omega\|_2=1}\omega^{\intercal}\frac{1}{n}E^{\intercal}E\omega\leq 1.1\cdot \lambda_{\max}(\Sigma_{E}).
 \label{eq: RE existing upper}
\end{equation}

Now we turn to $\omega^{\intercal}\frac{1}{n}R^{\intercal}R\omega.$ 
Fix $\Sb\subseteq [p]$ with $|\Sb|\leq k$. Then we have 
\begin{equation}
\begin{aligned}
\max_{\|\omega_{\Sb^c}\|_1\leq \Cq \cdot \|\omega_{\Sb}\|_1, \|\omega\|_2=1 }\omega^{\intercal}\frac{1}{n}R^{\intercal}R\omega &\leq \max_{\|\omega_{\Sb^c}\|_1\leq \Cq \cdot \|\omega_{\Sb}\|_1, \|\omega\|_2=1 }\max_{1\leq l\leq n}(\sum_{j=1}^{p}R_{l,j}\omega_{j})^2\\
&\leq \max_{\|\omega_{\Sb^c}\|_1\leq \Cq \cdot \|\omega_{\Sb}\|_1, \|\omega\|_2=1 } (\|R\|_{\infty} \|\omega\|_1)^2\\
&\leq \max_{\|\omega_{\Sb^c}\|_1\leq \Cq \cdot \|\omega_{\Sb}\|_1, \|\omega\|_2=1 } (\|R\|_{\infty} (1+\Cq)\|\omega_{\Sb}\|_1)^2\\
&\leq \max_{\|\omega_{\Sb^c}\|_1\leq \Cq \cdot \|\omega_{\Sb}\|_1, \|\omega\|_2=1 } (\|R\|_{\infty} (1+\Cq)\sqrt{k}\|\omega_{\Sb}\|_2)^2\\
&\lesssim  \Mq^2\cdot k\|R\|_{\infty}^2.
\end{aligned}
\label{eq: RE simplification}
\end{equation}

By combining \eqref{eq: RE existing upper} and \eqref{eq: RE simplification}, we establish 
\begin{equation}
\begin{aligned}
&\max_{\|\omega_{\Sb^c}\|_1\leq \Cq \cdot \|\omega_{\Sb}\|_1, \|\omega\|_2=1 }\omega^{\intercal}\frac{1}{n}R^{\intercal}E\omega\\
&\leq \max_{\|\omega_{\Sb^c}\|_1\leq \Cq \cdot \|\omega_{\Sb}\|_1, \|\omega\|_2=1} \sqrt{\frac{1}{n}\omega^{\intercal}R^{\intercal}R\omega}\cdot \sqrt{\frac{1}{n}\omega^{\intercal}E^{\intercal}E\omega}\\
&\leq  \sqrt{\max_{\|\omega_{\Sb^c}\|_1\leq \Cq \cdot \|\omega_{\Sb}\|_1, \|\omega\|_2=1} \frac{1}{n}\omega^{\intercal}R^{\intercal}R\omega}\cdot \sqrt{\max_{\|\omega_{\Sb^c}\|_1\leq \Cq \cdot \|\omega_{\Sb}\|_1, \|\omega\|_2=1} \frac{1}{n}\omega^{\intercal}E^{\intercal}E\omega}\\
&\lesssim \sqrt{\Mq^2\cdot k\|R\|_{\infty}^2}.
\end{aligned}
\label{eq: RE inter}
\end{equation}
By the decomposition in \eqref{eq: RE decomposition} and the bounds in \eqref{eq: RE existing lower}, \eqref{eq: RE simplification} and \eqref{eq: RE inter}, we establish  
\begin{equation*}
\begin{aligned}
&\min_{\|\omega_{\Sb^c}\|_1\leq \Cq\cdot  \|\omega_{\Sb}\|_1, \|\omega\|_2=1} \omega^{\intercal}\left(\frac{1}{n}(\widetilde{X}^{\rm PCA})^{\intercal}\widetilde{X}^{\rm PCA}-\Sigma_{E}\right)\omega\\
&\geq 0.9\cdot \lambda_{\min}(\Sigma_{E})-C\sqrt{\Mq^2 \cdot k\|R\|_{\infty}^2}-C\Mq^2\cdot k\|R\|_{\infty}^2 ,
\end{aligned}
\end{equation*}
where $C$ is a positive constant independent of $n$ and $p$. If 
\begin{equation}
n\gtrsim \Mq^2\cdot\frac{k \log p}{n} \quad \text{and}\quad \Mq \cdot \sqrt{k} \|R\|_{\infty} \rightarrow 0,
\label{eq: dim condition}
\end{equation}
we establish that, for a sufficiently large $n,$ there exists a small positive constant $0<c<0.9$ independent of $n$ and $p$ such that 
$${\rm RE}\left(\frac{1}{n}(\widetilde{X}^{\rm PCA})^{\intercal}\widetilde{X}^{\rm PCA}\right)\geq c \lambda_{\min}(\Sigma_{E}).$$

By the Weyl's inequality for singular values, $$\left|\lambda_{l}(X)-\lambda_{l}(D)\right|=\left|\lambda_{l}(D+E)-\lambda_{l}(D)\right|\leq \|E\|_2 \quad \text{for}\quad 1\leq l\leq p.$$
We then apply Theorem 5.39 of \citep{vershynin2010introduction} and establish that, with probability larger than $1-p^{-c}$ for some positive constant $c>0,$
\begin{equation*}
\left|\lambda_{l}(X)-\lambda_{l}(D)\right|\leq \|E\|_2 \lesssim \sqrt{n}+\sqrt{p} \quad \text{for}\quad 1\leq l\leq p.
\end{equation*}
Since $D$ is of rank $q$, then $\lambda_{q+1}(D)=0$ and hence 
\begin{equation}
\left|\lambda_{q+1}(X)\right|  \lesssim \sqrt{n}+\sqrt{p} \quad\text{and}\quad \left|\lambda_{q+1}\left(\frac{1}{n}X X^{\intercal}\right)\right|  \lesssim \max\left\{1,\frac{p}{n}\right\}.
\label{eq: bound on q+1 eigenvalue}
\end{equation}

Recall that $X=\sum_{j=1}^{m} \Lambda_{j,j} U_{\cdot,j} V_{\cdot,j}^{\intercal}.$
For any $\omega\in \R^{p}$ and $\lfloor \rho m \rfloor\geq q+1$, we have 
\begin{equation*}
\begin{aligned}
\omega^{\intercal}X^{\intercal}\mathcal{Q}^2X\omega&= \omega^{\intercal}\sum_{j=1}^{\lfloor \rho m \rfloor} \Lambda^2_{\lfloor \rho m \rfloor,\lfloor \rho m \rfloor}V_{\cdot,j} V_{\cdot,j}^{\intercal}\omega+\omega^{\intercal}\sum_{j=\lfloor \rho m \rfloor+1}^{m} \Lambda^2_{j,j}V_{\cdot,j} V_{\cdot,j}^{\intercal}\omega\\
&\geq \omega^{\intercal}\sum_{j=q+1}^{\lfloor \rho m \rfloor} \Lambda^2_{\lfloor \rho m \rfloor,\lfloor \rho m \rfloor}V_{\cdot,j} V_{\cdot,j}^{\intercal}\omega+\omega^{\intercal}\sum_{j=\lfloor \rho m \rfloor+1}^{m} \Lambda^2_{j,j}V_{\cdot,j} V_{\cdot,j}^{\intercal}\omega\\
&\geq \frac{\lambda_{\lfloor \rho m \rfloor}(\frac{1}{n} X X^{\intercal})}{\lambda_{q+1}(\frac{1}{n} X X^{\intercal})}\left(\omega^{\intercal}\sum_{j=q+1}^{\lfloor \rho m \rfloor} \Lambda^2_{j,j}V_{\cdot,j} V_{\cdot,j}^{\intercal}\omega+\omega^{\intercal}\sum_{j=\lfloor \rho m \rfloor+1}^{m} \Lambda^2_{j,j}V_{\cdot,j} V_{\cdot,j}^{\intercal}\omega\right).
\end{aligned}
\end{equation*}
If $\lambda_{\lfloor \rho m\rfloor}(\frac{1}{n}X X^{\intercal}) \geq c  \max\{1,p/n\},$ together with \eqref{eq: bound on q+1 eigenvalue}, 
we establish that, there exists some positive constant $c'>0$ such that, with probability larger than $1-p^{-c},$ 
$$\frac{\lambda_{\lfloor \rho m \rfloor}(\frac{1}{n} X X^{\intercal})}{\lambda_{q+1}(\frac{1}{n} X X^{\intercal})}\geq c'.$$
 This leads to 
\begin{equation*}
\omega^{\intercal}X^{\intercal}\mathcal{Q}^2X\omega\gtrsim \omega^{\intercal}(\widetilde{X}^{\rm PCA})^{\intercal}\widetilde{X}^{\rm PCA}\omega
\end{equation*}
for any $\omega\in \R^{p}$ and hence with probability larger than $1-p^{-c},$
$${\rm RE}\left(\frac{1}{n}X^{\intercal}\mathcal{Q}^2X\right)\gtrsim {\rm RE}\left(\frac{1}{n}(\widetilde{X}^{\rm PCA})^{\intercal}\widetilde{X}^{\rm PCA}\right).$$

To complete the proof, we shall apply the following lemma to verify the dimension condition \eqref{eq: dim condition}. The proof of the following lemma is presented at Section \ref{eq: factor analysis}. 

\begin{Lemma}
Suppose that assumptions (A1) and (A3) hold, $H_{i,\cdot}$ is a sub-Gaussian random vector, $q+\log p\lesssim \sqrt{n},$ $k=\|\beta\|_0$ satisfies $k q^2\log p\log n/n\rightarrow 0$. The loading matrix $\Psi\in \R^{q\times p}$ satisfies 
$\max_{1\leq i\leq q , 1\leq j\leq p}|\Psi_{i,j}|\lesssim \sqrt{\log (qp)},$ $\lambda_{1}(\Psi)/\lambda_{q}(\Psi)\leq C$ for some positive constant $C>0$ and \eqref{eq: strong factor}. Then with probability larger than  $1-p^{-c}-\exp(-cn)$ for some positive constant $c>0,$ 
\begin{equation}
\begin{aligned}
\|R\|_{\infty}
\lesssim &\sqrt{\frac{q\log p}{n}}\sqrt{q\log (qn)}+\frac{q^{\frac{9}{2}}(\log N)^{\frac{7}{2}}}{\min\{n,p\}}\cdot \left(\frac{p}{\lambda_{q}^2(\Psi)}\right)^2\sqrt{q\log (qp)}\\
&+\left(\sqrt{\frac{\log p}{n}}+\frac{q \log N}{\sqrt{p}}\right)\cdot \frac{p}{\lambda_{q}^2(\Psi)}\sqrt{q\log (qn)}.
\end{aligned}
\label{eq: R infinity bound}
\end{equation}
\label{lem: R bound}
\end{Lemma}

Hence the dimension condition $$\frac{\Mq^2\cdot k q^2 \log p\log n}{n}\rightarrow 0$$ together with \eqref{eq: strong factor} implies \eqref{eq: dim condition}.

{
Furthermore, in Section \ref{sec: lower bound}, we provide theoretical justification on the lower bound $\lambda_{\lfloor \rho m\rfloor}(\frac{1}{n}XX^{\intercal}).$
\subsection{Lower bounds for $\lambda_{\lfloor \rho m\rfloor}(\frac{1}{n}XX^{\intercal})$}
\label{sec: lower bound}
\begin{Lemma}
Suppose that assumptions (A1) and (A3) hold and $H_{i,\cdot}$ is a sub-Gaussian random vector.
{With probability larger than $1-p^{-c}$ for some positive constant $c>0$, if either of the following two assumptions hold for $Z_{i,\cdot}=\Sigma_{X}^{-1/2} X_{i,\cdot}$:
\begin{enumerate}
\item $p/n\rightarrow c^*\in [0, \infty)$ and $\frac{1}{p}\left(Z_{i,\cdot}^{\intercal} A Z_{i,\cdot}-{\rm Tr}(A)\right)\cip 0$ as $p\rightarrow \infty$ for all sequences of complex matrices $A\in \R^{p\times p}$ with uniformly bounded spectral norms $\|A\|_2.$ 
\item $p/n\rightarrow \infty$ and the entries of $Z_{i,\cdot}$ are independent.
\end{enumerate}
then $\lambda_{\lfloor \rho m\rfloor}(\frac{1}{n}X X^{\intercal})\gtrsim  \max\{1,p/n\}$ for {$n$ sufficiently large}.}
\label{lem: lower bound quantile}
\end{Lemma}
The condition $\frac{1}{p}\left(Z_{i,\cdot}^{\intercal} A Z_{i,\cdot}-{\rm Tr}(A)\right)\cip 0$ is implied by the forth order moment condition: for $1\leq i\leq n,$ \begin{equation}
E[Z_{i,j_1} Z_{i,j_2} Z_{i,j_3} Z_{i,j_4}] = 0\ \mbox{for all}\ j_1 \notin \{j_2,j_3,j_4\}.
\label{eq: moment condition}
\end{equation}
The moment condition \eqref{eq: moment condition} is {substantially weaker than assuming independent} 
entries of $Z_{i,\cdot}$. Both conditions {1. and 2.} are imposed only for technical reasons so that we can directly apply the lower bounds for the median (or smallest) singular values established in  \citep{yaskov2016short,vershynin2010introduction,rudelson2009smallest}. 

We now apply \eqref{eq: moment condition} to establish $\frac{1}{p}\left(Z_{i,\cdot}^{\intercal} A Z_{i,\cdot}-{\rm Tr}(A)\right)\cip 0$. 
Note that
\begin{align*}
\E \left|Z_{i,\cdot}^{\intercal} A Z_{i,\cdot}-{\rm Tr}(A)\right|^2&= \E \left|Z_{i,\cdot}^{\intercal} A Z_{i,\cdot}\right|^2-|{\rm Tr}(A)|^2\\
&\leq \sum_{1\leq j\neq l\leq p} \E Z_{i,j}^2 Z_{i,l}^2 |A_{j,l}|^2+\sum_{1\leq j\neq l\leq p} \E Z_{i,j}^2 Z_{i,l}^2 |A_{j,l}| |A_{l,j}|\\
&\lesssim \sum_{1\leq j\neq l\leq p} |A_{j,l}|^2\lesssim p
\end{align*}
where the {first equality uses that $\E[Z_{i,.}^T A Z_{i.}] = {\rm Tr}(A)$, the} first inequality follows from \eqref{eq: moment condition} and the last inequality follows from the bounded spectrum norm condition. Then we apply Markov's inequality to establish $\frac{1}{p}\left(Z_{i,\cdot}^{\intercal} A Z_{i,\cdot}-{\rm Tr}(A)\right)\cip 0$ as $p\rightarrow \infty.$} 

We now present the proof of Lemma \ref{lem: lower bound quantile}.
With $Z=X\Sigma_{X}^{-\frac{1}{2}}$, we have
\begin{equation}
\lambda_{\min}\left( Z Z^{\intercal}\right)=\lambda_{\min}\left(X\Sigma_{X}^{-1} X^{\intercal}\right)\leq \frac{1}{\lambda_{\min}(\Sigma_{X})} \lambda_{\min}( XX^{\intercal}).
\label{eq: relation}
\end{equation}
Note that 
$Z_{i,\cdot}=\Sigma_{X}^{-\frac{1}{2}} \Psi^{\intercal} H_{i\cdot}+\Sigma_{X}^{-\frac{1}{2}} E_{i\cdot}$. For any $v\in \R^{p}$ and $\|v\|_2\leq 1,$ the random variable  $v^{\intercal}Z_{i,\cdot}$ has sub-Gaussian norm upper bounded by $C\left(\|v^{\intercal}\Sigma_{X}^{-\frac{1}{2}} \Psi^{\intercal}\|_2+\|\Sigma_{X}^{-\frac{1}{2}}v\|_2\right)$ for some positive constant $C>0.$ By \eqref{eq: inter matrix expression}, we show that $v^{\intercal}Z_{i,\cdot}$ has a bounded sub-Gaussian norm and hence $Z_{i,\cdot}$ is sub-Gaussian.

We now establish the lower bound for $\lambda_{\lfloor \rho m\rfloor}(\frac{1}{n}XX^{\intercal})$ by considering two cases.\\
\noindent
{\bf Case 1: $p/n\rightarrow c_*\in (0,\infty).$} 
For any set $B\subseteq \R,$ define $\mu_{p}(B)=\frac{1}{p}\sum_{i=1}^{p}{\bf 1}(\lambda_j\in B)$ where $\{\lambda_{j}\}_{1\leq j\leq p}$ are eigenvalues of $\frac{1}{n}Z Z^{\intercal}.$ Let $\mu_{c_*}$ denote the Marchenko Pastur law: for any set $B\subseteq \R,$
\begin{equation*}
\mu_{c_*}(B)=\begin{cases}
(1-1/c_*)\cdot {\bf 1}(0\in B)+\int_{a}^{b} \frac{\sqrt{(b-t)(t-a)}}{2\pi c_* t} \cdot {\bf 1}(t\in B) dt & \text{if}\quad c_*>1\\
\int_{a}^{b} \frac{\sqrt{(b-t)(t-a)}}{2\pi c_* t} \cdot {\bf 1}(t\in B) dt  & \text{if}\quad 0< c_*\leq 1
\end{cases}
\end{equation*}
where $a=(1-\sqrt{c_*})^2$ and $b=(1+\sqrt{c_*})^2.$

In the following, we shall apply Theorem 1 of \citep{yaskov2016short} and establish 
\begin{equation}
\mu_{p}\cid \mu_{c_*} \quad \text{almost surely}.
\label{eq: MP law}
\end{equation}
Note that Theorem 1 of \citep{yaskov2016short} holds under the condition that $$\frac{1}{p}\left(Z_{i,\cdot}^{\intercal} A Z_{i,\cdot}-{\rm Tr}(A)\right)\cip 0$$ as $p\rightarrow \infty$ for any sequence of complex matrices $A\in \R^{p\times p}$ with uniformly bounded spectral norms $\|A\|_2.$ 

We now apply \eqref{eq: MP law}.
When $c_*\neq 1,$ \eqref{eq: MP law} implies that
\begin{equation}
\liminf_{n\rightarrow \infty} \lambda_{\min}\left(\frac{1}{n}ZZ^{\intercal}\right)\geq \left(1-\sqrt{c_*}\right)^2\quad \text{almost surely}.
\label{eq: case 2-a}
\end{equation}
When $c_*=1,$ we need to calculate the median (or more general quantiles) of the distribution with the density function $\frac{\sqrt{(4-t)t}}{2\pi t}.$ For $\rho=1/2$, the median is within the range between $0.65$ and $0.66$, which, together with \eqref{eq: MP law}, lead to 
\begin{equation}
\liminf_{n\rightarrow \infty} \lambda_{\lfloor m/2\rfloor}\left(\frac{1}{n}ZZ^{\intercal}\right)\geq 0.65 \quad \text{almost surely}.
\label{eq: case 2-b}
\end{equation}
We combine \eqref{eq: relation}, \eqref{eq: case 2-a} and \eqref{eq: case 2-b} and show that there exists some constant $c>0$ such that  
\begin{equation}
\liminf_{n\rightarrow \infty} \lambda_{\lfloor m/2\rfloor}\left(\frac{1}{n}XX^{\intercal}\right)\geq c \lambda_{\min}(\Sigma_{X}) \quad \text{almost surely}.
\label{eq: case 2}
\end{equation}

\noindent
{\bf Case 2: $p/n\geq C$ for some positive constant $C>0$ and the entries of $Z_{i,\cdot}$ are independent.} Theorem 5.39 of \citep{vershynin2010introduction} implies that with probability larger than $1-p^{-c},$
$\lambda_{m}(Z)\geq \sqrt{p}-C\sqrt{n}-\sqrt{\log p},$ where $C$ is the constant defined in \citep{vershynin2010introduction} and  independent of $n$ and $p.$ Combined with \eqref{eq: relation}, we establish that, with probability larger than $1-p^{-c},$ 
\begin{equation}
\lambda_{\lfloor \rho m \rfloor}(\frac{1}{n} X X^{\intercal})\geq \lambda_{m}(\frac{1}{n} X X^{\intercal})\gtrsim \frac{p}{n}\lambda_{\min}(\Sigma_{X}).
\label{eq: case 1}
\end{equation}

\subsection{Proof of Lemma \ref{lem: R bound}}
\label{eq: factor analysis}

We prove the lemma through a finite-sample analysis of the factor model \eqref{eq: matrix factor}.
The proof idea follows from that in \citep{bai2003inferential} and \citep{bai2002determining}, who establish the limiting distribution for any single entry of the matrix $R=\widetilde{D}-M$; see Theorem 3 in  \citep{bai2003inferential} for details. In our following proof, the main difference is to establish the rate of convergence of  $\|R\|_{\infty}$ using finite-sample concentration bounds. We also relax the strong factor assumption $\lambda_{q}(\Psi)\asymp \sqrt{p}$ in \citep{bai2002determining} to the weaker condition \eqref{eq: strong factor}.

Define $\widehat{\Lambda}^{2}\in \R^{q\times q}$ to be the diagonal matrix consisting of the top $q$ eigenvalues of the matrix  $\frac{1}{np}XX^{\intercal}.$ Define 
\begin{equation}
\T=(\Psi\Psi^{\intercal}/p)({H}^{\intercal} \widetilde{H}/n)\widehat{\Lambda}^{-2} \in \R^{q\times q}.
\label{eq: rotation random}
\end{equation} 

Define $N=\max\{n,p\}.$ 
Define the events
\begin{equation*}
\begin{aligned}
\mathcal{G}_1&=\left\{\|\frac{1}{n}\sum_{i=1}^{n} {H}_{i,\cdot} {H}^{\intercal}_{i,\cdot}-{\rm I}\|_2\lesssim \sqrt{\frac{q+{\log p}}{n}}\right\}\\
\mathcal{G}_2&=\left\{\max_{1\leq i\leq n}\|{H}_{i,\cdot}\|_2\lesssim \sqrt{q \log (nq)}\right\}\\
\mathcal{G}_3&=\left\{\max_{1\leq t\leq n}\|\Psi^{\intercal} H_{t,\cdot}/p\|_{2}\lesssim \frac{\sqrt{q}\sqrt{\log (pq)}\sqrt{\log (np)}}{\sqrt{p}}\right\}\\
\mathcal{G}_4&=\left\{\max_{1\leq t\leq n}\max_{1\leq j\leq q} \frac{1}{\|\Psi_{j,\cdot}\|_2}\left|\Psi_{j,\cdot}^{\intercal} E_{t,\cdot}\right|\lesssim \sqrt{\log N}\right\}\\
\mathcal{G}_5&=\left\{ \max_{1\leq i\leq n} E_{i,\cdot}^{\intercal} E_{i,\cdot}/p \lesssim \log (np)\right\}\\
\mathcal{G}_6&=\left\{ \max_{1\leq t\neq i\leq n}\left|E_{i,\cdot}^{\intercal} E_{t,\cdot}/p\right| \lesssim \frac{\sqrt{\log p}\sqrt{\log (np)}}{\sqrt{p}}\right\}\\
\mathcal{G}_7&=\left\{\frac{\|H^{\intercal}E\Psi^{\intercal}\|_2}{np}=\left\|\frac{1}{n}\sum_{i=1}^{n} H_{i\cdot} \frac{1}{p} E_{i\cdot}^{\intercal}\Psi^{\intercal}\right\|_2  \lesssim \sqrt{\frac{q+\log p}{n}}\cdot\frac{\lambda_{\max}(\Psi)}{p}\right\}\\
\mathcal{G}_{8}&=\left\{\|H\|_2\lesssim \sqrt{n}, \|E\|_2\lesssim \sqrt{n}+\sqrt{p}\right\}\\
\mathcal{G}_{9}&=\left\{\frac{\|H^{\intercal}E\|_2}{np}=\left\|\frac{1}{n}\sum_{i=1}^{n} H_{i\cdot} \frac{1}{p} E_{i\cdot}^{\intercal}\right\|_2 \lesssim \frac{1}{p}+\frac{1}{\sqrt{np}}\right\}\\
\mathcal{G}_{10}&=\left\{ c \frac{\lambda_{q}(\Psi)}{\sqrt{p}}\leq \lambda_{\min}(\widehat{\Lambda})\leq \lambda_{1}(\widehat{\Lambda})\leq C\frac{\lambda_{1}(\Psi)}{\sqrt{p}},\; \lambda_{\max}(\T)\leq C \right\}\\
\mathcal{G}_{11}&=\left\{\max_{1\leq j,l\leq p} \left|\frac{1}{n}\sum_{t=1}^{n} E_{t,j}E_{t,l}-(\Sigma_{E})_{j,l}\right|\lesssim \sqrt{\frac{\log p}{n}}\right\}\\
\mathcal{G}_{12}&=\left\{\max_{1\leq j\leq q,1\leq l\leq p} \left|\frac{1}{n}\sum_{t=1}^{n}H_{t,j}E_{t,l}\right|\lesssim \sqrt{\frac{\log p}{n}} \right\}\\
\end{aligned}
\end{equation*}
where $C>0$ and $c>0$ are some positive constants. 
Define $$\mathcal{G}=\cap_{j=1}^{12}\mathcal{G}_j.$$

On the event $\mathcal{G}_4,$ we have
\begin{equation}
 \max_{1\leq t\leq n}\|\frac{1}{p} \Psi E_{t,\cdot}\|_2
\lesssim \sqrt{q}\cdot \max_{1\leq j\leq q} \frac{\|\Psi_{j,\cdot}\|_2\sqrt{\log N}}{p} \lesssim\frac{{q}{\log N}}{\sqrt{p}}.
\label{eq: inter bound 1}
\end{equation}

The following lemma shows that the event $\mathcal{G}$ happens with a high probability, whose proof can be found in Section \ref{sec: high prob event}.
\begin{Lemma}
Suppose that the conditions of Lemma \ref{lem: R bound} hold, then we have 
\begin{equation}
\PP(\mathcal{G})\geq 1-p^{-c}-\exp(-cn)
\end{equation}
for some positive constant $c>0.$
\label{lem: high prob event}
\end{Lemma}

The following lemma characterizes the accuracy of the loading estimation, which can be viewed as the finite sample version of Theorem 1 in \citep{bai2002determining} and Theorem 1 in \citep{bai2003inferential}. The proof of the following lemma can be found in Section \ref{sec: factor proof}.   
\begin{Lemma}
On the event $\mathcal{G},$
\begin{equation}
\max_{1\leq t\leq n}\|\widetilde{H}_{t,\cdot}-\T^{\intercal} {H}_{t,\cdot}\|_2\lesssim \frac{p}{\lambda_{q}^2(\Psi)}\left(\frac{q^2 (\log N)^{3/2}}{\sqrt{p}}+\sqrt{\frac{q \log N}{n}}\right) 
\label{eq: bound factor}
\end{equation}
with $N=\max\{n,p\}.$
Furthermore, with probability larger than $1-n^{-c}-p^{-c}$ for some constant $c>0,$
\begin{equation}
\left|\widetilde{H}_{t,\cdot}-\T^{\intercal}{H}_{t,\cdot}-\widehat{\Lambda}^{-2} \frac{1}{n}\sum_{i=1}^{n}{H}_{i,\cdot} \frac{1}{p}{H}_{i,\cdot}^{\intercal} \Psi E_{t,\cdot}\right|\lesssim \left(\frac{p}{\lambda_{q}^2(\Psi)}\right)^2\frac{q^{\frac{7}{2}}(\log N)^3}{\min\{n,p\}}
\label{eq: refined bound factor}
\end{equation}
and 
\begin{equation}
\left\|\widehat{\Lambda}^{-2} \frac{1}{n}\sum_{i=1}^{n} {H}_{i,\cdot} \frac{1}{p}{H}_{i,\cdot}^{\intercal} \Psi E_{t,\cdot}\right\|_2\leq \|\widehat{\Lambda}^{-2} \|_2\cdot\|\frac{1}{n}\sum_{i=1}^{n} {H}_{i,\cdot}{H}_{i,\cdot}^{\intercal}\|_2 \cdot \|\frac{1}{p} \Psi E_{t,\cdot}\|_2 \lesssim \frac{p}{\lambda_{q}^2(\Psi)} \cdot \frac{{q}{\log N}}{\sqrt{p}}.
\label{eq: upper bound 1a}
\end{equation}
\label{lem: factor bound}
\end{Lemma}

The following lemma characterizes the accuracy of the loading estimation, which can be viewed as the finite sample version of Theorem 2 in \citep{bai2003inferential}. The proof of the following lemma can be found in Section \ref{sec: loading proof}.

\begin{Lemma}
On the event $\mathcal{G},$
\begin{equation}
\max_{1\leq l\leq p}\left\|\widetilde{\Psi}_{\cdot,l}-\T^{-1}{\Psi}_{\cdot,l}\right\|\lesssim \frac{q^{\frac{9}{2}}(\log N)^{\frac{7}{2}}}{\min\{n,p\}}\cdot \left(\frac{p}{\lambda_{q}^2(\Psi)}\right)^2+\left(\sqrt{\frac{\log p}{n}}+\frac{1}{\sqrt{p}}\right)\cdot \frac{p}{\lambda_{q}^2(\Psi)}+\sqrt{\frac{q\log p}{n}}.
\label{eq: loading bound}
\end{equation}
\label{lem: loading bound}
\end{Lemma}

For $1\leq t\leq n$ and $1\leq l\leq p,$ we have the the following decomposition for $\widetilde{D}_{t,l}-M_{t,l}$
\begin{equation}
\begin{aligned}
&\widetilde{H}_{t,\cdot}^{\intercal}\widetilde{\Psi}_{\cdot,l}-{H}_{t,\cdot}^{\intercal}{\Psi}_{\cdot,l}\\
&=\widetilde{H}_{t,\cdot}^{\intercal}\widetilde{\Psi}_{\cdot,l}-(\T^{\intercal}{H}_{t,\cdot})^{\intercal}\T^{-1}{\Psi}_{\cdot,l}\\
&=(\widetilde{H}_{t,\cdot}-\T^{\intercal}{H}_{t,\cdot})^{\intercal}\widetilde{\Psi}_{\cdot,l}+(\T^{\intercal}{H}_{t,\cdot})^{\intercal}(\widetilde{\Psi}_{\cdot,l}-\T^{-1}{\Psi}_{\cdot,l})\\
&=(\widetilde{H}_{t,\cdot}-\T^{\intercal}{H}_{t,\cdot})^{\intercal}\T^{-1}\Psi_{\cdot,l}+(\T^{\intercal}{H}_{t,\cdot})^{\intercal}(\widetilde{\Psi}_{\cdot,l}-\T^{-1}{\Psi}_{\cdot,l})+(\widetilde{H}_{t,\cdot}-\T^{\intercal}{H}_{t,\cdot})^{\intercal}(\widetilde{\Psi}_{\cdot,l}-\T^{-1}\Psi_{\cdot,l}).
\end{aligned}
\label{eq: decomposition M}
\end{equation}

On the event $\mathcal{G}_{2}\cap\mathcal{G}_{10},$ we have 
$$\|\T^{-1}\Psi_{\cdot,l}\|_2\lesssim \sqrt{q \log(qp)} \quad \text{and}\quad \|\T^{\intercal}{H}_{t,\cdot}\|_2\lesssim \sqrt{q \log(nq)}.$$
Note that
$$\|R\|_{\infty}=\max_{1\leq t\leq n, 1\leq l\leq p}|\|\widetilde{H}_{t,\cdot}^{\intercal}\widetilde{\Psi}_{\cdot,l}-{H}_{t,\cdot}^{\intercal}{\Psi}_{\cdot,l}\|_2.$$
By applying Lemmas \ref{lem: factor bound} and \ref{lem: loading bound} to the decomposition \eqref{eq: decomposition M}, we establish that \eqref{eq: R infinity bound} holds on the event $\mathcal{G}.$

\subsection{Proof of Lemma \ref{lem: factor bound}}
\label{sec: factor proof}
Recall that  $\widehat{\Lambda}^{2}\in \R^{q\times q}$ denotes the diagonal matrix consisting of the top $q$ eigenvalues of the matrix  $\frac{1}{np}XX^{\intercal}.$
By the definition of $\widetilde{H}$ in \eqref{eq: factor def},  we have 
\begin{equation*}
\widetilde{H}=\frac{1}{np}XX^{\intercal}\widetilde{H}\widehat{\Lambda}^{-2}.
\end{equation*}
With the above expression, we establish the following decomposition of $\widetilde{H}_{t,\cdot}-\T^{\intercal}{H}_{t,\cdot}\in \R^{q}$ for $1\leq t\leq n,$
\begin{equation}
\begin{aligned}
&\widetilde{H}_{t,\cdot}-\T^{\intercal}{H}_{t,\cdot}=\frac{1}{np}\widehat{\Lambda}^{-2}\widetilde{H}^{\intercal}X {X}_{t,\cdot}-\T^{\intercal}{H}_{t,\cdot}\\
&=\frac{1}{np}\widehat{\Lambda}^{-2}\widetilde{H}^{\intercal}(H\Psi+E) (\Psi^{\intercal} H_{t,\cdot}+E_{t,\cdot})-\T^{\intercal}{H}_{t,\cdot}\\
&=\frac{1}{np}\widehat{\Lambda}^{-2} \widetilde{H}^{\intercal} H\Psi E_{t,\cdot}+\frac{1}{np}\widehat{\Lambda}^{-2} \widetilde{H}^{\intercal} E \Psi^{\intercal} H_{t,\cdot}+\frac{1}{np}\widehat{\Lambda}^{-2} \widetilde{H}^{\intercal} E E_{t,\cdot}\\
&=\widehat{\Lambda}^{-2} \left(\frac{1}{n}\sum_{i=1}^{n}\widetilde{H}_{i,\cdot} \frac{1}{p}{H}_{i,\cdot}^{\intercal} \Psi E_{t,\cdot}+\frac{1}{n} \sum_{i=1}^{n} \widetilde{H}_{i,\cdot} \frac{1}{p}E_{i,\cdot}^{\intercal} \Psi^{\intercal} H_{t,\cdot}+\frac{1}{n} \sum_{i=1}^{n} \widetilde{H}_{i,\cdot} \frac{1}{p} E_{i,\cdot}^{\intercal} E_{t,\cdot}\right).
\end{aligned}
\label{eq: key decomp 0}
\end{equation}
\noindent
\underline{Proof of \eqref{eq: bound factor}.} By \eqref{eq: key decomp 0}, we have  
\begin{equation}
\begin{aligned}
&\|\widetilde{H}_{t,\cdot}-\T^{\intercal} {H}_{t,\cdot}\|_2\\
\leq &\|\widehat{\Lambda}^{-2}\|_2 \left(\|\frac{1}{n}\sum_{i=1}^{n}\widetilde{H}_{i,\cdot} \frac{1}{p}{H}_{i,\cdot}^{\intercal} \Psi E_{t,\cdot}\|_2+\|\frac{1}{n} \sum_{i=1}^{n} \widetilde{H}_{i,\cdot} \frac{1}{p}E_{i,\cdot}^{\intercal} \Psi^{\intercal} H_{t,\cdot}\|_2+\|\frac{1}{n} \sum_{i=1}^{n} \widetilde{H}_{i,\cdot} \frac{1}{p} E_{i,\cdot}^{\intercal} E_{t,\cdot}\|_2\right).
\end{aligned}
\label{eq: decomp 1}
\end{equation}

We upper bound the three terms on the right hand side of \eqref{eq: decomp 1} as
\begin{equation}
\begin{aligned}
\|\frac{1}{n}\sum_{i=1}^{n}\widetilde{H}_{i,\cdot} \frac{1}{p}{H}_{i,\cdot}^{\intercal} \Psi E_{t,\cdot}\|_2&\leq \frac{1}{n}\sum_{i=1}^{n}\|\widetilde{H}_{i,\cdot}\|_2 |\frac{1}{p}{H}_{i,\cdot}^{\intercal} \Psi E_{t,\cdot}|\\&\leq \sqrt{\frac{1}{n}\sum_{i=1}^{n}\|\widetilde{H}_{i,\cdot}\|_2^2}\cdot \sqrt{\frac{1}{n}\sum_{i=1}^{n}|\frac{1}{p}{H}_{i,\cdot}^{\intercal} \Psi E_{t,\cdot}|^2};
\label{eq: temp upper 1}
\end{aligned}
\end{equation}
\begin{equation}
\begin{aligned}
\|\frac{1}{n} \sum_{i=1}^{n} \widetilde{H}_{i,\cdot} \frac{1}{p}E_{i,\cdot}^{\intercal} \Psi^{\intercal} H_{t,\cdot}\|_2&\leq \frac{1}{n}\sum_{i=1}^{n}\|\widetilde{H}_{i,\cdot}\|_2 |\frac{1}{p}{E}_{i,\cdot}^{\intercal} \Psi H_{t,\cdot}|\\
&\leq \sqrt{\frac{1}{n}\sum_{i=1}^{n}\|\widetilde{H}_{i,\cdot}\|_2^2}\cdot \sqrt{\frac{1}{n}\sum_{i=1}^{n}|\frac{1}{p}{E}_{i,\cdot}^{\intercal} \Psi H_{t,\cdot}|^2};
\label{eq: temp upper 2}
\end{aligned}
\end{equation}
\begin{equation}
\begin{aligned}
\|\frac{1}{n} \sum_{i=1}^{n} \widetilde{H}_{i,\cdot} \frac{1}{p} E_{i,\cdot}^{\intercal} E_{t,\cdot}\|_2&\leq \frac{1}{n}\sum_{i=1}^{n}\|\widetilde{H}_{i,\cdot}\|_2 |\frac{1}{p}{E}_{i,\cdot}^{\intercal} E_{t,\cdot}|\\
&\leq \sqrt{\frac{1}{n}\sum_{i=1}^{n}\|\widetilde{H}_{i,\cdot}\|_2^2}\cdot \sqrt{\frac{1}{n}\sum_{i=1}^{n}|\frac{1}{p}{E}_{i,\cdot}^{\intercal} E_{t,\cdot}|^2}.
\label{eq: temp upper 3}
\end{aligned}
\end{equation}
 Note that 
$$\max_{1\leq t\leq n}\sqrt{\frac{1}{n}\sum_{i=1}^{n}|\frac{1}{p}{H}_{i,\cdot}^{\intercal} \Psi E_{t,\cdot}|^2}\leq \max_{1\leq i\leq n}\|{H}_{i,\cdot}\|_2 \max_{1\leq t\leq n}\|\Psi E_{t,\cdot}/p\|_2,$$
$$\max_{1\leq t\leq n}\sqrt{\frac{1}{n}\sum_{i=1}^{n}|\frac{1}{p}{E}_{i,\cdot}^{\intercal} \Psi H_{t,\cdot}|^2}\leq \max_{1\leq t\leq n}\|{H}_{t,\cdot}\|_2 \max_{1\leq i\leq n}\|\Psi E_{i,\cdot}/p\|_2.$$

Together with \eqref{eq: inter bound 1}, we establish that, on the event $\mathcal{G}_2,$
$$\max\left\{\max_{1\leq t\leq n}\sqrt{\frac{1}{n}\sum_{i=1}^{n}|\frac{1}{p}{H}_{i,\cdot}^{\intercal} \Psi E_{t,\cdot}|^2},\max_{1\leq t\leq n}\sqrt{\frac{1}{n}\sum_{i=1}^{n}|\frac{1}{p}{E}_{i,\cdot}^{\intercal} \Psi H_{t,\cdot}|^2}\right\}\lesssim \frac{q^{\frac{3}{2}}(\log N)^{\frac{3}{2}}}{\sqrt{p}}.$$
Note that $\widetilde{H}^{\intercal}\widetilde{H}/n={\rm I}$ implies $\frac{1}{n}\sum_{i=1}^{n}\|\widetilde{H}_{i,\cdot}\|_2^2=q.$
Combined with \eqref{eq: temp upper 1} and \eqref{eq: temp upper 2}, we establish that, on the event $\mathcal{G},$ 
\begin{equation}
\max\left\{\|\frac{1}{n}\sum_{i=1}^{n}\widetilde{H}_{i,\cdot} \frac{1}{p}{H}_{i,\cdot}^{\intercal} \Psi E_{t,\cdot}\|_2\|\frac{1}{n},\sum_{i=1}^{n} \widetilde{H}_{i,\cdot} \frac{1}{p}E_{i,\cdot}^{\intercal} \Psi^{\intercal} H_{t,\cdot}\|_2\right\}\lesssim \frac{q^{2}(\log N)^{\frac{3}{2}}}{\sqrt{p}}.
\label{eq: first upper bound}
\end{equation}
Note that 
$$\sqrt{\frac{1}{n}\sum_{i=1}^{n}|\frac{1}{p}{E}_{i,\cdot}^{\intercal} E_{t,\cdot}|^2}=\sqrt{\frac{1}{n}\sum_{t\neq i}|\frac{1}{p}{E}_{i,\cdot}^{\intercal} E_{t,\cdot}|^2+\frac{1}{n}|\frac{1}{p}{E}_{t,\cdot}^{\intercal} E_{t,\cdot}|^2}.$$
On the event $\mathcal{G}_5\cap \mathcal{G}_6,$ we have
\begin{equation*}
\max_{1\leq t\leq n}\sqrt{\frac{1}{n}\sum_{i=1}^{n}|\frac{1}{p}{E}_{i,\cdot}^{\intercal} E_{t,\cdot}|^2}\lesssim \sqrt{\frac{{q}{\log p}\log (np)}{{p}}+\frac{\log (np)}{n}}.
\end{equation*}
Combined with \eqref{eq: temp upper 3}, we establish
\begin{equation}
\|\frac{1}{n} \sum_{i=1}^{n} \widetilde{H}_{i,\cdot} \frac{1}{p} E_{i,\cdot}^{\intercal} E_{t,\cdot}\|_2\lesssim \frac{q \log N}{\sqrt{p}}+\sqrt{\frac{q \log N}{n}}. 
\label{eq: second upper bound}
 \end{equation}
Together with \eqref{eq: first upper bound}, \eqref{eq: second upper bound} and the definition of $\mathcal{G}_{10}$, we apply the decomposition \eqref{eq: decomp 1} and establish \eqref{eq: bound factor}.

\noindent
\underline{Proof of \eqref{eq: refined bound factor}.}
We shall establish the bound by applying \eqref{eq: key decomp 0} and the bound \eqref{eq: bound factor}.
Note the following three decompositions
\begin{equation*}
\begin{aligned}
\frac{1}{n}\sum_{i=1}^{n}\widetilde{H}_{i,\cdot} \frac{1}{p}{H}_{i,\cdot}^{\intercal} \Psi E_{t,\cdot}=
\frac{1}{n}\sum_{i=1}^{n}(\widetilde{H}_{i,\cdot}-\T^{\intercal} {H}_{i,\cdot}) \frac{1}{p}{H}_{i,\cdot}^{\intercal} \Psi E_{t,\cdot}+\T^{\intercal} \frac{1}{n}\sum_{i=1}^{n} {H}_{i,\cdot} \frac{1}{p}{H}_{i,\cdot}^{\intercal} \Psi E_{t,\cdot}
\end{aligned}
\label{eq: bound decomp 1}
\end{equation*}
\begin{equation*}
\begin{aligned}
\frac{1}{n} \sum_{i=1}^{n} \widetilde{H}_{i,\cdot} \frac{1}{p}E_{i,\cdot}^{\intercal} \Psi^{\intercal} H_{t,\cdot}=
\frac{1}{n} \sum_{i=1}^{n} (\widetilde{H}_{i,\cdot}-\T^{\intercal} {H}_{i,\cdot}) \frac{1}{p}E_{i,\cdot}^{\intercal} \Psi^{\intercal} H_{t,\cdot}+
\T^{\intercal} \frac{1}{n} \sum_{i=1}^{n} {H}_{i,\cdot}\frac{1}{p}E_{i,\cdot}^{\intercal} \Psi^{\intercal} H_{t,\cdot}
\end{aligned}
\label{eq: bound decomp 2}
\end{equation*}
\begin{equation*}
\begin{aligned}
\frac{1}{n} \sum_{i=1}^{n} \widetilde{H}_{i,\cdot} \frac{1}{p} E_{i,\cdot}^{\intercal} E_{t,\cdot}=
\frac{1}{n} \sum_{i=1}^{n}(\widetilde{H}_{i,\cdot}-\T^{\intercal} {H}_{i,\cdot})\frac{1}{p} E_{i,\cdot}^{\intercal} E_{t,\cdot}
+\T^{\intercal} \frac{1}{n} \sum_{i=1}^{n} {H}_{i,\cdot} \frac{1}{p} E_{i,\cdot}^{\intercal} E_{t,\cdot}
\end{aligned}
\label{eq: bound decomp 3}
\end{equation*}
By applying \eqref{eq: key decomp 0} and the above three decompositions, we establish 
\begin{equation}
\begin{aligned}
&\widehat{\Lambda}^{2}(\frac{1}{np}\widehat{\Lambda}^{-2}\widetilde{H}^{\intercal}X {X}_{t,\cdot}-\T^{\intercal}{H}_{t,\cdot}-\widehat{\Lambda}^{-2} \frac{1}{n}\sum_{i=1}^{n}{H}_{i,\cdot} \frac{1}{p}{H}_{i,\cdot}^{\intercal} \Psi E_{t,\cdot})\\
&=\frac{1}{n}\sum_{i=1}^{n}(\widetilde{H}_{i,\cdot}-\T^{\intercal} {H}_{i,\cdot}) \frac{1}{p}{H}_{i,\cdot}^{\intercal} \Psi E_{t,\cdot}+\frac{1}{n} \sum_{i=1}^{n} (\widetilde{H}_{i,\cdot}-\T^{\intercal} {H}_{i,\cdot}) \frac{1}{p}E_{i,\cdot}^{\intercal} \Psi^{\intercal} H_{t,\cdot}\\
&+\T^{\intercal} \frac{1}{n} \sum_{i=1}^{n} {H}_{i,\cdot}\frac{1}{p}E_{i,\cdot}^{\intercal} \Psi^{\intercal} H_{t,\cdot}
+\frac{1}{n} \sum_{i=1}^{n}(\widetilde{H}_{i,\cdot}-\T^{\intercal} {H}_{i,\cdot})\frac{1}{p} E_{i,\cdot}^{\intercal} E_{t,\cdot}
+\T^{\intercal} \frac{1}{n} \sum_{i=1}^{n} {H}_{i,\cdot} \frac{1}{p} E_{i,\cdot}^{\intercal} E_{t,\cdot}
\end{aligned}
\label{eq: key decomp 1}
\end{equation}

Note that 
$$\left|\frac{1}{n}\sum_{i=1}^{n}(\widetilde{H}_{i,\cdot}-\T^{\intercal} {H}_{i,\cdot}) \frac{1}{p}{H}_{i,\cdot}^{\intercal} \Psi E_{t,\cdot}\right|\leq \sqrt{\frac{1}{n}\sum_{i=1}^{n}(\widetilde{H}_{i,\cdot}-\T^{\intercal} {H}_{i,\cdot})^2 }\sqrt{\frac{1}{n}\sum_{i=1}^{n}({H}_{i,\cdot}^{\intercal} \Psi E_{t,\cdot}/p)^2}$$
and
$$\left|\frac{1}{n} \sum_{i=1}^{n} (\widetilde{H}_{i,\cdot}-\T^{\intercal} {H}_{i,\cdot}) \frac{1}{p}E_{i,\cdot}^{\intercal} \Psi^{\intercal} H_{t,\cdot}\right|\leq \sqrt{\frac{1}{n} \sum_{i=1}^{n} (\widetilde{H}_{i,\cdot}-\T^{\intercal} {H}_{i,\cdot})^2}\sqrt{\frac{1}{n} \sum_{i=1}^{n}(E_{i,\cdot}^{\intercal} \Psi^{\intercal} H_{t,\cdot}/p)^2}.$$
On the event $\mathcal{G},$ we have \eqref{eq: inter bound 1} and then
$$
\max_{1\leq t\leq n}\max_{1\leq i\leq n}\left|{H}_{i,\cdot}^{\intercal} \Psi E_{t,\cdot}/p\right|\leq \max_{1\leq i\leq n}\|{H}_{i,\cdot}\|_2 \max_{1\leq t\leq n}\|\Psi E_{t,\cdot}/p\|_2 \lesssim \frac{(q\log N)^{3/2} }{\sqrt{p}}.
$$
With the above three inequalities, we apply \eqref{eq: bound factor} and establish 
\begin{equation}
\begin{aligned}
&\left|\frac{1}{n}\sum_{i=1}^{n}(\widetilde{H}_{i,\cdot}-\T^{\intercal} {H}_{i,\cdot}) \frac{1}{p}{H}_{i,\cdot}^{\intercal} \Psi E_{t,\cdot}\right|+\left|\frac{1}{n} \sum_{i=1}^{n} (\widetilde{H}_{i,\cdot}-\T^{\intercal} {H}_{i,\cdot}) \frac{1}{p}E_{i,\cdot}^{\intercal} \Psi^{\intercal} H_{t,\cdot}\right|\\
&\lesssim \frac{p}{\lambda_{q}^2(\Psi)}\frac{q^{\frac{7}{2}}(\log N)^3}{\sqrt{p}\sqrt{\min\{n,p\}}}
\end{aligned}
\label{eq: separate bound 1}
\end{equation}
Note that 
$$\frac{1}{n} \sum_{i=1}^{n}(\widetilde{H}_{i,\cdot}-\T^{\intercal} {H}_{i,\cdot})\frac{1}{p} E_{i,\cdot}^{\intercal} E_{t,\cdot}
\leq \sqrt{\frac{1}{n} \sum_{i=1}^{n}(\widetilde{H}_{i,\cdot}-\T^{\intercal} {H}_{i,\cdot})^2 }\sqrt{\frac{1}{n} \sum_{i=1}^{n} (E_{i,\cdot}^{\intercal} E_{t,\cdot}/p)^2
}.$$
On the event $\mathcal{G}$, we apply \eqref{eq: bound factor} and \eqref{eq: second upper bound} and establish  
\begin{equation}
\left|\frac{1}{n} \sum_{i=1}^{n} (\widetilde{H}_{i,\cdot}-\T^{\intercal} {H}_{i,\cdot}) \frac{1}{p}E_{i,\cdot}^{\intercal} E_{t,\cdot}\right|\leq \frac{p}{\lambda_{q}^2(\Psi)}\cdot \frac{q^{3}(\log N)^{\frac{5}{2}}}{\min\{n,p\}}.
\label{eq: separate bound 2}
\end{equation}

We now turn to the upper bound for 
$\T^{\intercal} \frac{1}{n} \sum_{i=1}^{n} {H}_{i,\cdot}\frac{1}{p}E_{i,\cdot}^{\intercal} \Psi^{\intercal} H_{t,\cdot}$ and first consider the setting $i\neq t.$
Note that 
\begin{equation}
\left\|\frac{1}{n} \sum_{i\neq t} {H}_{i,\cdot}\frac{1}{p}E_{i,\cdot}^{\intercal} \Psi^{\intercal} H_{t,\cdot}\right\|_2 \leq \sqrt{q} \max_{1\leq j\leq q} \left|\frac{1}{n} \sum_{i\neq t} {H}_{i,j}\frac{1}{p}E_{i,\cdot}^{\intercal} \Psi^{\intercal} H_{t,\cdot}\right|.
\end{equation}
Conditioning on $H_{t,\cdot},$ the random variable ${H}_{i,j}\frac{1}{p}E_{i,\cdot}^{\intercal} \Psi^{\intercal} H_{t,\cdot}$ is of zero mean and sub-exponential with sub-exponential norm upper bounded by $C \|\Psi^{\intercal} H_{t,\cdot}/p\|_{2}.$ By Proposition 5.16 of \citep{vershynin2010introduction}, we establish 
\begin{equation*}
\PP\left(\max_{1\leq j\leq q} \left|\frac{1}{n} \sum_{i\neq t} {H}_{i,j}\frac{1}{p}E_{i,\cdot}^{\intercal} \Psi^{\intercal} H_{t,\cdot}\right|\geq C \|\Psi^{\intercal} H_{t,\cdot}/p\|_{2}\sqrt{\frac{\log n}{n}}\mid H_{t,\cdot}\right)\leq n^{-c}.
\end{equation*}
Together with the definition of the event $\mathcal{G}_3,$ we establish that, with probability larger than $(1-n^{-c})\cdot \PP(\mathcal{G}_3),$
\begin{equation}
\max_{1\leq j\leq q} \left|\frac{1}{n} \sum_{i\neq t} {H}_{i,j}\frac{1}{p}E_{i,\cdot}^{\intercal} \Psi^{\intercal} H_{t,\cdot}\right|\lesssim \frac{\sqrt{q}(\log N)^{\frac{3}{2}}}{\sqrt{np}}.
\label{eq: separate bound 3 inter}
\end{equation}
On the event $\mathcal{G}_2\cap \mathcal{G}_4$, we apply \eqref{eq: inter bound 1} and establish that for any $1\leq t\leq n,$ $$\left\|\frac{1}{n} {H}_{t,\cdot}\frac{1}{p}E_{t,\cdot}^{\intercal} \Psi^{\intercal} H_{t,\cdot}\right\|_2\leq \frac{1}{n}\|H_{t,\cdot}\|_2^2 \|\frac{1}{p}E_{t,\cdot}^{\intercal} \Psi^{\intercal}\|_2\lesssim \frac{(q \log N)^2}{n\sqrt{p}}.
$$
Together with \eqref{eq: separate bound 3 inter}, we establish 
\begin{equation}
\left\|\frac{1}{n} \sum_{i=1}^{n} {H}_{i,\cdot}\frac{1}{p}E_{i,\cdot}^{\intercal} \Psi^{\intercal} H_{t,\cdot}\right\|_2\lesssim \frac{{q}(\log N)^{\frac{3}{2}}}{\sqrt{np}}+\frac{(q \log N)^2}{n\sqrt{p}}.
\label{eq: separate bound 3}
\end{equation}

We now consider the upper bound for 
$\T^{\intercal} \frac{1}{n} \sum_{i=1}^{n} {H}_{i,\cdot} \frac{1}{p} E_{i,\cdot}^{\intercal} E_{t,\cdot}$ and consider the setting $i\neq t.$ Note that 
\begin{equation}
\left\|\frac{1}{n} \sum_{i\neq t} {H}_{i,\cdot} \frac{1}{p} E_{i,\cdot}^{\intercal} E_{t,\cdot}\right\|_2\leq \sqrt{q} \max_{1\leq j\leq q}\left\|\frac{1}{n} \sum_{i\neq t} {H}_{i,j} \frac{1}{p} E_{i,\cdot}^{\intercal} E_{t,\cdot}\right\|_2
\end{equation}

Conditioning on $E_{t,\cdot},$ the random variable ${H}_{i,j}\frac{1}{p}E_{i,\cdot}^{\intercal} E_{t,\cdot}$ is of zero mean and sub-exponential  with sub-exponential norm upper bounded by $C \|E_{t,\cdot}/p\|_{2}.$ By  Proposition 5.16 of \citep{vershynin2010introduction}, we establish 
\begin{equation}
\PP\left(\max_{1\leq j\leq q} \left|\frac{1}{n} \sum_{i\neq t} {H}_{i,j}\frac{1}{p}E_{i,\cdot}^{\intercal}  E_{t,\cdot}\right|\geq C \|E_{t,\cdot}/p\|_{2}\sqrt{\frac{\log n}{n}}\mid E_{t,\cdot}\right)\leq n^{-c}.
\end{equation}
Together with the definition of $\mathcal{G}_5,$ we show that, with probability larger than $(1-n^{-c})\cdot \PP(\mathcal{G}_5),$
\begin{equation}
\max_{1\leq j\leq q} \left|\frac{1}{n} \sum_{i\neq t} {H}_{i,j}\frac{1}{p}E_{i,\cdot}^{\intercal}  E_{t,\cdot}\right|\lesssim \frac{\log N}{\sqrt{np}} \quad \text{and}\quad \left\|\frac{1}{n} \sum_{i\neq t} {H}_{i,\cdot} \frac{1}{p} E_{i,\cdot}^{\intercal} E_{t,\cdot}\right\|_2\lesssim \frac{\sqrt{q}\log N}{\sqrt{np}}.
\label{eq: separate bound 4 inter}
\end{equation}

On the event $\mathcal{G}_2\cap \mathcal{G}_5$, we have $$\left\|\frac{1}{n} {H}_{t,\cdot}\frac{1}{p}E_{t,\cdot}^{\intercal} E_{t,\cdot}\right\|_2\leq \frac{1}{n}\|H_{t,\cdot}\|_2 \frac{1}{p}E_{t,\cdot}^{\intercal} E_{t,\cdot}\lesssim \frac{\sqrt{q} (\log N)^{\frac{3}{2}}}{n}.
$$
Together with \eqref{eq: separate bound 4 inter}, we establish 
\begin{equation}
\left\|\frac{1}{n} \sum_{i=1}^{n} {H}_{i,\cdot} \frac{1}{p} E_{i,\cdot}^{\intercal} E_{t,\cdot}\right\|_2\lesssim \frac{\sqrt{q}\log N}{\sqrt{np}}+\frac{\sqrt{q} (\log N)^{\frac{3}{2}}}{n}
\label{eq: separate bound 4}
\end{equation}
On the event $\mathcal{G},$ we apply the decomposition \eqref{eq: key decomp 1} with the error bounds
\eqref{eq: separate bound 1},\eqref{eq: separate bound 2},\eqref{eq: separate bound 3}, \eqref{eq: separate bound 4} and then establish \eqref{eq: refined bound factor}. The upper bound in \eqref{eq: upper bound 1a} follows from the definition of $\mathcal{G}_2$ and \eqref{eq: inter bound 1}.

\subsection{Proof of Lemma \ref{lem: loading bound}}
\label{sec: loading proof}
By the definition of $\widetilde{\Psi}$ in \eqref{eq: factor def}, we now control the estimation error of $\widetilde{\Psi}_{\cdot,l}=\frac{1}{n}\widetilde{H}^{\intercal} X_{\cdot,l} \in \R^{q}.$
We start with the following decomposition,
\begin{equation}
\begin{aligned}
\widetilde{\Psi}_{\cdot,l}-\T^{-1}{\Psi}_{\cdot,l}&=\frac{1}{n}\widetilde{H}^{\intercal} (H \Psi_{\cdot,l}+E_{\cdot,l})-\T^{-1}{\Psi}_{\cdot,l}\\
&=\frac{1}{n}\widetilde{H}^{\intercal} (H \Psi_{\cdot,l}+E_{\cdot,l})-\T^{-1}{\Psi}_{\cdot,l}\\
&=\frac{1}{n}\widetilde{H}^{\intercal} \left((H\T-\widetilde{H}+\widetilde{H}) \T^{-1} \Psi_{\cdot,l}+E_{\cdot,l}\right)-\T^{-1}{\Psi}_{\cdot,l}\\
&=\frac{1}{n}\widetilde{H}^{\intercal} (H-\widetilde{H}\T^{-1})  \Psi_{\cdot,l}+\frac{1}{n}\T^{\intercal} H^{\intercal}  E_{\cdot,l}+\frac{1}{n}(\widetilde{H}-H\T)^{\intercal}  E_{\cdot,l}.
\end{aligned}
\label{eq: loading decomposition}
\end{equation}

To establish \eqref{eq: loading bound}, we control all three terms on the right-hand-side of \eqref{eq: loading decomposition}. 

\noindent
\underline{Control of $\frac{1}{n}\widetilde{H}^{\intercal} (H-\widetilde{H}\T^{-1})  \Psi_{\cdot,l}.$}
Note that 
\begin{equation}
\begin{aligned}
&\frac{1}{n}\widetilde{H}^{\intercal} (H-\widetilde{H}\T^{-1})  \Psi_{\cdot,l}\\
=&\frac{1}{n}\widetilde{H}^{\intercal} (H\T-\widetilde{H})  \T^{-1}\Psi_{\cdot,l}\\
=&\frac{1}{n}\T^{\intercal} H^{\intercal} (H\T-\widetilde{H})  \T^{-1}\Psi_{\cdot,l}+\frac{1}{n}(\widetilde{H}-H\T)^{\intercal} (H\T-\widetilde{H})  \T^{-1}\Psi_{\cdot,l}.\\
\end{aligned}
\end{equation}
On the event $\mathcal{G}$, it follows from \eqref{eq: bound factor} that 
\begin{equation}
\begin{aligned}
\|\frac{1}{n}(\widetilde{H}-H\T)^{\intercal} (H\T-\widetilde{H})  \T^{-1}\Psi_{\cdot,l}\|_2
&\leq  \frac{1}{n}\|\widetilde{H}-H\T\|_2^2 \|\T^{-1}\Psi_{\cdot,l}\|_2\\
&\lesssim \frac{q^{\frac{9}{2}}(\log N)^{\frac{7}{2}}}{\min\{n,p\}}\cdot \left(\frac{p}{\lambda_{q}^2(\Psi)}\right)^2.
\end{aligned}
\label{eq: loading separate bound 1}
\end{equation}
Since
\begin{equation*}
\frac{1}{n}\T^{\intercal} H^{\intercal} (H\T-\widetilde{H})  \T^{-1}\Psi_{\cdot,l}
=\frac{1}{n}\T^{\intercal} \sum_{t=1}^{n}H_{t,\cdot} (\T^{\intercal} H_{t,\cdot}-\widetilde{H}_{t,\cdot})^{\intercal}  \T^{-1}\Psi_{\cdot,l},
\end{equation*}
then on the event $\mathcal{G},$ we have 
\begin{equation}
\|\frac{1}{n}\T^{\intercal} H^{\intercal} (H\T-\widetilde{H})  \T^{-1}\Psi_{\cdot,l}\|_2\lesssim \|\frac{1}{n}\sum_{t=1}^{n}H_{t,\cdot} (\T^{\intercal} H_{t,\cdot}-\widetilde{H}_{t,\cdot})^{\intercal} \|_2 \sqrt{q\log N}.
\label{eq: simple upper bound}
\end{equation}
In the following, we shall control 
\begin{equation*}
\|\frac{1}{n}\sum_{t=1}^{n}H_{t,\cdot} (\T^{\intercal} H_{t,\cdot}-\widetilde{H}_{t,\cdot})^{\intercal} \|_2=\|\frac{1}{n}\sum_{t=1}^{n}(\T^{\intercal} H_{t,\cdot}-\widetilde{H}_{t,\cdot})H_{t,\cdot}^{\intercal}\|_2.
\end{equation*}
It follows from \eqref{eq: key decomp 0} that $\widehat{\Lambda}^{2}\frac{1}{n}\sum_{t=1}^{n}(\T^{\intercal} H_{t,\cdot}-\widetilde{H}_{t,\cdot})H_{t,\cdot}^{\intercal}
$ can be decomposed as
\begin{equation}
\begin{aligned}
 &\frac{1}{n}\sum_{t=1}^{n}\left(\frac{1}{n}\sum_{i=1}^{n}\widetilde{H}_{i,\cdot} \frac{1}{p}{H}_{i,\cdot}^{\intercal} \Psi E_{t,\cdot}+\frac{1}{n} \sum_{i=1}^{n} \widetilde{H}_{i,\cdot} \frac{1}{p}E_{i,\cdot}^{\intercal} \Psi^{\intercal} H_{t,\cdot}+\frac{1}{n} \sum_{i=1}^{n} \widetilde{H}_{i,\cdot} \frac{1}{p} E_{i,\cdot}^{\intercal} E_{t,\cdot}\right)H_{t,\cdot}^{\intercal}\\
 &=\left(\frac{1}{n}\sum_{i=1}^{n}\widetilde{H}_{i,\cdot} {H}_{i,\cdot}^{\intercal} \right)\left(\frac{1}{np}\sum_{t=1}^{n}\Psi E_{t,\cdot}H_{t,\cdot}^{\intercal}\right)+\left(\frac{1}{n} \sum_{i=1}^{n} \widetilde{H}_{i,\cdot} \frac{1}{p}E_{i,\cdot}^{\intercal} \Psi^{\intercal}\right) \left(\frac{1}{n}\sum_{t=1}^{n}H_{t,\cdot}H_{t,\cdot}^{\intercal}\right)\\
 &+\left(\frac{1}{n} \sum_{i=1}^{n} \widetilde{H}_{i,\cdot} \frac{1}{p} E_{i,\cdot}^{\intercal}\right)\left(\frac{1}{n}\sum_{t=1}^{n} E_{t,\cdot}H_{t,\cdot}^{\intercal}\right).
 \end{aligned}
\label{eq: loading decomposition 1}
\end{equation}

Note that 
\begin{equation*}
\frac{1}{n}\sum_{i=1}^{n}\widetilde{H}_{i,\cdot} {H}_{i,\cdot}^{\intercal}=\frac{1}{n}\sum_{i=1}^{n}\left(\widetilde{H}_{i,\cdot}-\T^{\intercal} H_{i,\cdot}\right) {H}_{i,\cdot}^{\intercal}
+\T^{\intercal} \frac{1}{n}\sum_{i=1}^{n}{H}_{i,\cdot} {H}_{i,\cdot}^{\intercal}
\end{equation*}
\begin{equation*}
\frac{1}{n} \sum_{i=1}^{n} \widetilde{H}_{i,\cdot} \frac{1}{p}E_{i,\cdot}^{\intercal} \Psi^{\intercal}=\frac{1}{n}\sum_{i=1}^{n}\left(\widetilde{H}_{i,\cdot}-\T^{\intercal} H_{i,\cdot}\right) \frac{1}{p}E_{i,\cdot}^{\intercal} \Psi^{\intercal}
+\T^{\intercal} \frac{1}{n}\sum_{i=1}^{n}{H}_{i,\cdot} \frac{1}{p}E_{i,\cdot}^{\intercal} \Psi^{\intercal}
\end{equation*}
\begin{equation*}
\frac{1}{n} \sum_{i=1}^{n} \widetilde{H}_{i,\cdot} \frac{1}{p} E_{i,\cdot}^{\intercal}=\frac{1}{n}\sum_{i=1}^{n}\left(\widetilde{H}_{i,\cdot}-\T^{\intercal} H_{i,\cdot}\right) \frac{1}{p} E_{i,\cdot}^{\intercal}
+\T^{\intercal} \frac{1}{n}\sum_{i=1}^{n}{H}_{i,\cdot} \frac{1}{p} E_{i,\cdot}^{\intercal}
\end{equation*}
On the event $\mathcal{G},$ we apply \eqref{eq: bound factor} and \eqref{eq: inter bound 1} and establish 
\begin{equation*}
\|\frac{1}{n}\sum_{i=1}^{n}\left(\widetilde{H}_{i,\cdot}-\T^{\intercal} H_{i,\cdot}\right) {H}_{i,\cdot}^{\intercal}\|_2\lesssim \frac{q^{5/2}(\log N)^2}{\sqrt{\min\{n,p\}}}\cdot \frac{p}{\lambda_{q}^2(\Psi)},
\end{equation*}
\begin{equation*}
\|\frac{1}{n}\sum_{i=1}^{n}\left(\widetilde{H}_{i,\cdot}-\T^{\intercal} H_{i,\cdot}\right) \frac{1}{p}E_{i,\cdot}^{\intercal} \Psi^{\intercal}\|_2\lesssim \frac{q^3 (\log N)^{\frac{5}{2}}}{\sqrt{\min\{n,p\}}\sqrt{p}}\cdot \frac{p}{\lambda_{q}^2(\Psi)},
\end{equation*}
\begin{equation*}
\|\frac{1}{n}\sum_{i=1}^{n}\left(\widetilde{H}_{i,\cdot}-\T^{\intercal} H_{i,\cdot}\right) \frac{1}{p}E_{i,\cdot}^{\intercal} \|_2\lesssim \frac{q^{2} (\log N)^{\frac{5}{2}}}{\sqrt{\min\{n,p\}}\sqrt{p}} \cdot \frac{p}{\lambda_{q}^2(\Psi)}. 
\end{equation*}
On the event $\mathcal{G},$ we have
\begin{equation}
\left\|\frac{1}{n}\sum_{i=1}^{n}\widetilde{H}_{i,\cdot} {H}_{i,\cdot}^{\intercal}\right\|_2\lesssim 1+\frac{q^{5/2}(\log N)^2}{\sqrt{\min\{n,p\}}}\cdot \frac{p}{\lambda_{q}^2(\Psi)}.
\label{eq: matrix approx 1}
\end{equation}
\begin{equation}
\max\left\{\left\|\frac{1}{n} \sum_{i=1}^{n} \widetilde{H}_{i,\cdot} \frac{1}{p}E_{i,\cdot}^{\intercal} \Psi^{\intercal}\right\|_2,\left\|\frac{1}{n} \sum_{i=1}^{n} \widetilde{H}_{i,\cdot} \frac{1}{p} E_{i,\cdot}^{\intercal}\right\|_2\right\}\lesssim \frac{q^3 (\log N)^{\frac{5}{2}}}{\sqrt{\min\{n,p\}}\sqrt{p}}\cdot \frac{p}{\lambda_{q}^2(\Psi)}.
\label{eq: matrix approx 2}
\end{equation}
Then on the event $\mathcal{G}$, we have established that 
\begin{equation*}
\|\widehat{\Lambda}^{2}\frac{1}{n}\sum_{t=1}^{n}(\T^{\intercal} H_{t,\cdot}-\widetilde{H}_{t,\cdot})H_{t,\cdot}^{\intercal}\|_2 \lesssim \frac{q^{3} (\log N)^{\frac{5}{2}}}{\sqrt{\min\{n,p\}}\sqrt{p}}\cdot \frac{p}{\lambda_{q}^2(\Psi)}\cdot (1+\sqrt{p/n}).
\end{equation*}

Together with \eqref{eq: loading separate bound 1} and \eqref{eq: simple upper bound}, we have 
\begin{equation}
\left\|\frac{1}{n}\widetilde{H}^{\intercal} (H-\widetilde{H}\T^{-1})  \Psi_{\cdot,l}\right\|_2\lesssim \frac{q^{\frac{9}{2}}(\log N)^{\frac{7}{2}}}{\min\{n,p\}}\cdot \left(\frac{p}{\lambda_{q}^2(\Psi)}\right)^2.
\label{eq: loading accuracy 1} 
\end{equation}

\noindent
\underline{Control of $\frac{1}{n}\T^{\intercal} H^{\intercal}  E_{\cdot,l}$} Note that 
$$\frac{1}{n}\T^{\intercal} H^{\intercal}  E_{\cdot,l}=\T^{\intercal}\frac{1}{n}\sum_{i=1}^{n} H_{i,\cdot} E_{i,l}.$$
On the event $\mathcal{G}_{10}\cap \mathcal{G}_{12},$ we have 
\begin{equation}
\left\|\frac{1}{n}\T^{\intercal} H^{\intercal}  E_{\cdot,l}\right\|_2\lesssim \sqrt{\frac{q \log p}{n}}.
\label{eq: loading accuracy 3}
\end{equation}

\noindent
\underline{Control of $\frac{1}{n}(\widetilde{H}-H\T)^{\intercal}  E_{\cdot,l}$.}
It follows from \eqref{eq: key decomp 0} that
the term $\widehat{\Lambda}^{2}\frac{1}{n}(\widetilde{H}-H\T)^{\intercal}  E_{\cdot,l}$ can be decomposed as 

\begin{equation}
\begin{aligned}
& \frac{1}{n}\sum_{t=1}^{n}\left(\frac{1}{n}\sum_{i=1}^{n}\widetilde{H}_{i,\cdot} \frac{1}{p}{H}_{i,\cdot}^{\intercal} \Psi E_{t,\cdot}+\frac{1}{n} \sum_{i=1}^{n} \widetilde{H}_{i,\cdot} \frac{1}{p}E_{i,\cdot}^{\intercal} \Psi^{\intercal} H_{t,\cdot}+\frac{1}{n} \sum_{i=1}^{n} \widetilde{H}_{i,\cdot} \frac{1}{p} E_{i,\cdot}^{\intercal} E_{t,\cdot}\right)E_{t,l}\\
 &=\left(\frac{1}{n}\sum_{i=1}^{n}\widetilde{H}_{i,\cdot} {H}_{i,\cdot}^{\intercal} \right)\frac{1}{p}\Psi \left(\frac{1}{n}\sum_{t=1}^{n}E_{t,\cdot}E_{t,l}-(\Sigma_{E})_{\cdot,l}\right)+\left(\frac{1}{n}\sum_{i=1}^{n}\widetilde{H}_{i,\cdot} {H}_{i,\cdot}^{\intercal} \right)\frac{1}{p}\Psi(\Sigma_{E})_{\cdot,l}\\
 &+\left(\frac{1}{n} \sum_{i=1}^{n} \widetilde{H}_{i,\cdot} \frac{1}{p}E_{i,\cdot}^{\intercal} \Psi^{\intercal}\right) \left(\frac{1}{n}\sum_{t=1}^{n}H_{t,\cdot}E_{t,l}\right)\\
 &+\left(\frac{1}{n} \sum_{i=1}^{n} \widetilde{H}_{i,\cdot} \frac{1}{p} E_{i,\cdot}^{\intercal}\right)\left(\frac{1}{n}\sum_{t=1}^{n} E_{t,\cdot}E_{t,l}-(\Sigma_{E})_{\cdot,l}\right)+\left(\frac{1}{n} \sum_{i=1}^{n} \widetilde{H}_{i,\cdot} \frac{1}{p} E_{i,\cdot}^{\intercal}\right)(\Sigma_{E})_{\cdot,l}.
 \end{aligned}
\label{eq: loading decomposition 2}
\end{equation}
On the event $\mathcal{G}_{11}\cap \mathcal{G}_{12}$, together with the fact that $\lambda_{\max}(\Sigma_{E})\leq C$ for some positive constant $C>0,$ we apply \eqref{eq: matrix approx 1} and  \eqref{eq: matrix approx 2} and establish 
\begin{align*}
&\left|\left(\frac{1}{n}\sum_{i=1}^{n}\widetilde{H}_{i,\cdot} {H}_{i,\cdot}^{\intercal} \right)\frac{1}{p}\Psi \left(\frac{1}{n}\sum_{t=1}^{n}E_{t,\cdot}E_{t,l}-(\Sigma_{E})_{\cdot,l}\right)\right|+\left|\left(\frac{1}{n}\sum_{i=1}^{n}\widetilde{H}_{i,\cdot} {H}_{i,\cdot}^{\intercal} \right)\frac{1}{p}\Psi(\Sigma_{E})_{\cdot,l}\right|\\
&\lesssim \frac{\lambda_{1}(\Psi)}{\sqrt{p}}\left(\sqrt{\frac{\log p}{n}}+\frac{1}{\sqrt{p}}\right)\left(1+\frac{q^{5/2}(\log N)^2}{\sqrt{\min\{n,p\}}}\cdot \frac{p}{\lambda_{q}^2(\Psi)}\right),
\end{align*}
$$\left|\left(\frac{1}{n} \sum_{i=1}^{n} \widetilde{H}_{i,\cdot} \frac{1}{p}E_{i,\cdot}^{\intercal} \Psi^{\intercal}\right) \left(\frac{1}{n}\sum_{t=1}^{n}H_{t,\cdot}E_{t,l}\right)\right| \lesssim \frac{q^{\frac{7}{2}} (\log N)^{\frac{5}{2}}}{\sqrt{\min\{n,p\}}\sqrt{p}}\sqrt{\frac{\log p}{n}}\cdot \frac{p}{\lambda_{q}^2(\Psi)}.$$
\begin{equation*}
\begin{aligned}
&\left|\left(\frac{1}{n} \sum_{i=1}^{n} \widetilde{H}_{i,\cdot} \frac{1}{p} E_{i,\cdot}^{\intercal}\right)\left(\frac{1}{n}\sum_{t=1}^{n} E_{t,\cdot}E_{t,l}-(\Sigma_{E})_{\cdot,l}\right)+\left(\frac{1}{n} \sum_{i=1}^{n} \widetilde{H}_{i,\cdot} \frac{1}{p} E_{i,\cdot}^{\intercal}\right)(\Sigma_{E})_{\cdot,l}\right|\\
&\lesssim \frac{q^3 (\log N)^{\frac{5}{2}}}{\sqrt{\min\{n,p\}}\sqrt{p}}\left(\sqrt{\frac{p\log p}{n}}+1\right)\cdot \frac{p}{\lambda_{q}^2(\Psi)}.
\end{aligned}
\end{equation*}
By the above bounds, we apply the decomposition
\eqref{eq: loading decomposition 2} and establish 
\begin{equation}
\begin{aligned}
\left|\widehat{\Lambda}^{2}\frac{1}{n}(\widetilde{H}-H\T)^{\intercal}  E_{\cdot,l}\right|\lesssim \sqrt{\frac{\log p}{n}}+\frac{1}{\sqrt{p}}+\frac{q^3 (\log N)^{\frac{5}{2}}}{\sqrt{\min\{n,p\}}\sqrt{p}}\left(\sqrt{\frac{p\log p}{n}}+1\right)\cdot \frac{p}{\lambda_{q}^2(\Psi)}.
\end{aligned}
\label{eq: loading accuracy 2}
\end{equation}


A combination of \eqref{eq: loading accuracy 1}, \eqref{eq: loading accuracy 2} and \eqref{eq: loading accuracy 3} leads to  \eqref{eq: loading bound}.

\subsection{Proof of Lemma \ref{lem: high prob event}}
\label{sec: high prob event}
\noindent \underline{Control of $\mathcal{G}_1\cap \mathcal{G}_2\cap \mathcal{G}_3$.}
By the equation (5.23) of \citep{vershynin2010introduction}, with probability larger than $1-p^{-c},$ the event $\mathcal{G}_1$ holds. Note that 
$$\max_{1\leq i\leq n} \|H_{i,\cdot}\|_2\leq \sqrt{q} \max_{1\leq i\leq n, 1\leq j\leq q}|H_{i,j}|\quad \text{and}\quad \max_{1\leq t\leq n}\|\Psi^{\intercal} H_{t,\cdot}/p\|_{2}\leq \frac{1}{\sqrt{p}} \max_{1\leq t\leq n, 1\leq j\leq p}\left|\Psi_{j}^{\intercal}H_{t,\cdot}\right|$$
Since $\{{H}_{i,\cdot}\}_{1\leq i\leq n}$ are i.i.d. sub-Gaussian vectors, with probability larger than $1-(p)^{-c},$$$\max_{1\leq i\leq n, 1\leq j\leq q}|H_{i,j}|\lesssim \sqrt{\log (nq)}$$
and  
$$\max_{1\leq t\leq n, 1\leq j\leq p}\left|\Psi_{j}^{\intercal}H_{t,\cdot}\right|\lesssim \|\Psi_{j}\|_2\cdot \sqrt{\log (np)}\lesssim \sqrt{q}\sqrt{\log (pq) \log (np)}$$
Hence, we establish 
$$\PP\left(\mathcal{G}_1\cap \mathcal{G}_2\cap \mathcal{G}_3\right)\geq 1-p^{-c}.$$ 

\noindent \underline{Control of $\mathcal{G}_4\cap \mathcal{G}_5\cap \mathcal{G}_6$.}
For any $1\leq j\leq q,$ $\frac{1}{\|\Psi_{j,\cdot}\|_2}\Psi_{j,\cdot}^{\intercal} E_{t,\cdot}$ is sub-Gaussian random variable and this leads to $\PP(\mathcal{G}_4)\geq 1-p^{-c}.$ We also have 
$\PP\left(\max_{t,j} |E_{t,j}|\lesssim \sqrt{\log (np)}\right)\geq 1-(np)^{-c},$ which leads to $\PP(\mathcal{G}_5)\geq 1-p^{-c}.$ 

We fix $1\leq t\leq n$ and consider $i\neq t$. Conditioning on $E_{t,\cdot},$ the random variable $E_{i,\cdot}^{\intercal} E_{t,\cdot}/p$ is a zero-mean sub-Gaussian random variable with sub-Gaussian norm $\|E_{t,\cdot}\|_2/p.$ On the event $\mathcal{G}_5$, we establish 
\begin{equation}
\PP\left(\max_{i\neq t}|E_{i,\cdot}^{\intercal} E_{t,\cdot}/p|\lesssim \sqrt{\log p}\|E_{t,\cdot}\|_2/p\mid E_{t,\cdot}\right)\geq 1-p^{-c}.
\label{eq: inter prob 1}
\end{equation}
Note that 
\begin{equation}
\begin{aligned}
&\PP\left(\max_{i\neq t}|E_{i,\cdot}^{\intercal} E_{t,\cdot}/p|\lesssim \sqrt{\log p}\sqrt{p \log (np)}/p\right)\\
&\geq \PP\left(\max_{i\neq t}|E_{i,\cdot}^{\intercal} E_{t,\cdot}/p|\lesssim \sqrt{\log p}\|E_{t,\cdot}\|_2/p, \; \|E_{t,\cdot}\|_2\lesssim \sqrt{p\log (np)}\right)\\
&\geq \int \PP\left(\max_{i\neq t}|E_{i,\cdot}^{\intercal} E_{t,\cdot}/p|\lesssim \sqrt{\log p}\|E_{t,\cdot}\|_2/p\mid E_{t,\cdot}\right) {\bf 1}(\|E_{t,\cdot}\|_2\lesssim \sqrt{p\log (np)}) \mu(E_{t,\cdot})
\end{aligned}
\end{equation}
where $\mu(E_{t,\cdot})$ denotes the measure of $E_{t,\cdot}.$
Combined with \eqref{eq: inter prob 1}, we establish that, for a given $1\leq t\leq n,$ 
$$\PP\left(\max_{i\neq t}|E_{i,\cdot}^{\intercal} E_{t,\cdot}/p|\lesssim \sqrt{\log p}\sqrt{p \log (np)}/p\right)\geq (1-p^{-c})\cdot \PP(\mathcal{G}_5)\geq 1-p^{-c},$$
where $c>1$ is some positive constant. By applying another union bound, we establish $\PP(\mathcal{G}_6)\geq 1-p^{-(c-1)}.$
Hence, we establish
$$\PP\left(\mathcal{G}_4\cap \mathcal{G}_5\cap \mathcal{G}_6\right)\geq 1-p^{-c}.$$ 

\noindent \underline{Control of $\mathcal{G}_7$.}
For any vector $u\in \R^{q}$ and $v\in \R^{q},$ we have 
$$\left\|\frac{1}{n}\sum_{i=1}^{n} H_{i\cdot} \frac{1}{p} E_{i\cdot}^{\intercal}\Psi^{\intercal}\right\|_2=\sup_{u,v\in \R^{q}, \|u\|_2=1, \|v\|_2=1} u^{\intercal} \frac{1}{n}\sum_{i=1}^{n} H_{i\cdot} \frac{1}{p} E_{i\cdot}^{\intercal}\Psi^{\intercal}v$$
Since $H_{i\cdot}$ and $E_{i,\cdot}$ are sub-Gaussian random vectors, the random variable $u^{\intercal}  H_{i\cdot} \frac{1}{p} E_{i\cdot}^{\intercal}\Psi^{\intercal}v$ is zero-mean with sub-exponential norm upper bounded by $C\frac{\|\Psi^{\intercal}v\|_2}{p}\lesssim \frac{\lambda_1(\Psi)}{{p}}.$ We apply Corollary 5.17 of \citep{vershynin2010introduction} and establish that, for $t\leq \sqrt{n},$
$$\PP\left(\left|u^{\intercal} \frac{1}{n}\sum_{i=1}^{n} H_{i\cdot} \frac{1}{p} E_{i\cdot}^{\intercal}\Psi^{\intercal}v\right|\gtrsim \frac{t}{\sqrt{n}}\cdot \frac{\lambda_1(\Psi)}{{p}} \right)\leq \exp(-ct^2).$$
We shall use $\mathcal{N}_{q}$ to denote the $\epsilon$-net of the unit ball in $\R^{q}$; see the definition of $\epsilon$-net in Definition 5.1 in \citep{vershynin2010introduction}.
Taking the union bound over all vectors $u,v \in \mathcal{N}_{q},$ we have 
\begin{equation}
\PP\left(\max_{u,v\in \mathcal{N}_q}\left|u^{\intercal} \frac{1}{n}\sum_{i=1}^{n} H_{i\cdot} \frac{1}{p} E_{i\cdot}^{\intercal}\Psi^{\intercal}v\right|\gtrsim \frac{t}{\sqrt{n}}\cdot \frac{\lambda_1(\Psi)}{{p}}\right)\leq |\mathcal{N}_{q}|^2\exp(-ct^2).
\label{eq: net bound 1}
\end{equation}
where $c>0$ is some positive constant. 
We choose $t^2=C \log (|\mathcal{N}_{q}|^2\cdot p)\leq \sqrt{n}$ for some positive constant $C>0$ such that $|\mathcal{N}_{q}|^2\exp(-ct^2)\leq p^{-c'}$ for some positive constant $c'>0.$ By Lemmas 5.2 and 5.3 of \citep{vershynin2010introduction}, we take $|\mathcal{N}_{q}|^2=C^{2q}$ and apply \eqref{eq: net bound 1} to establish that $\PP(\mathcal{G}_7)\geq 1-p^{-c}.$

\noindent \underline{Control of $\mathcal{G}_8\cap \mathcal{G}_9$.} By Theorem 5.39 of \citep{vershynin2010introduction}, we establish that $\PP(\mathcal{G}_8)\geq 1-\exp(-c\min\{n,p\}).$
Since $\|H^{\intercal}E\|_2\leq \|H\|_2\|E\|_2,$ on the event $\mathcal{G}_8$, the event $\mathcal{G}_9$ holds. That is, we establish that $\PP(\mathcal{G}_8\cap \mathcal{G}_9)\geq 1-\exp(-c\min\{n,p\}).$

\noindent \underline{Control of $\mathcal{G}_{10}$.}
We start with the decomposition
\begin{equation*}
\frac{1}{np}XX^{\intercal}-\frac{1}{np} H\Psi\Psi^{\intercal}H^{\intercal}=\frac{1}{np} H\Psi E^{\intercal} +\frac{1}{np} E \Psi^{\intercal}H^{\intercal}+\frac{1}{np} EE^{\intercal} 
\end{equation*}
On the event $\mathcal{G}_8$, we have 
$$\|\frac{1}{np} E^{\intercal}E\|_2\leq \frac{1}{np}\|E\|_2^2\lesssim \frac{1}{n}+\frac{1}{p}$$
$$\left\|\frac{1}{np} H\Psi E^{\intercal}\right\|_2\leq \frac{1}{np}\|H\|_2\|\Psi\|_2\|E\|_2\lesssim \frac{1}{\sqrt{n}}+\frac{1}{\sqrt{p}}.$$
Then we have 
\begin{equation}
\left\|\frac{1}{np}XX^{\intercal}-\frac{1}{np} H\Psi\Psi^{\intercal}H^{\intercal}\right\|_2\lesssim \frac{1}{\sqrt{n}}+\frac{1}{\sqrt{p}}.
\label{eq: spectrum bound 1}
\end{equation}
Note that the top $q$ eigenvalues of $\frac{1}{np} H\Psi\Psi^{\intercal}H^{\intercal}$ are the same as the top $q$ eigenvalues of $\frac{1}{np} \Psi^{\intercal}H^{\intercal} H\Psi.$ We have 
\begin{equation}
\frac{1}{np} \Psi^{\intercal}H^{\intercal} H\Psi=\frac{1}{p}\Psi^{\intercal}(H^{\intercal} H/n-{\rm I}) \Psi+\frac{1}{p}\Psi^{\intercal}\Psi
\end{equation}
On the event $\mathcal{G}_1,$ we have 
$$\|\frac{1}{p}\Psi^{\intercal}(H^{\intercal} H/n-{\rm I}) \Psi\|_2\lesssim \sqrt{\frac{q+\log p}{n}}\cdot\frac{\lambda_1^2(\Psi)}{p}.$$
Note that the top $q$ eigenvalues of $\frac{1}{p}\Psi^{\intercal}\Psi$ are the same as the top $q$
 eigenvalues of $\frac{1}{p}\Psi\Psi^{\intercal}.$ 
 Hence, we have 
 \begin{equation}
\max_{1\leq i\leq q}\left|\lambda_{i}\left(\frac{1}{np} \Psi^{\intercal}H^{\intercal} H\Psi\right)-\lambda_{i}\left(\frac{1}{p}\Psi\Psi^{\intercal}\right)\right|\lesssim \sqrt{\frac{q+\log p}{n}}\cdot\frac{\lambda_1^2(\Psi)}{p}.
\label{eq: spectrum bound 2}
 \end{equation}
 A combination of \eqref{eq: spectrum bound 1} and \eqref{eq: spectrum bound 2} leads to 
 \begin{equation*}
 \max_{1\leq i\leq q}\left|\lambda_{i}\left( \frac{1}{np}XX^{\intercal}\right)-\lambda_{i}\left(\frac{1}{p}\Psi\Psi^{\intercal}\right)\right|\lesssim \sqrt{\frac{q+\log p}{n}}\cdot\frac{\lambda_1^2(\Psi)}{p}+\frac{1}{\sqrt{p}}
 \end{equation*}
 By \eqref{eq: strong factor}, there exists positive constants $C\geq c>0$ such that
 $$c \frac{\lambda_q^2(\Psi)}{p} \leq \lambda_{\min}(\widehat{\Lambda}^2)\leq \lambda_{\max}(\widehat{\Lambda}^2)\leq C\frac{\lambda_1^2(\Psi)}{p}.$$
 By the definition of $\T$ in \eqref{eq: rotation random}, we have 
 $$\|\T\|_2 \leq \|\Psi\Psi^{\intercal}/p\|_2 \|{H}\|_2 \|\widetilde{H}\|_2\frac{1}{n}\|\widehat{\Lambda}^{-2}\|_2.$$
With probability larger than $1-p^{-c}-\exp(-cn),$ $$\|\widehat{\Lambda}^{-2}\|_2\lesssim \frac{p}{\lambda_q^2(\Psi)} \quad \text{and} \quad \|H\|_2\lesssim \sqrt{n}.$$ Applying the above inequality together with the fact that $\|\Psi\Psi^{\intercal}/p\|_2\leq \lambda_1^2(\Psi)/p$, $\|\widetilde{H}\|_2=\sqrt{n}$ and $\lambda_1(\Psi)/\lambda_{q}(\Psi)\leq C,$ we establish that, with probability larger than $1-p^{-c}-\exp(-cn),$ $$\|\T\|_2\leq C',$$ for some positive constant $C'>0.$ That is, 
$\PP(\mathcal{G}_{10})\geq 1-p^{-c}-\exp(-cn).$ 

\noindent \underline{Control of $\mathcal{G}_{11}\cap \mathcal{G}_{12}$.} The proofs follows from the fact that $E_{t,j}E_{t,l}-(\Sigma_{E})_{j,l}$ and $H_{t,j}E_{t,l}$ are zero mean sub-exponential random variable. We apply Corollary 5.17 of \citep{vershynin2010introduction} and the union bound to establish $\PP(\mathcal{G}_{11}\cap \mathcal{G}_{12})\geq 1-p^{-c}$ for some positive constant $c>0.$

\section{Additional Proofs}
\label{sec:proofs}
\subsection{Proof of Proposition \ref{prop: noise consistency}}
\label{sec: var consistency proof}
We note that 
$$\mathcal{Q}y-\mathcal{Q} X \betainit=\mathcal{Q}e+\mathcal{Q}\Delta+\mathcal{Q} X (\beta-\betainit)+\mathcal{Q} X b,$$
where $\Delta_i=\psi^{\intercal}H_{i, \cdot}-b^{\intercal}X_{i, \cdot}$ for $1\leq i\leq n.$

Then we have 
\begin{equation}
\begin{aligned}
\widehat{\sigma}_{e}^2-\sigma_{e}^2&=\frac{\|\mathcal{Q}e\|_2^2}{{\rm Tr}(\mathcal{Q}^2)}-\sigma_{e}^2+\frac{1}{{\rm Tr}(\mathcal{Q}^2)}\|\mathcal{Q}\Delta+\mathcal{Q} X (\beta-\betainit)+\mathcal{Q} X b\|_2^2\\
&+\frac{1}{{\rm Tr}(\mathcal{Q}^2)}e^{\intercal}\mathcal{Q}^2 \Delta+\frac{1}{{\rm Tr}(\mathcal{Q}^2)}e^{\intercal}\mathcal{Q}^2 X (\beta-\betainit)+\frac{1}{{\rm Tr}(\mathcal{Q}^2)}e^{\intercal}\mathcal{Q}^2 X b.
\end{aligned}
\label{eq: decomposition}
\end{equation}
The following analysis is to study the above decomposition term by term. First note that 
\begin{equation*}
\frac{\|\mathcal{Q}e\|_2^2}{{\rm Tr}(\mathcal{Q}^2)}-\sigma_{e}^2=\frac{e^{\intercal}U S^2 U^{\intercal}e}{{\rm Tr}(\mathcal{Q}^2)}-\sigma_{e}^2.
\end{equation*}
By Lemma \ref{lem: HW-bound} in Section \ref{sec: proof limiting}, we establish that with probability larger than $1-\exp(-ct^2)$ for $0<t\lesssim {\rm Tr}(\mathcal{Q}^4) \asymp n,$
\begin{equation}
\left|\frac{e^{\intercal}\mathcal{Q}e}{{\rm Tr}(\mathcal{Q}^2)}-\sigma_{e}^2\right|\lesssim t \frac{\sqrt{{\rm Tr}(\mathcal{Q}^4)}}{{\rm Tr}(\mathcal{Q}^2)}.
\label{eq: term a}
\end{equation}
By \eqref{eq: approximation expectation}, we show that
\begin{equation}
\mathbf{P}\left(\frac{1}{n}\|\Delta\|_2^2\lesssim q{\log p}/p\right)\geq 1-(\log p)^{-1/2}.
\label{eq: l2 control}
\end{equation}

Since $e_i$ is independent of $X_{i, \cdot}$ and $H_{i, \cdot}$, the term $\frac{1}{{\rm Tr}(\mathcal{Q}^2)}e^{\intercal}\mathcal{Q}^2 \Delta$ is of mean zero and variance 
$$\frac{1}{{\rm Tr}^2(\mathcal{Q}^2)}\sigma_{e}^2\|\mathcal{Q}^2 \Delta\|_2^2\lesssim \frac{\sigma_e^2}{n^2}\|\Delta\|_2^2,$$
where the inequality follows from ${\rm Tr}(\mathcal{Q}^2)\asymp m\asymp n$ and$\|\mathcal{Q}^2\Delta\|_2\leq \|\Delta\|_2.$ 
Together with \eqref{eq: l2 control}, we establish that, with probability larger than $1-(\log p)^{-1/2}-\frac{1}{t^2}$ for some $t>0,$
\begin{equation}
\left|\frac{1}{{\rm Tr}(\mathcal{Q}^2)}e^{\intercal}\mathcal{Q}^2 \Delta\right|\lesssim t \sqrt{\frac{q\log p}{np}} \sigma_e.
\label{eq: approx variance bound}
\end{equation}
Since ${\rm Tr}(\mathcal{Q}^2)\asymp m\asymp n$ and$\|\mathcal{Q}\Delta\|_2\leq \|\Delta\|_2$, we have 
\begin{equation}
\begin{aligned}
\frac{1}{{\rm Tr}(\mathcal{Q}^2)}\|\mathcal{Q}\Delta+\mathcal{Q} X (\beta-\betainit)+\mathcal{Q} X b\|_2^2&\lesssim\frac{1}{n}\|\mathcal{Q}\Delta\|_2^2+\frac{1}{n}\|\mathcal{Q} X (\beta-\betainit)\|_2^2+\frac{1}{n}\|\mathcal{Q} X b\|_2^2\\
&\lesssim \frac{q\log p}{p}+\Mq^2\frac{k \log p}{n}+\frac{1}{n}\|\mathcal{Q} X b\|_2^2\\
\end{aligned}
\label{eq: term b}
\end{equation}
with probability larger than $1-(\log p)^{-1/2}.$

Recall that $\widetilde{W}\in \R^{p\times p}$ as a diagonal matrix with  diagonal entries as $\widetilde{W}_{l,l}={\|\mathcal{Q} X_{\cdot,l}\|_2}/{\sqrt{n}}$ for $1\leq l \leq p.$ 
We establish that 
\begin{equation}
\left|\frac{1}{{\rm Tr}(\mathcal{Q}^2)}e^{\intercal}\mathcal{Q}^2 X (\beta-\betainit)\right|\lesssim \|\frac{1}{n}e^{\intercal}\mathcal{Q}^2 X\widetilde{W}^{-1}\|_\infty\|\widetilde{W}(\beta-\betainit)\|_{1}\lesssim \Mq^2 k\lambda^2+\left(\frac{\|\mathcal{Q} X b\|_2}{\sqrt{n}}\right)^2,
\label{eq: term c}
\end{equation}
where the last inequality follows from \eqref{eq: accuracy-beta-1} and \eqref{eq: tuning beta}.

Finally, we control $\frac{1}{{\rm Tr}(\mathcal{Q}^2)}e^{\intercal}\mathcal{Q}^2 X b$, which has mean zero and variance
$$\E\left(\frac{1}{{\rm Tr}(\mathcal{Q}^2)}e^{\intercal}\mathcal{Q}^2 X b\right)^2 \lesssim \frac{1}{n^2}\sigma_{e}^2 b^{\intercal}X^{\intercal}\mathcal{Q}^4 X b\leq \frac{\sigma_e^2}{n^2}\|X^{\intercal}\mathcal{Q}^2 X\|_2 \|b\|_2^2$$
and hence with probability larger than $1-\frac{1}{t^2}$ for some $t>0,$
\begin{equation}
\frac{1}{n}e^{\intercal}\mathcal{Q}^2 X b \lesssim \frac{t}{\sqrt{n}}\frac{1}{\sqrt{n}}\|\mathcal{Q}X\|_2 \|b\|_2.
\label{eq: term d}
\end{equation}
A combination of the decomposition \eqref{eq: decomposition} and the error bounds \eqref{eq: term a}, \eqref{eq: approx variance bound}, \eqref{eq: term b}, \eqref{eq: term c}, \eqref{eq: term d} and \eqref{eq: approx proj error Q} leads to Proposition \ref{prop: noise consistency}.

\subsection{Proof of Proposition \ref{prop: fund-spectral}}
\label{sec: proof spectral}
By the Wely's inequality, we have that, for $1\leq l\leq m,$
\begin{equation}
\left|\lambda_{l}(X)-\lambda_{l}(H\Psi)\right|=\left|\lambda_{l}(H\Psi+E)-\lambda_{l}(H\Psi)\right|\leq \lambda_{1}(E).
\label{eq: wely singular}
\end{equation}
By Theorem 5.39 and equation (5.26) in \citep{vershynin2010introduction} and $\lambda_{\max}(\Sigma_{E})\leq C_0,$ with probability larger than $1-\exp(-cn)$ for some $c>0,$ $$\lambda_{\max}(E)\lesssim \max\{\sqrt{n},\sqrt{p}\}.$$
Note that $\lambda_{l}\left(\frac{1}{n}X X^{\intercal}\right)=\frac{1}{n}\lambda^2_{l}(X).$ Since $\lambda_{q+1}(H\Psi)=0,$ we establish the proposition by applying \eqref{eq: wely singular}.

\subsection{Proof of Lemma \ref{lem: decomposition lemma}}
\label{sec: proof decomp}
We express the model \eqref{eq: confounder model} as 
$$X_{1,j}=\Psi_{j}^{\intercal}H_{1,\cdot}+E_{1,j}, \quad X_{1,-j}=\Psi_{-j}^{\intercal}H_{1,\cdot}+E_{1,-j},$$
where $\Psi_{j}\in \R^{q}$ denotes the $j$-th column of $\Psi_{j}$ and $\Psi_{-j} \in \R^{q \times (p-1)}$ denotes the sub-matrix of $\Psi$ except for the $j$-th column. We define $B=\E E_{1,-j}E_{1,-j}^{\intercal}.$
Since ${\rm Cov}(H_{i, \cdot})={\rm I}_{q\times q}$ and the components of $H_{i, \cdot}$ are uncorrelated with the components of $\E_{i,\cdot}$, then we have 
 \begin{equation}
 \begin{aligned}
\gamma= [\E(X_{1,-j}X_{1,-j}^{\intercal})]^{-1}\E(X_{1,-j} X_{1,j})=\left(\Psi_{-j}^{\intercal} \Psi_{-j}+\BM\right)^{-1}\left(\Psi_{-j}^{\intercal}\Psi_{j}+\E E_{1,j}E_{1,-j}\right).
\end{aligned}
\label{eq: gamma expression}
\end{equation}
 We apply Woodbury matrix identity and then have 
 \begin{equation}
\begin{aligned}
\left(\Psi_{-j}^{\intercal} \Psi_{-j}+\BM\right)^{-1}=\BM^{-1}-\BM^{-1}\Psi_{-j}^{\intercal}\left({\rm I}+\Psi_{-j} \BM^{-1}\Psi_{-j}^{\intercal}\right)^{-1}\Psi_{-j}\BM^{-1}.
\end{aligned}
\label{eq: inter Woodbury}
\end{equation}
We combine the above two equalities and establish the decomposition $\gamma=\gamma^{E}+\gamma^{A}$ with 
$$
\gamma^{E}=\BM^{-1}\E E_{1,j}E_{1,-j}
$$
and 
\begin{equation}
\gamma^{A}=\left(\Psi_{-j}^{\intercal} \Psi_{-j}+\BM\right)^{-1}\Psi_{-j}^{\intercal}\Psi_{j}-\BM^{-1}\Psi_{-j}^{\intercal}\left({\rm I}+\Psi_{-j}\BM^{-1}\Psi_{-j}^{\intercal} \right)^{-1}\Psi_{-j}\gamma^{E}.
\label{eq: approx expression}
\end{equation}
\noindent \underline{Proof of \eqref{eq: approximation decoup}.}
We define $D=\Psi_{-j} B^{-\frac{1}{2}} \in \R^{q \times (p-1)}$ and hence the first component on the right hand side of \eqref{eq: approx expression} can be expressed as 
$$\left(\Psi_{-j}^{\intercal} \Psi_{-j}+\BM\right)^{-1}\Psi_{-j}^{\intercal}\Psi_{j}=\BM^{-\frac{1}{2}}\left(D^{\intercal}D+{\rm I}\right)^{-1}D^{\intercal}\Psi_{j}.$$
By  Woodbury matrix identity, we have 
$$\left(D^{\intercal}D+{\rm I}\right)^{-1}D^{\intercal}=\left({\rm I}-D^{\intercal}({\rm I}+D D^{\intercal})^{-1}D\right)D^{\intercal}=D^{\intercal}({\rm I}+D D^{\intercal})^{-1}$$
and hence 
\begin{equation}
\left(\Psi_{-j}^{\intercal} \Psi_{-j}+\BM\right)^{-1}\Psi_{-j}^{\intercal}\Psi_{j}=\BM^{-\frac{1}{2}}D^{\intercal}({\rm I}+D D^{\intercal})^{-1}\Psi_{j}.
\label{eq: term 1}
\end{equation}
The second component on the right hand side of \eqref{eq: approx expression} can be expressed as 
$$B^{-\frac{1}{2}}D^{\intercal}\left({\rm I}+DD^{\intercal}\right)^{-1}\Psi_{-j}\gamma^{E}.$$
Together with \eqref{eq: term 1}, we simplify \eqref{eq: approx expression} as
\begin{equation}
\gamma^{A}=B^{-\frac{1}{2}}D^{\intercal}\left({\rm I}+DD^{\intercal}\right)^{-1}\left(\Psi_j+\Psi_{-j}\gamma^{E}\right).
\label{eq: re-exp}
\end{equation}
 Under the assumption that $c_0\leq \lambda_{\min}(\Omega_{E})\leq \lambda_{\max}(\Omega_{E})\leq C_0$, we introduce the SVD for D as $D=U(D)\Lambda(D) V(D)^{\intercal}$, where $U(D),\Lambda(D)\in \R^{q\times q}$ and $V(D)\in \R^{(p-1)\times q}$. Since
$$D^{\intercal}\left({\rm I}+DD^{\intercal}\right)^{-1}=V(D)\Lambda(D)(\Lambda(D)^2+{\rm I})^{-1} U(D)^{\intercal},$$
it follows from \eqref{eq: re-exp} that
\begin{equation}
\|\gamma^{A}\|_2
\leq \|B^{-\frac{1}{2}}\|_2 \max_{1\leq l\leq q} \frac{|\lambda_{l}(D)|}{\lambda_{l}^2(D)+1}\|\Psi_j+\Psi_{-j}\gamma^{E}\|_2,
\label{eq: inter bound}
\end{equation}
where $\lambda_{l}(D)$ is the $l$-th largest singular value of $D$ in absolute value. By the condition $c_0\leq \lambda_{\min}(\Omega_{E})\leq \lambda_{\max}(\Omega_{E})\leq C_0$, we have
$\frac{1}{C_0} {\rm I}\preceq B=\E E_{1,-j}E_{1,-j}^{\intercal} \preceq \frac{1}{c_0} {\rm I}.$  We further have ${c_{0}}\lambda_{l}^2(\Psi_{-j})\leq \lambda^2_{l}(D)\leq {C_{0}}\lambda_{l}^2(\Psi_{-j})$ for $1\leq l\leq q$ and establish the first inequality of \eqref{eq: approximation decoup}.  The second inequality of \eqref{eq: approximation decoup} follows from condition (A2).

\noindent
\underline{Proof of \eqref{eq: residue upper}}
We fix $1\leq i\leq n$ and $1\leq j\leq p.$ Recall that 
$$\eta_{i,j}=X_{i, j} - X^{\intercal}_{i, -j}\gamma=\Psi_{j}^{\intercal} H_{i,\cdot}-(\Psi_{-j}^{\intercal}H_{i,\cdot})^{\intercal}\gamma+E_{i,j}-E_{i,-j}^{\intercal}\gamma^{E}-E_{i,-j}^{\intercal}\gamma^{A},$$
$$\nu_{i,j}=E_{i,j}-E_{i,-j}^{\intercal}\gamma^{E},$$  
and $$\delta_{i,j}=\eta_{i,j}-\nu_{i,j}=\Psi_{j}^{\intercal} H_{i,\cdot}-(\Psi_{-j}^{\intercal}H_{i,\cdot})^{\intercal}\gamma-E_{i,-j}^{\intercal}\gamma^{A}.$$
Since $E_{i,\cdot}$ is uncorrelated with $H_{i,\cdot}$ and $\nu_{i,j}$ is uncorrelated with $E_{i,-j}$ and $H_{i,\cdot},$ we have $\nu_{i,j}$ to be uncorrelated with $\delta_{i,j}.$ Hence we have 
\begin{equation}
{\rm Var}(\eta_{i,j})={\rm Var}(\nu_{i,j})+{\rm Var}(\delta_{i,j}).
\label{eq: var decomposition}
\end{equation}
By the expression of $\gamma$ in \eqref{eq: gamma expression}, we express ${\rm Var}(\eta_{i,j})={\rm Var}(X_{i, j}) - {\rm Var}(X^{\intercal}_{i, -j}\gamma)$ as
\begin{equation}
\begin{aligned}
&\|\Psi_{j}\|_2^2+(\Sigma_{E})_{j,j}-\left(\Psi_{-j}^{\intercal}\Psi_{j}+\E E_{1,j}E_{1,-j}\right)^{\intercal}\left(\Psi_{-j}^{\intercal} \Psi_{-j}+\BM\right)^{-1}\left(\Psi_{-j}^{\intercal}\Psi_{j}+\E E_{1,j}E_{1,-j}\right)\\
=&\|\Psi_{j}\|_2^2+(\Sigma_{E})_{j,j}-\left(\Psi_{-j}^{\intercal}\Psi_{j}+\E E_{1,j}E_{1,-j}\right)^{\intercal}\BM^{-1}\left(\Psi_{-j}^{\intercal}\Psi_{j}+\E E_{1,j}E_{1,-j}\right)\\
+&\left(\Psi_{-j}^{\intercal}\Psi_{j}+\E E_{1,j}E_{1,-j}\right)^{\intercal}B^{-1}\Psi_{-j}^{\intercal}\left({\rm I}+\Psi_{-j} \BM^{-1}\Psi_{-j}^{\intercal}\right)^{-1}\Psi_{-j}\BM^{-1}\left(\Psi_{-j}^{\intercal}\Psi_{j}+\E E_{1,j}E_{1,-j}\right).
\end{aligned}
\label{eq: var decomp 1}
\end{equation}
where the equation follows from \eqref{eq: inter Woodbury}.
Note that 
$${\rm Var}(\nu_{i,j})=(\Sigma_{E})_{j,j}-(\E E_{1,j}E_{1,-j})^{\intercal}\BM^{-1}(\E E_{1,j}E_{1,-j}).$$
Together with \eqref{eq: var decomposition} and \eqref{eq: var decomp 1}, we obtain 
\begin{equation}
\begin{aligned}
&{\rm Var}(\delta_{i,j})=\|\Psi_{j}\|_2^2-\Psi_{j}^{\intercal}\Psi_{-j} \BM^{-1} \Psi_{-j}^{\intercal}\Psi_{j}-2\Psi_{j}^{\intercal}\Psi_{-j} \gamma^{E}\\
+&\left(\Psi_{-j}^{\intercal}\Psi_{j}+\E E_{1,j}E_{1,-j}\right)^{\intercal}B^{-1}\Psi_{-j}^{\intercal}\left({\rm I}+\Psi_{-j} \BM^{-1}\Psi_{-j}^{\intercal}\right)^{-1}\Psi_{-j}\BM^{-1}\left(\Psi_{-j}^{\intercal}\Psi_{j}+\E E_{1,j}E_{1,-j}\right)\\
&=\|\Psi_{j}\|_2^2-\Psi_{j}^{\intercal}\Psi_{-j} \BM^{-1} \Psi_{-j}^{\intercal}\Psi_{j}-2\Psi_{j}^{\intercal}\Psi_{-j} \gamma^{E}\\
&+\Psi_{j}^{\intercal}\Psi_{-j}B^{-1}\Psi_{-j}^{\intercal}\left({\rm I}+\Psi_{-j} \BM^{-1}\Psi_{-j}^{\intercal}\right)^{-1}\Psi_{-j}\BM^{-1}\Psi_{-j}^{\intercal}\Psi_{j}\\
&+2\Psi_{j}^{\intercal}\Psi_{-j}B^{-1}\Psi_{-j}^{\intercal}\left({\rm I}+\Psi_{-j} \BM^{-1}\Psi_{-j}^{\intercal}\right)^{-1}\Psi_{-j}\gamma^{E}+(\gamma^{E})^{\intercal}\Psi_{-j}^{\intercal}\left({\rm I}+\Psi_{-j} \BM^{-1}\Psi_{-j}^{\intercal}\right)^{-1}\Psi_{-j}\gamma^{E}.
\end{aligned}
\label{eq: var decomp 2}
\end{equation}
Note that 
$$\|\Psi_{j}\|_2^2=\Psi_{j}^{\intercal}\left({\rm I}+\Psi_{-j} \BM^{-1}\Psi_{-j}^{\intercal}\right)^{-1}\left({\rm I}+\Psi_{-j} \BM^{-1}\Psi_{-j}^{\intercal}\right)\Psi_{j}.
$$
We have 
\begin{equation}
\begin{aligned}
&\|\Psi_{j}\|_2^2+\Psi_{j}^{\intercal}\Psi_{-j}B^{-1}\Psi_{-j}^{\intercal}\left({\rm I}+\Psi_{-j} \BM^{-1}\Psi_{-j}^{\intercal}\right)^{-1}\Psi_{-j}\BM^{-1}\Psi_{-j}^{\intercal}\Psi_{j}\\
&=\Psi_{j}^{\intercal}\left({\rm I}+\Psi_{-j} \BM^{-1}\Psi_{-j}^{\intercal}\right)^{-1}\Psi_{j}+\Psi_{j}^{\intercal}\Psi_{-j}\BM^{-1}\Psi_{-j}^{\intercal}\Psi_{j}.
\end{aligned}
\label{eq: trick 1}
\end{equation}
Note that 
\begin{equation*}
\begin{aligned}
&\Psi_{j}^{\intercal}\Psi_{-j}B^{-1}\Psi_{-j}^{\intercal}\left({\rm I}+\Psi_{-j} \BM^{-1}\Psi_{-j}^{\intercal}\right)^{-1}\Psi_{-j}\gamma^{E}-\Psi_{j}^{\intercal}\Psi_{-j} \gamma^{E}\\
=&\Psi_{j}^{\intercal}\Psi_{-j}B^{-1}\Psi_{-j}^{\intercal}\left({\rm I}+\Psi_{-j} \BM^{-1}\Psi_{-j}^{\intercal}\right)^{-1}\Psi_{-j}\gamma^{E}\\
&-\Psi_{j}^{\intercal}\left({\rm I}+\Psi_{-j} \BM^{-1}\Psi_{-j}^{\intercal}\right)\left({\rm I}+\Psi_{-j} \BM^{-1}\Psi_{-j}^{\intercal}\right)^{-1}\Psi_{-j} \gamma^{E}\\
=&-\Psi_{j}^{\intercal}\left({\rm I}+\Psi_{-j} \BM^{-1}\Psi_{-j}^{\intercal}\right)^{-1}\Psi_{-j}\gamma^{E}.
\end{aligned}
\end{equation*}
Together with \eqref{eq: var decomp 2} and \eqref{eq: trick 1}, we establish 
\begin{equation*}
\begin{aligned}
{\rm Var}(\delta_{i,j})&=\Psi_{j}^{\intercal}\left({\rm I}+\Psi_{-j} \BM^{-1}\Psi_{-j}^{\intercal}\right)^{-1}\Psi_{j}+(\gamma^{E})^{\intercal}\Psi_{-j}^{\intercal}\left({\rm I}+\Psi_{-j} \BM^{-1}\Psi_{-j}^{\intercal}\right)^{-1}\Psi_{-j}\gamma^{E}\\
&-2\Psi_{j}^{\intercal}\left({\rm I}+\Psi_{-j} \BM^{-1}\Psi_{-j}^{\intercal}\right)^{-1}\Psi_{-j}\gamma^{E}\\
&=(\Psi_{j}-\Psi_{-j}\gamma^{E})^{\intercal}\left({\rm I}+\Psi_{-j} \BM^{-1}\Psi_{-j}^{\intercal}\right)^{-1}(\Psi_{j}-\Psi_{-j}\gamma^{E}).
\end{aligned}
\end{equation*}
We establish \eqref{eq: residue upper} by applying condition (A2) and the following inequality
 $$\lambda_{\min}({\rm I}+\Psi_{-j} \BM^{-1}\Psi_{-j}^{\intercal})\geq 1+C\lambda_{q}^2(\Psi_{-j}),$$
 for some positive constant $C>0.$

\subsection{Proof of Lemma \ref{lem: confounding error}}
\label{sec: proof confounding}
The proof of this lemma is similar to Lemma \ref{lem: decomposition lemma} in terms of controlling $\|b\|_2$. We start with the exact expression of $b$
\begin{equation*}
b=\Sigma_{X}^{-1}\Psi^{\intercal}\phi=(\Sigma_{E}+\Psi^{\intercal}\Psi)^{-1}\Psi^{\intercal}\phi.
\end{equation*}
By apply the Woodbury matrix inverse formula, we have 
\begin{equation*}
b=\Sigma_{E}^{-1}\Psi^{\intercal}\left({\rm I}+\Psi\Sigma_{E}^{-1}\Psi^{\intercal}\right)^{-1}\phi.
\end{equation*}
We define $D_{E}=\Psi \Sigma_{E}^{-1/2}\in \R^{q\times p}$ and hence we have 
\begin{equation*}
b=\Sigma_{E}^{-1/2} D_{E}^{\intercal}({\rm I}+D_{E} D_{E}^{\intercal})^{-1}\phi,
\end{equation*}
and 
\begin{equation}
b_j=(\Omega_{E})_{\cdot,j}^{\intercal}\Psi^{\intercal}({\rm I}+D_{E} D_{E}^{\intercal})^{-1}\phi.
\label{eq: single element}
\end{equation}
Hence, we control $\|b\|_2$ as
\begin{equation*}
\|b\|_2\leq \sqrt{C_0}\max_{1\leq l\leq q}\frac{\lambda_l(D_{E})}{1+\lambda_l^{2}(D_{E})}\|\phi\|_2 \lesssim \frac{\sqrt{q} (\log p)^{1/4}}{\lambda_{q}(\Psi)}.
\end{equation*}
where the last inequality follows from the fact ${c_{0}}\lambda_{j}^2(\Psi)\leq \lambda^2_{j}(D_{E})\leq {C_{0}}\lambda_{j}^2(\Psi)$ and the condition {\rm (A2)}. Similarly, we apply condition {\rm (A2)} and control $|b_j|$ as
\begin{equation*}
|b_j|\leq \|\Psi(\Omega_{E})_{\cdot,j}\|_2 \frac{1}{1+\lambda_{q}^2(D_{E})}\|\phi\|_2\lesssim \frac{q\sqrt{\log p}}{{\lambda_{q}^2(\Psi)}}.
\end{equation*}

It follows from Woodbury matrix inverse formula that
\begin{equation}
\Psi\Sigma_{X}^{-1}\Psi^{\intercal}=\Psi^{\intercal}\Sigma_{E}^{-1}\Psi({\rm I}_q+\Psi \Sigma_{E}^{-1}\Psi^{\intercal})^{-1},
\label{eq: inter matrix expression}
\end{equation}
and hence
\begin{equation*}
\sigma_{\epsilon}^2-\sigma_{e}^2=\phi^{\intercal}\left({\rm I}_q-\Psi\Sigma_{X}^{-1}\Psi^{\intercal}\right)\phi=\phi^{\intercal}({\rm I}_{q}+\Psi\Sigma_{E}^{-1}\Psi^{\intercal})^{-1}\phi.
\end{equation*}
We establish \eqref{eq: var-diff} by applying condition (A2) and the following inequality
 $$\lambda_{\min}({\rm I}+\Psi\Sigma_{E}^{-1}\Psi^{\intercal})\geq 1+C\lambda_{q}^2(\Psi),$$
  for some positive constant $C>0.$

\subsection{Proof of Proposition \ref{prop: estimation-gamma}}
\label{sec: proof estimation}
Define $W\in \R^{p\times p}$ as a diagonal matrix with  diagonal entries as $W_{l,l}={\|\NT X_{\cdot,l}\|_2}/{\sqrt{n}}$ for $1\leq l \leq p.$ 
For the vector $a\in \R^{p-1}$, we define the weighted $\ell_1$ norm $\|a\|_{1,w}=\sum_{l\neq j}\frac{\|\NT X_{\cdot,l}\|_2}{\sqrt{n}}|a_l|=\|(W_{-l,-l})a\|_1.$ Define the event 
\begin{equation}
\mathcal{A}_0=\left\{c \leq \frac{\|\NT X_{\cdot,l}\|_2}{\sqrt{n}}\leq C \Mq \quad \text{for} \; 1\leq l\leq p\right\},
\label{eq: event A0}
\end{equation}
for some positive constants $C>0$ and $c>0.$
On the event $\mathcal{A}_0$, we have 
\begin{equation}
c\|a\|_1\leq \|a\|_{1,w}\leq C \Mq \|a\|_1.
\label{eq: norm relation}
\end{equation} We now show that 
$
\mathbb{P}(\mathcal{A}_0)\geq 1-p^{-c}-\exp(-cn),
$
for some positive constant $c>0.$
By the construction of $\NT $, we have $$\frac{\|\NT X_{\cdot,l}\|_2}{\sqrt{n}}\leq \frac{\|X_{\cdot,l}\|_2}{\sqrt{n}}.$$ Following from the fact that $X_{i,l}$ is of sub-Gaussian norm  $\Mq$, we apply the Corollary 5.17 in \citep{vershynin2010introduction} and establish that, with probability larger than $1-p^{-c}-\exp(-cn)$,
\begin{equation}
\frac{\|X_{\cdot,l}\|_2}{\sqrt{n}}\lesssim \sqrt{{\rm Var}(X_{1,l})}(1+\Mq\sqrt{{\log p}/{n}})\lesssim \Mq,
\label{eq: upper control}
\end{equation}
where the last inequality follows from the definition of sub-Gaussian norm and $\Mq\sqrt{{\log p}/{n}}\leq C$ for some positive constant $C>0.$
It follows from condition {\rm (A4)} that 
\begin{equation}
\min_{l\neq j}\frac{\|\NT  X_{\cdot,l}\|_2}{\sqrt{n}}\geq \sqrt{\tau_*}.
\label{eq: lower control}
\end{equation}
Recall the definitions $$\eta_j=(\eta_{1,j},\ldots,\eta_{n,j})^{\intercal}\in \R^{n}, \quad  \nu_j=(\nu_{1,j},\ldots,\nu_{n,j})^{\intercal}\in \R^{n} \quad \text{and}\quad \delta_j=\eta_j-\nu_j.$$ In the following, we shall choose the tuning parameter $\lambda_0$ such that $$\lambda_0\geq \|\frac{1}{n}\eta_{j}^{\intercal}(\NT) ^2X_{-j} (W_{-j,-j})^{-1}\|_{\infty}.$$ 
Since $\nu_{i,j}=E_{i,j}-(\gamma^{E})^{\intercal}E_{i,-j}$ is sub-Gaussian and independent of $X_{i,-j}$, we apply Proposition 5.10 in \citep{vershynin2010introduction} and the maximum inequality to establish
\begin{equation}
\P\left(\|\frac{1}{n}\nu_{j}^{\intercal}(\NT) ^2X_{-j} (W_{-j,-j})^{-1}\|_{\infty}\geq {A}_0\sigma_{j}\sqrt{{\log p}/{n}}\right)\leq e\cdot p^{1-c(A_0/C_1)^2}
\label{eq: tuning concentration}
\end{equation}
for some positive constants $A_0>0$ and $c>0$. We then control  $\|\frac{1}{n}\delta_{j}^{\intercal}(\NT) ^2X_{-j} (W_{-j,-j})^{-1}\|_{\infty}$ by the inequality $$\|\frac{1}{n}\delta_{j}^{\intercal}(\NT) ^2X_{-j} (W_{-j,-j})^{-1}\|_{\infty} \leq \frac{1}{\sqrt{n}}\|\delta_{j}\|_2$$ and the upper bound for $\frac{1}{n}\E \|\delta_{j}\|_2^2$ in \eqref{eq: residue upper}.
 As a consequence, we have
\begin{equation*}
\P\left(\|\frac{1}{n}\delta_{j}^{\intercal}(\NT) ^2X_{-j} (W_{-j,-j})^{-1}\|_{\infty}\geq \frac{1}{1+c}\sqrt{\frac{q \log p}{1+\lambda_{q}^2(\Psi_{-j})}}
\right)\lesssim (\log p)^{-1/2}
\end{equation*}
for any positive constant $c>0.$
Together with \eqref{eq: tuning concentration}, we then choose $$\lambda_0=A_0 \sigma_{j} \sqrt{\frac{\log p}{n}}+ \frac{1}{1+c}\sqrt{\frac{q \log p}{1+\lambda_{q}^2(\Psi_{-j})}}\quad \text{and}\quad \lambda_j\geq (1+c)\lambda_0,$$
and have
\begin{equation}
\P\left(\|\frac{1}{n}\eta_{j}^{\intercal}(\NT) ^2X_{-j} (W_{-j,-j})^{-1}\|_{\infty}\leq \lambda_0\right)\geq 1- C(\log p)^{-1/2},
\label{eq: tuning inequality scaled}
\end{equation}
for some positive constant $C>0.$ 

By the definition of the estimator $\widehat{\gamma}$, we have the following basic inequality,
\begin{equation}
\frac{1}{2n}\|\NT (X_j-X_{-j}\widehat{\gamma})\|_{2}^2+\lambda_j\|\widehat{\gamma}\|_{1,w}\leq \frac{1}{2n}\|\NT \left(X_j-X_{-j}\gamma^{E}\right)\|_{2}^2+\lambda_j\|\gamma^{E}\|_{1,w}.
\label{eq: basic}
\end{equation}
 By decomposing $X_j-X_{-j}\widehat{\gamma}=X_{-j}\gamma^{A}+\eta_{j}+X_{-j}\left(\gamma^{E}-\widehat{\gamma}\right)$, we simplify \eqref{eq: basic} as
\begin{equation}
\begin{aligned}
&\frac{1}{2n}\|\NT X_{-j}\left(\gamma^{E}-\widehat{\gamma}\right)\|_{2}^2+\lambda_j \|\widehat{\gamma}\|_{1,w}\leq \lambda_j\|\gamma^{E}\|_{1,w}\\
&-\frac{1}{n}\eta_{j}^{\intercal}(\NT) ^2X_{-j}\left(\gamma^{E}-\widehat{\gamma}\right)-\frac{1}{n}\left(\NT X_{-j}\gamma^{A}\right)^{\intercal}\NT  X_{-j}\left(\gamma^{E}-\widehat{\gamma}\right).
\end{aligned}
\label{eq: basic 1}
\end{equation}
Regarding the right hand side of the above inequality, we apply \eqref{eq: tuning inequality scaled} and establish that, with probability larger than $1- C(\log p)^{-1/2}$ for some positive constant $C>0,$ 
\begin{equation*}
\begin{aligned}
\left|\frac{1}{n}\eta_{j}^{\intercal}(\NT) ^2X_{-j}\left(\gamma^{E}-\widehat{\gamma}\right)\right|\leq &\|\frac{1}{n}\eta_{j}^{\intercal}(\NT) ^2X_{-j}(W_{-j,-j})^{-1}\|_{\infty} \|W_{-j,-j}(\gamma^{E}-\widehat{\gamma})\|_{1}\\
\leq &\lambda_0 \|\gamma^{E}-\widehat{\gamma}\|_{1,w}.
\end{aligned}
\end{equation*}
Additionally, we have $$\left|\frac{1}{n}\left(\NT X_{-j}\gamma^{A}\right)^{\intercal}\NT  X_{-j}\left(\gamma^{E}-\widehat{\gamma}\right)\right|\leq \|\frac{1}{\sqrt{n}}\NT X_{-j}\gamma^{A}\|_2\|\frac{1}{\sqrt{n}}\NT  X_{-j}\left(\gamma^{E}-\widehat{\gamma}\right)\|_2.$$
Then we further simply \eqref{eq: basic 1} as 
\begin{equation*}
\begin{aligned}
\frac{1}{2n}\|\NT X_{-j}\left(\gamma^{E}-\widehat{\gamma}\right)\|_{2}^2+\lambda_j \|\widehat{\gamma}\|_{1,w}&\leq \lambda_j\|\gamma^{E}\|_{1,w}+\lambda_0 \|\gamma^{E}-\widehat{\gamma}\|_{1,w}\\
&+\|\frac{1}{\sqrt{n}}\NT X_{-j}\gamma^{A}\|_2\|\frac{1}{\sqrt{n}}\NT  X_{-j}\left(\gamma^{E}-\widehat{\gamma}\right)\|_2.
\end{aligned}
\end{equation*}
Let $\Sg$ denote the support of $\gamma^{E}$. 
By the fact that $\|\gamma^{E}_{\Sg}\|_{1,w}-\|\widehat{\gamma}_{\Sg}\|_{1,w} \leq \|\gamma^{E}_{\Sg}-\widehat{\gamma}_{\Sg}\|_{1,w}$ and $\|\widehat{\gamma}_{\Sg^{c}}\|_{1,w}=\|\gamma^{E}_{\Sg^{c}}-\widehat{\gamma}_{\Sg^{c}}\|_{1,w},$ then we establish 
\begin{equation}
\begin{aligned}
&\frac{1}{2n}\|\NT X_{-j}\left(\gamma^{E}-\widehat{\gamma}\right)\|_{2}^2+\left(\lambda_j-\lambda_0\right) \|\gamma^{E}_{\Sg^{c}}-\widehat{\gamma}_{\Sg^{c}}\|_{1,w}\\
&\leq \left(\lambda_j+\lambda_0\right) \|\gamma^{E}_{\Sg}-\widehat{\gamma}_{\Sg}\|_{1,w}+\|\frac{1}{\sqrt{n}}\NT X_{-j}\gamma^{A}\|_2\|\frac{1}{\sqrt{n}}\NT  X_{-j}\left(\gamma^{E}-\widehat{\gamma}\right)\|_2.
\end{aligned}
\label{eq: basic 2}
\end{equation}
The following analysis is based on \eqref{eq: basic 2} and divided into two cases depending on the dominating term on the right hand side of \eqref{eq: basic 2}.\\
\noindent {\bf Case 1}: We consider $$ \left(\lambda_j+\lambda_0\right)\|\gamma^{E}_{\Sg}-\widehat{\gamma}_{\Sg}\|_{1,w}\geq \|\frac{1}{\sqrt{n}}\NT X_{-j}\gamma^{A}\|_2\|\frac{1}{\sqrt{n}}\NT  X_{-j}\left(\gamma^{E}-\widehat{\gamma}\right)\|_2$$ and then simplify \eqref{eq: basic 2} as 
\begin{equation}
\frac{1}{2n}\|\NT X_{-j}\left(\gamma^{E}-\widehat{\gamma}\right)\|_{2}^2+\left(\lambda_j-\lambda_0\right) \|\gamma^{E}_{\Sg^{c}}-\widehat{\gamma}_{\Sg^{c}}\|_{1,w}\leq 2\left(\lambda_j+\lambda_0\right) \|\gamma^{E}_{\Sg}-\widehat{\gamma}_{\Sg}\|_{1,w}.
\label{eq: final basic}
\end{equation}
It follows from \eqref{eq: final basic} that $$\|\gamma^{E}_{\Sg^{c}}-\widehat{\gamma}_{\Sg^{c}}\|_{1,w}\leq \frac{\lambda_j+\lambda_0}{\lambda_j-\lambda_0}\|\gamma^{E}_{\Sg}-\widehat{\gamma}_{\Sg}\|_{1,w}.$$ By the choices of $\lambda_j$ and $\lambda_0$, on the event $\mathcal{A}_0$, we establish 
\begin{equation*}
\|\gamma^{E}_{\Sg^{c}}-\widehat{\gamma}_{\Sg^{c}}\|_{1}\leq C \Mq \|\gamma^{E}_{\Sg}-\widehat{\gamma}_{\Sg}\|_{1},
\label{eq: rescaled relation}
\end{equation*}
for some positive constant $C>0$. By the restricted eigenvalue condition \eqref{eq: RE-b}, we have 
$$\frac{1}{2n}\|\NT X_{-j}\left(\gamma^{E}-\widehat{\gamma}\right)\|_{2}^2\geq \frac{\tau_*}{2}\|\gamma^{E}_{\Sg}-\widehat{\gamma}_{\Sg}\|_2^2.$$
Together with \eqref{eq: final basic} and \eqref{eq: norm relation}, we have 
\begin{equation*}
\begin{aligned}
\frac{\tau_*}{2}\|\gamma^{E}_{\Sg}-\widehat{\gamma}_{\Sg}\|_2^2&\leq 2\left(\lambda_j+\lambda_0\right) \|\gamma^{E}_{\Sg}-\widehat{\gamma}_{\Sg}\|_{1,w}\\
&\leq 2 C \Mq \left(\lambda_j+\lambda_0\right) \|\gamma^{E}_{\Sg}-\widehat{\gamma}_{\Sg}\|_{1}\\
&\leq 2 C \Mq \sqrt{|\Sg|}\left(\lambda_j+\lambda_0\right)\|\gamma^{E}_{\Sg}-\widehat{\gamma}_{\Sg}\|_2,
\end{aligned}
\end{equation*}
which leads to 
\begin{equation*}
\|\gamma^{E}_{\Sg}-\widehat{\gamma}_{\Sg}\|_2\lesssim \frac{\Mq}{\tau_{*}}\sqrt{|\Sg|}\left(\lambda_j+\lambda_0\right) \quad \text{and}\quad \|\gamma^{E}_{\Sg}-\widehat{\gamma}_{\Sg}\|_1\lesssim 
\frac{\Mq}{\tau_{*}}{|\Sg|}\left(\lambda_j+\lambda_0\right).
\label{eq: important inter}
\end{equation*}
On the event $\mathcal{A}_0,$ the above inequality implies that 
\begin{equation}
\|\gamma^{E}_{\Sg^{c}}-\widehat{\gamma}_{\Sg^{c}}\|_{1}
\lesssim \|\gamma^{E}_{\Sg^{c}}-\widehat{\gamma}_{\Sg^{c}}\|_{1,w}\lesssim \|\gamma^{E}_{\Sg}-\widehat{\gamma}_{\Sg}\|_{1,w}\lesssim \Mq \|\gamma^{E}_{\Sg}-\widehat{\gamma}_{\Sg}\|_{1} 
\lesssim \frac{\Mq^2}{\tau_{*}} {|\Sg|}\left(\lambda_j+\lambda_0\right).
\label{eq: l1 error a}
\end{equation}

Together with \eqref{eq: final basic}, \eqref{eq: l1 error a} implies that
\begin{equation}
\frac{1}{2n}\|\NT X_{-j}\left(\gamma^{E}-\widehat{\gamma}\right)\|_{2}^2\lesssim \frac{\Mq^2}{\tau_{*}}{|\Sg|}\left(\lambda_j+\lambda_0\right)^2.
\label{eq: prediction error a}
\end{equation}
We apply the restricted eigenvalue condition \eqref{eq: RE-b} again to establish 
\begin{equation}
\|\gamma^{E}-\widehat{\gamma}\|_2\lesssim \Mq\sqrt{|\Sg|}\left(\lambda_j+\lambda_0\right).
\label{eq: l2 error a}
\end{equation}

\noindent {\bf Case 2}: We consider $$ \left(\lambda_j+\lambda_0\right)\|\gamma^{E}_{\Sg}-\widehat{\gamma}_{\Sg}\|_{1,w}\leq \|\frac{1}{\sqrt{n}}\NT X_{-j}\gamma^{A}\|_2\|\frac{1}{\sqrt{n}}\NT  X_{-j}\left(\gamma^{E}-\widehat{\gamma}\right)\|_2,$$ and then simplify \eqref{eq: basic 2} as 
\begin{equation*}
\begin{aligned}
\frac{1}{2n}\|\NT &X_{-j}\left(\gamma^{E}-\widehat{\gamma}\right)\|_{2}^2+\left(\lambda_j-\lambda_0\right) \|\gamma^{E}_{\Sg^{c}}-\widehat{\gamma}_{\Sg^{c}}\|_{1,w}\\
&\leq \|\frac{1}{\sqrt{n}}\NT X_{-j}\gamma^{A}\|_2\|\frac{1}{\sqrt{n}}\NT  X_{-j}\left(\gamma^{E}-\widehat{\gamma}\right)\|_2.
\end{aligned}
\end{equation*}
Then we derive 
\begin{equation}
\frac{1}{\sqrt{n}}\|\NT X_{-j}\left(\gamma^{E}-\widehat{\gamma}\right)\|_{2}\lesssim \|\frac{1}{\sqrt{n}}\NT X_{-j}\gamma^{A}\|_2,
\label{eq: prediction error b}
\end{equation}
\begin{equation}
\|\gamma^{E}_{\Sg}-\widehat{\gamma}_{\Sg}\|_{1,w} \lesssim \frac{\|\frac{1}{{n}}\NT X_{-j}\gamma^{A}\|_2^2}{\lambda_j+\lambda_0} \quad \text{and}\quad \|\gamma^{E}_{\Sg^{c}}-\widehat{\gamma}_{\Sg^{c}}\|_{1,w}\lesssim \frac{\|\frac{1}{{n}}\NT X_{-j}\gamma^{A}\|_2^2}{\lambda_j-\lambda_0}.
\label{eq: l1 error b}
\end{equation}
Then, on the event $\mathcal{A}_0,$ we have 
\begin{equation}
\|\gamma^{E}-\widehat{\gamma}\|_2\leq \|\gamma^{E}-\widehat{\gamma}\|_1\lesssim \|\gamma^{E}-\widehat{\gamma}\|_{1,w} \lesssim \frac{\|\frac{1}{{n}}\NT X_{-j}\gamma^{A}\|_2^2}{\lambda_j+\lambda_0}+\frac{\|\frac{1}{{n}}\NT X_{-j}\gamma^{A}\|_2^2}{\lambda_j-\lambda_0}.
\label{eq: l2 error b}
\end{equation}
Finally, we establish \eqref{eq: accuracy-gamma-1} by combining \eqref{eq: l1 error a} and \eqref{eq: l1 error b}; 
establish \eqref{eq: accuracy-gamma-2} by combining \eqref{eq: l2 error a} and \eqref{eq: l2 error b};  
establish \eqref{eq: prediction-accuracy-gamma} by combining \eqref{eq: prediction error a} and \eqref{eq: prediction error b}; 

\subsection{Proof of Proposition \ref{prop: estimation-beta-refined}}
\label{sec: proof estimation refined}
The proof of Proposition \ref{prop: estimation-beta-refined} is similar to the proof of Proposition \ref{prop: estimation-gamma} in Section \ref{sec: proof estimation}. In the following, we prove Proposition \ref{prop: estimation-beta-refined} and mainly highlight its difference from the proof of Proposition \ref{prop: estimation-gamma} in Section \ref{sec: proof estimation}. 

Define $\widetilde{W}\in \R^{p\times p}$ as a diagonal matrix with  diagonal entries as $\widetilde{W}_{l,l}={\|\mathcal{Q} X_{\cdot,l}\|_2}/{\sqrt{n}}$ for $1\leq l \leq p.$
With a slight abuse of notation, for $a\in \R^{p}$, we define 
$\|a\|_{1,w}=\sum_{l=1}^{p}\frac{\|\mathcal{Q} X_{\cdot,l}\|_2}{\sqrt{n}}|a_l|.$ Define the  event 
$$
\mathcal{A}_1=\left\{c \leq \frac{\|\mathcal{Q} X_{\cdot,l}\|_2}{\sqrt{n}}\leq C\Mq \quad \text{for} \; 1\leq l\leq p\right\},
$$
for some positive constants $C>c>0.$ 
On the event $\mathcal{A}_1$, we have \eqref{eq: norm relation}. Similar to the control of $\mathcal{A}_0$ defined in \eqref{eq: event A0}, we can show that $\mathbb{P}(\mathcal{A}_1)\geq 1-p^{-c}-\exp(-cn)$
for some positive constant $c>0.$

The main part of the proof is to calculate the tuning parameter $\lambda$ such that
$$\lambda\geq (1+c)\|\frac{1}{n}\epsilon^{\intercal}\mathcal{Q}^2X \widetilde{W}^{-1}\|_{\infty}$$ 
for a small positive constant $c>0.$
Note that $\epsilon=\err+\Delta$ with $\Delta_i=\psi^{\intercal}H_{i, \cdot}-b^{\intercal}X_{i, \cdot}.$ Since $e_i$ is independent of $X_{i,\cdot}$, we apply Proposition 5.10 in \citep{vershynin2010introduction} and the maximum inequality to establish 
\begin{equation}
\P\left(\|\frac{1}{n}\err^{\intercal}\mathcal{Q}^2X\widetilde{W}^{-1}\|_{\infty}\geq A_0 \sigma_{e} \sqrt{\log p/n}\right)\leq e\cdot p^{-c(A_0/C_1)^2},
\label{eq: upper 1}
\end{equation}
for some positive constants $c>0$ and $A_0>0$.
We then control the other part $\|\frac{1}{n}\Delta^{\intercal}\mathcal{Q}^2X \widetilde{W}^{-1}\|_{\infty}$ by the inequality $$\|\frac{1}{n}\Delta^{\intercal}\mathcal{Q}^2X \widetilde{W}^{-1}\|_{\infty}\leq \frac{1}{\sqrt{n}}\|\Delta\|_2$$ and the upper bound for $\frac{1}{n}\E \|\Delta\|_2^2$ in \eqref{eq: approximation expectation}.
 As a consequence, we have
\begin{equation}
\P\left(\|\frac{1}{n}\Delta^{\intercal}\mathcal{Q}^2X\widetilde{W}^{-1}\|_{\infty}\geq \frac{1}{1+c}  \sqrt{\frac{q \log p}{1+\lambda_{q}^2(\Psi)}}\right)\lesssim (\log p)^{-1/2},
\label{eq: upper 2}
\end{equation}
for any positive constant $c>0.$
We then choose $$\lambda\geq A \sigma_{e} \sqrt{\frac{\log p}{n}}+ \sqrt{\frac{q \log p}{1+\lambda_{q}^2(\Psi)}} \quad \text{with}\quad A=(1+c)A_0.$$ We  combine \eqref{eq: upper 1} and \eqref{eq: upper 2} and establish that 
\begin{equation}
\P\left((1+c_0)\|\frac{1}{n}\epsilon^{\intercal}\mathcal{Q}^2XW^{-1}\|_{\infty}\leq \lambda\right)\geq 1-C(\log p)^{-1/2}-p^{-c},
\label{eq: tuning beta}
\end{equation} 
for some positive constant $C>0.$

By the definition of $\betainit$, we establish the basic inequality in a similar fashion to \eqref{eq: basic}
\begin{equation}
\frac{1}{2n}\|\mathcal{Q} (Y-X\betainit)\|_{2}^2+\lambda\|\betainit\|_{1,w}\leq \frac{1}{2n}\|\mathcal{Q}\left(Y-X\beta\right)\|_{2}^2+\lambda\|\beta\|_{1,w}.
\label{eq: basic refined}
\end{equation}
We can apply the similar argument from \eqref{eq: basic} to \eqref{eq: l2 error b} by replacing $\NT$, $X_j$, $X_{-j}$, $\widehat{\gamma}$, $\gamma^{E}$, $\gamma^{A}$
 with $\mathcal{Q}$, $Y$, $X$, $\betainit$, $\beta$, $b$, respectively. We replace the tuning parameters $\lambda_j$ and $\lambda_0$ by $\lambda$ and $\frac{1}{1+c_0}\lambda$, respectively. Then we establish Proposition \ref{prop: estimation-beta-refined}.

\subsection{Proof of Lemma \ref{lem: limiting dist}}
\label{sec: proof limiting}
We introduce the following lemma about the concentration of quadratic forms, which is Theorem 1.1 in \citep{rudelson2013hanson}. 
\begin{Lemma} {\rm (Hanson-Wright inequality)} Let $\xi\in \R^{n}$ be a random vector with independent sub-Gaussian components $\xi_i$ with zero mean and sub-Gaussian norm $K$. Let $A$ be an $n\times n$ matrix. Then for every $t\geq 0$, 
\begin{equation}
\mathbf{P}\left(|\xi^{\intercal}A \xi- \E \xi^{\intercal} A\xi|>t\right)\leq 2\exp\left[-c\min\left(\frac{t^2}{K^{4}\|A\|_{F}^2}, \frac{t}{K^2\|A\|_2}\right)\right].
\label{eq: HW-bound}
\end{equation}
\label{lem: HW-bound}
\end{Lemma}

For the high-dimensional setting where ${p}/{n}\rightarrow c^{*}\in (0,\infty]$, we have $m\asymp n$ for $m=\min\{n,p-1\}$. We also note ${\rm Tr}[(\NT) ^{l}]\asymp m$ for $l=2,4,8$.

\subsubsection{Proof of \eqref{eq: variance limit 1}}
We decompose $\frac{1}{n}(\NT Z_{j})^{\intercal}\NT X_j$ as 
\begin{equation}
\begin{aligned}
&\frac{1}{n}(\NT Z_{j})^{\intercal}\NT X_j=\frac{1}{n}(\NT \eta_j)^{\intercal}\NT \eta_j+\frac{1}{n}(\NT Z_{j})^{\intercal}\NT X_{-j}\gamma\\
&-\frac{1}{n}(\NT X_{-j}(\widehat{\gamma}-\gamma^{E}))^{\intercal}\NT \eta_j+\frac{1}{n}(\NT X_{-j}\gamma^{A})^{\intercal}\NT \eta_j,
\end{aligned}
\label{eq: first var decomposition}
\end{equation}
where $\eta_j=(\eta_{1,j},\ldots,\eta_{n,j})^{\intercal}\in \R^{n}.$

In the following, we control the right hand side of \eqref{eq: first var decomposition} term by term. Since $\eta_j=\nu_j+\delta_j$, we have 
$$\frac{1}{n}(\NT \eta_j)^{\intercal}\NT \eta_j=\frac{1}{n} \nu_j^{\intercal}(\NT)^{2} \nu_j+\frac{2}{n}\nu_j^{\intercal}(\NT)^{2} \delta_j+\frac{1}{n}\delta_j^{\intercal}(\NT)^{2} \delta_j.$$
By applying \eqref{eq: HW-bound} with $A=(\NT) ^2$, then with probability larger than $1-2\exp(-ct^2)$ for $0<t\lesssim {\rm Tr}[(\NT)^4]\asymp n,$
\begin{equation}
\left|\frac{1}{n}\nu_j^{\intercal}(\NT)^{2} \nu_j-{\rm Tr}[(\NT) ^2]\cdot \frac{\sigma_{j}^2}{n}\right|\lesssim {t}\frac{{\sqrt{{\rm Tr}[(\NT) ^4]}}}{n}\lesssim t \frac{\sqrt{m}}{n}.
\label{eq: quad concen}
\end{equation}
Since $\left|\delta_j^{\intercal}(\NT)^{2} \delta_j\right|\leq \|\delta_j\|_2^2,$ we apply the upper bound \eqref{eq: residue upper} for $\frac{1}{n}\E \|\delta_j\|_2^2$ and the Markov inequality to establish 
\begin{equation}
\PP\left(\frac{1}{n}\|\delta_j\|_2^2\gtrsim \frac{q\log p}{1+\lambda_{q}^2(\Psi_{-j})}\right)\leq (\log p)^{-1/2}.
\label{eq: approx square}
\end{equation}
Hence, we have, with probability larger than $1-2\exp(-ct^2)-(\log p)^{-1/2},$
\begin{align*}
\left|\frac{2}{n}\nu_j^{\intercal}(\NT)^{2} \delta_j\right|&\leq 2 \sqrt{\frac{1}{n}\nu_j^{\intercal}(\NT)^{2}\nu_j}\sqrt{\frac{1}{n}\delta_j^{\intercal}(\NT)^{2}\delta_j}\\
&\lesssim \sqrt{\left({\rm Tr}[(\NT) ^2]\cdot \frac{\sigma_{j}^2}{n}+Ct \frac{\sqrt{m}}{n}\right)\frac{q\log p}{1+\lambda_{q}^2(\Psi_{-j})}},
\end{align*}
for some positive constant $C>0.$ Combined with \eqref{eq: quad concen} and \eqref{eq: approx square}, we apply the fact that ${\rm Tr}[(\NT) ^2]\asymp n$ and establish that, with probability larger than $1-2\exp(-ct^2)-(\log p)^{-1/2}$ for $0<t\lesssim n,$
\begin{equation}
\left|\frac{1}{n}\eta_j^{\intercal}(\NT)^{2} \eta_j-{\rm Tr}[(\NT) ^2]\cdot \frac{\sigma_{j}^2}{n}\right|\lesssim t \frac{\sqrt{m}}{n}+\sqrt{\frac{q\log p}{1+\lambda_{q}^2(\Psi_{-j})}}.
\label{eq: projection limit}
\end{equation}

By the KKT condition of \eqref{eq: est-gamma}, we establish  
$$
\left|\frac{1}{n}(\NT Z_{j})^{\intercal}\NT X_{-j}\gamma\right|\leq \|\gamma\|_1\|\frac{1}{n}(\NT Z_{j})^{\intercal}\NT X_{-j}\|_{\infty}\leq \lambda_j \max_{l\neq j}\frac{\|\NT X_{\cdot,l}\|_2}{\sqrt{n}}\|\gamma\|_1,$$
where $\lambda_j$ is defined in \eqref{eq: tuning gamma}. We control the right hand side as 
\begin{equation*}
 \left(\|\gamma^{E}\|_1+\|\gamma^{A}\|_1\right)\lambda_j
\leq \sqrt{s} \|\gamma^{E}\|_2\lambda_j+\sqrt{p} \|\gamma^{A}\|_2\lambda_j.
\end{equation*}
On the event $\mathcal{A}_0$ defined in \eqref{eq: event A0}, we obtain 
\begin{equation}
\begin{aligned}
\left|\frac{1}{n}(\NT Z_{j})^{\intercal}\NT X_{-j}\gamma\right|
&\leq \Mq\cdot(\sqrt{s} \|\gamma^{E}\|_2\lambda_j+\sqrt{p} \|\gamma^{A}\|_2\lambda_j)\\
&\lesssim  \Mq\cdot\left(\sqrt{s} \|\gamma^{E}\|_2\lambda_j+\sqrt{\frac{p\log p}{n}+\frac{qp\log p}{\lambda^2_{q}(\Psi_{-j})}}\cdot \frac{\sqrt{q} (\log p)^{1/4}}{\lambda_{q}(\Psi_{-j})}\right).
\end{aligned}
\label{eq: bound1-c}
\end{equation}
where the last bound follows from the definition of  $\lambda_j$ in \eqref{eq: tuning gamma} and the upper bound for $\|\gamma^{A}\|_2$ in \eqref{eq: approximation decoup}.  
We apply H{\"o}lder's inequality and establish
\begin{equation}
\begin{aligned}
\left|\frac{1}{n}(\NT X_{-j}(\widehat{\gamma}-\gamma^{E}))^{\intercal}\NT \eta_j\right|&\leq \| W_{-j,-j}(\widehat{\gamma}-\gamma^{E})\|_1\|\frac{1}{n}\eta_j^{\intercal}(\NT) ^2X_{-j}(W_{-j,-j})^{-1}\|_{\infty}\\
&\lesssim \frac{\Mq^2}{\tau_*}s\lambda^2_j+\frac{\|\NT  X_{-j}\gamma^{A}\|_2^2}{{n}}\\
&\lesssim \frac{\Mq^2}{\tau_*}s\lambda^2_j+\frac{q \sqrt{\log p}}{1+\lambda^2_{q}(\Psi_{-j})}\cdot \max\left\{1,\frac{p}{n}\right\}
\end{aligned}
\label{eq: bound1-a}
\end{equation}
where the second inequality follows from \eqref{eq: accuracy-gamma-1} and \eqref{eq: tuning inequality} and the last inequality follows from \eqref{eq: approx proj error}.
Since $\nu_j$ is independent of $X_{-j},$ we show that 
$\frac{1}{n}(\NT X_{-j}\gamma^{A})^{\intercal}\NT \nu_j
$ has mean zero and variance 
\begin{equation}
\frac{\sigma_{j}^2}{n^2}(\gamma^{A})^{\intercal}X_{-j} (\NT )^{4}X_{-j}^{\intercal} \gamma^{A}\lesssim \frac{\|\gamma^{A}\|^2_2}{n} \|\frac{1}{n}X_{-j} (\NT) ^{4}X_{-j}^{\intercal} \|_2\lesssim \max\left\{1,\frac{p}{n}\right\}\cdot \frac{q \sqrt{\log p}}{n\lambda^2_{q}(\Psi_{-j})},
\label{eq: inter upper bound}
\end{equation} 
where the last inequality follows from the upper bound for $\|\gamma^{A}\|_2$ in \eqref{eq: approximation decoup} together with the property ${\rm (P1)}.$
Then with probability larger than $1-\frac{1}{t^2}$,
\begin{equation}
\left|\frac{1}{n}(\NT X_{-j}\gamma^{A})^{\intercal}\NT \nu_j\right|\lesssim \frac{t}{\sqrt{n}} \cdot \sqrt{ \max\left\{1,\frac{p}{n}\right\}\cdot \frac{q (\log p)^{1/2}}{\lambda^2_{q}(\Psi_{-j})}}. 
\label{eq: bound1-b}
\end{equation}
Note that, with probability larger than $1-(\log p)^{-1/2},$
\begin{equation}
\begin{aligned}
\left|\frac{1}{n}(\NT X_{-j}\gamma^{A})^{\intercal}\NT \delta_j\right|&\leq \|\frac{1}{n}(\NT)^{2} X_{-j}\gamma^{A}\|_2\|\delta_j\|_2\\
&\lesssim \sqrt{\max\left\{1,\frac{p}{n}\right\}\cdot \frac{q \sqrt{\log p}}{n\lambda^2_{q}(\Psi_{-j})}}\cdot \sqrt{\frac{q\log p}{1+\lambda_{q}^2(\Psi_{-j})}}
\end{aligned}
\label{eq: bound1-d}
\end{equation}
where the last inequality follows from \eqref{eq: inter upper bound} and \eqref{eq: approx square}.

By \eqref{eq: first var decomposition}, we combine the fact that ${\rm Tr}[(\NT) ^2]\cdot \frac{\sigma_{j}^2}{n}$ is of a constant order and the upper bounds \eqref{eq: projection limit}, \eqref{eq: bound1-c}, \eqref{eq: bound1-a}, \eqref{eq: bound1-b} and \eqref{eq: bound1-d}. We establish \eqref{eq: variance limit 1} under the conditions $s\lambda_j^2\Mq^2\rightarrow 0$ and 
\begin{equation*}
\lambda_{q}(\Psi_{-j})\gg \max\left\{(1+\Mq)\cdot \sqrt{\frac{qp}{n}}(\log p)^{3/4}, \sqrt{q (1+\Mq)}p^{1/4}(\log p)^{{3}/{8}}\right\}.
\label{eq: dim condition a}
\end{equation*}
Note that the above conditions are implied by \eqref{eq: factor condition} and $s\ll n/[\Mq^2 \log p]$. 

\subsubsection{Proof of \eqref{eq: variance limit 2}}

Note that 
\begin{equation}
\begin{aligned}
\frac{1}{n}Z_{j}^{\intercal}(\NT) ^4 Z_{j}&=\frac{1}{n}\eta_j^{\intercal}(\NT) ^4\eta_j +2 \eta_j^{\intercal}(\NT) ^4X_{-j}(\gamma^{E}-\widehat{\gamma}+\gamma^{A})\\
&+\frac{1}{n}\|(\NT) ^2X_{-j}(\gamma^{E}-\widehat{\gamma}+\gamma^{A})\|_2^2.
\end{aligned}
\label{eq: var 2 decomposition}
\end{equation}
By applying \eqref{eq: HW-bound} with $A=(\NT) ^4$, then with probability larger than $1-2\exp(-ct^2)$ for $0<t\lesssim {\rm Tr}[(\NT) ^8]\asymp n,$
\begin{equation*}
\left|\frac{1}{n}\nu_j^{\intercal}(\NT) ^4\nu_j-{\rm Tr}[(\NT) ^4]\cdot \frac{\sigma_{j}^2}{n}\right|\lesssim {t}\frac{\sqrt{{\rm Tr}[(\NT) ^8]}}{n}\lesssim t\frac{\sqrt{m}}{n}.
\label{eq: projection limit-2}
\end{equation*}
By a similar argument as in \eqref{eq: projection limit}, we establish that, with probability larger than $1-2\exp(-ct^2)-(\log p)^{-1/2}$ for $0<t\lesssim n,$
\begin{equation}
\left|\frac{1}{n}\eta_j^{\intercal}(\NT)^{4} \eta_j-{\rm Tr}[(\NT)^4]\cdot \frac{\sigma_{j}^2}{n}\right|\lesssim t \frac{\sqrt{m}}{n}+\sqrt{\frac{q\log p}{1+\lambda_{q}^2(\Psi_{-j})}}.
\label{eq: bound2-a}
\end{equation}
By a similar argument as \eqref{eq: bound1-a}, we have
\begin{equation}\left|\frac{1}{n}\eta_j^{\intercal}(\NT) ^4X_{-j}(\widehat{\gamma}-\gamma^{E})\right|\lesssim \frac{\Mq^2}{\tau_*}s\lambda^2_j+\frac{q \sqrt{\log p}}{1+\lambda^2_{q}(\Psi_{-j})}\cdot\max\left\{1,\frac{p}{n}\right\}.
\label{eq: bound2-b}
\end{equation}
In addition, 
$\frac{1}{n}\nu_j^{\intercal}(\NT) ^4X_{-j}\gamma^{A}
$ has mean zero and variance 
$$\frac{\sigma_{j}^2}{n^2}(\gamma^{A})^{\intercal}X_{-j} (\NT) ^{8}X_{-j}^{\intercal} \gamma^{A}\lesssim \frac{\|\gamma^{A}\|^2_2}{n} \|\frac{1}{n}X_{-j} (\NT) ^{8}X_{-j}^{\intercal} \|_2,$$
and hence with probability larger than $1-\frac{1}{t^2}$ for any $t>0$,
\begin{equation}
\left|\frac{1}{n}\nu_j^{\intercal}(\NT) ^4X_{-j}\gamma^{A}
\right|\lesssim \frac{t\|\gamma^{A}\|_2}{\sqrt{n}} \sqrt{\|\frac{1}{n}X_{-j} (\NT) ^{4}X_{-j}^{\intercal} \|_2}\lesssim t \sqrt{\max\left\{1,\frac{p}{n}\right\}\cdot \frac{q \sqrt{\log p}}{n\lambda^2_{q}(\Psi_{-j})}},
\label{eq: bound2-bp}
\end{equation}
where the last inequality follows from the upper bound for $\|\gamma^{A}\|_2$ in \eqref{eq: approximation decoup} together with the property ${\rm (P1)}.$
Note that, with probability larger than $1-(\log p)^{-1/2},$
\begin{equation}
\begin{aligned}
\left|\frac{1}{n}\delta_j^{\intercal}(\NT) ^4X_{-j}\gamma^{A}\right|&\leq \|\frac{1}{n}(\NT)^{4} X_{-j}\gamma^{A}\|_2\|\delta_j\|_2\\
&\lesssim \sqrt{\max\left\{1,\frac{p}{n}\right\}\cdot \frac{q \sqrt{\log p}}{n\lambda^2_{q}(\Psi_{-j})}}\cdot \sqrt{\frac{q\log p}{1+\lambda_{q}^2(\Psi_{-j})}}
\end{aligned}
\label{eq: bound2-c}
\end{equation}
where the last inequality follows from \eqref{eq: inter upper bound} and \eqref{eq: approx square}.

Note that 
\begin{equation*}
\begin{aligned}
&\frac{1}{n}\|(\NT) ^2X_{-j}(\widehat{\gamma}-\gamma^{E}-\gamma^{A})\|_2^2\leq \frac{1}{n}\|\NT X_{-j}(\widehat{\gamma}-\gamma^{E}-\gamma^{A})\|_2^2\\
&\lesssim \frac{1}{n}\|\NT X_{-j}(\widehat{\gamma}-\gamma^{E})\|_2^2+\frac{1}{n}\|\NT X_{-j}\gamma^{A}\|_2^2. 
\end{aligned}
\end{equation*}
By applying \eqref{eq: prediction-accuracy-gamma} and \eqref{eq: approx proj error}, 
we establish that, with probability larger than $1-e\cdot p^{1-c(A_0/C_1)^2}-\exp(-c n)-(\log p)^{-1/2}$ for some positive constant $c>0$,
\begin{equation}
\begin{aligned}
\frac{1}{n}\|(\NT) ^2X_{-j}(\widehat{\gamma}-\gamma^{E}-\gamma^{A})\|_2^2\lesssim &\left(\frac{\Mq}{\tau_*} \sqrt{s}\lambda_j+\sqrt{\max\left\{1,\frac{p}{n}\right\}\cdot \frac{q \sqrt{\log p}}{n\lambda^2_{q}(\Psi_{-j})}}\right)^2.
\end{aligned}
\label{eq: bound2-d}
\end{equation}

By \eqref{eq: var 2 decomposition}, we combine the fact that ${\rm Tr}[(\NT) ^2]\cdot \frac{\sigma_{j}^4}{n}$ is of a constant order and the upper bounds \eqref{eq: bound2-a}, \eqref{eq: bound2-b}, \eqref{eq: bound2-bp}, \eqref{eq: bound2-c} and \eqref{eq: bound2-d}. We establish \eqref{eq: variance limit 2} under the condition 
\begin{equation*}
\lambda_{q}(\Psi_{-j})\gg \sqrt{q}(\log p)^{1/4} \max\left\{\sqrt{\frac{{p}}{n}},(\log p)^{1/4}\right\}  \quad \text{and}\quad s\lambda_j^2\Mq^2\rightarrow 0.
\label{eq: dim condition b}
\end{equation*}
Note that the above condition is implied by \eqref{eq: factor condition} and $s\ll n/[\Mq^2 \log p]$. 

\subsubsection{Proof of \eqref{eq: bias control}}
Note that \begin{equation*}
|B_{\beta}|=\left|\frac{1}{\sqrt{V}}\frac{(\NT Z_{j})^{\intercal}\NT X_{-j}(\beta_{-j}-\betainit_{-j})}{(\NT Z_{j})^{\intercal}\NT X_j}\right|\leq \frac{\left|(\NT Z_{j})^{\intercal}\NT X_{-j}(\beta_{-j}-\betainit_{-j})\right|}{\sqrt{\sigma_{e}^2\cdot Z_j^{\intercal} (\NT) ^4 Z_{j}}}.
\end{equation*}
It follows from H{\"o}lder's inequality and also the KKT condition of \eqref{eq: est-gamma} that
\begin{equation*}
\begin{aligned}
\left|\frac{1}{n}(\NT Z_j)^{\intercal}\NT X_{-j}(\beta_{-j}-\betainit_{-j})\right|&\leq \|\beta_{-j}-\betainit_{-j}\|_1\|\frac{1}{n}(\NT Z_j)^{\intercal}\NT X_{-j}\|_{\infty}\\
&\leq\lambda_j \max_{l\neq j}\frac{\|\NT X_{\cdot,l}\|_2}{\sqrt{n}} \|\beta_{-j}-\betainit_{-j}\|_1.
\end{aligned}
\end{equation*}
By the definition of the event $\mathcal{A}_0$ in \eqref{eq: event A0} and the upper bound for $\|\betainit-\beta\|_1$ in \eqref{eq: accuracy-beta-1} and \eqref{eq: approx proj error Q}, with probability larger than $1-e\cdot p^{1-c(A/C_1)^2}-\exp(-c n)-(\log p)^{-1/2}$ for some positive constant $c>0$,
$$
\left|\frac{1}{n}(\NT Z_j)^{\intercal}\NT X_{-j}(\beta_{-j}-\betainit_{-j})\right|\lesssim \Mq \left(\frac{\Mq^2}{\tau_*}k\lambda_j\lambda+\frac{\lambda_j}{\lambda}\frac{q \sqrt{\log p}}{1+\lambda^2_{q}(\Psi)}\right).$$
Together with \eqref{eq: variance limit 2}, we establish $B_{\beta}\cip 0$ under the condition 
\begin{equation}
\lambda_{q}(\Psi)\gg [q \Mq]^{1/2} (n \log p)^{1/4}  \quad \text{and}\quad \sqrt{n}k\lambda_j\lambda [\Mq]^3\rightarrow 0.
\label{eq: dim condition c}
\end{equation}
Note that the above condition is implied by \eqref{eq: factor condition} and  $k\ll \sqrt{n}/[\Mq^3\log p].$

Now we control the other bias component 
\begin{equation*}
|B_{b}|= \left|\frac{1}{\sqrt{V}}\frac{(\NT Z_{j})^{\intercal}\NT X_{-j}b_{-j}}{(\NT Z_{j})^{\intercal}\NT X_j}+\frac{1}{\sqrt{V}}b_j\right|\leq \frac{\left|(\NT Z_{j})^{\intercal}\NT X_{-j}b_{-j}\right|}{\sqrt{\sigma_{e}^2\cdot Z_j^{\intercal} (\NT) ^4 Z_{j}}}+\left|\frac{1}{\sqrt{V}}b_j\right|
\end{equation*}

We investigate $\frac{1}{n}(\NT Z_{j})^{\intercal}\NT X_{-j}b_{-j}$:
\begin{equation}
\begin{aligned}
\frac{1}{n}(\NT Z_{j})^{\intercal}\NT X_{-j}b_{-j}&=\frac{1}{n}(\NT \nu_j)^{\intercal}\NT X_{-j}b_{-j}+\frac{1}{n}(\NT \delta_j)^{\intercal}\NT X_{-j}b_{-j}\\
&+\frac{1}{n}(\NT X_{-j}(\widehat{\gamma}-\gamma^{E}-\gamma^{A}))^{\intercal}\NT X_{-j}b_{-j}.
\end{aligned}
\label{eq: bias 2 decomposition}
\end{equation}
{
Note that $\frac{1}{n}(\NT \nu_j)^{\intercal}\NT X_{-j}b_{-j}$ has mean zero and variance 
$$\frac{\sigma_{j}^2}{n^2}(b_{-j})^{\intercal}X_{-j} (\NT) ^{4}X_{-j}^{\intercal} b_{-j}\lesssim \frac{1}{n} \|\frac{1}{n}X_{-j} (\NT) ^{4}X_{-j}^{\intercal} \|_2 \|b_{-j}\|^2_2\lesssim \max\left\{1,\frac{p}{n}\right\}\cdot \frac{q \sqrt{\log p}}{n\lambda^2_{q}(\Psi)},
$$ 
where the last inequality follows from the upper bound for $\|b_{-j}\|_2$ in \eqref{eq: b approximation upper}
 together with the property ${\rm (P1)}.$
Hence with probability larger than $1-\frac{1}{t^2}$ for some $t>0,$
\begin{equation}
\left|\frac{1}{n}(\NT \nu_j)^{\intercal}\NT X_{-j}b_{-j}\right| \lesssim t \sqrt{\max\left\{1,\frac{p}{n}\right\}\cdot \frac{q \sqrt{\log p}}{n\lambda^2_{q}(\Psi)}}.
\label{eq: bound3-a}
\end{equation}
}
Note that, with probability larger than $1-(\log p)^{-1/2},$
\begin{equation}
\begin{aligned}
\left|\frac{1}{n}(\NT \delta_j)^{\intercal}\NT X_{-j}b_{-j}\right|&\leq \|\frac{1}{n}(\NT)^{2} X_{-j}\|_2\|b_{-j}\|_2\|\delta_j\|_2\\
&\lesssim \sqrt{\max\left\{1,\frac{p}{n}\right\}\cdot \frac{q \sqrt{\log p}}{\lambda^2_{q}(\Psi)}}\cdot \sqrt{\frac{q\log p}{1+\lambda_{q}^2(\Psi_{-j})}},
\end{aligned}
\label{eq: bound3-b}
\end{equation}
where the last inequality follows from the upper bound for $\|\delta_j\|_2$ in \eqref{eq: approx square}, the upper bound for $\|b_{-j}\|_2$ in \eqref{eq: b approximation upper}
 together with the property ${\rm (P1)}.$
In addition, we note the following two inequalities 
\begin{equation}
\begin{aligned}
\left|\frac{1}{n}(\NT X_{-j}\gamma^{A})^{\intercal}\NT X_{-j}b_{-j}\right|&\leq \|\gamma^{A}\|_2\|b_{-j}\|_2\|\frac{1}{n}X_{-j} (\NT) ^{2}X_{-j}^{\intercal} \|_2\\ 
&\lesssim \max\left\{1,\frac{p}{n}\right\}\cdot \frac{q \sqrt{\log p}}{\lambda_{q}(\Psi)\cdot \lambda_{q}(\Psi_{-j})},
\end{aligned}
\label{eq: bound3-c}
\end{equation}
where the last inequality follows from the upper bound for $\|\gamma^{A}\|_2$ in \eqref{eq: approximation decoup}, the upper bound for $\|b_{-j}\|_2$ in \eqref{eq: b approximation upper}
 together with the property ${\rm (P1)}.$
 
Note that
$$\left|\frac{1}{n}(\NT X_{-j}(\widehat{\gamma}-\gamma^{E}))^{\intercal}\NT X_{-j}b_{-j}\right|\\
\leq \frac{1}{\sqrt{n}}\|\NT X_{-j}(\widehat{\gamma}-\gamma^{E})\|_2\|\frac{1}{\sqrt{n}}\NT X_{-j}\|_2\|b_{-j}\|_2.$$
Furthermore, we apply the upper bound \eqref{eq: prediction-accuracy-gamma} and establish that, with probability larger than $1-e\cdot p^{1-c(A_0/C_1)^2}-\exp(-c n)-(\log p)^{-1/2}$ for some positive constant $c>0$,
\begin{equation}
\begin{aligned}
&\left|\frac{1}{n}(\NT X_{-j}(\widehat{\gamma}-\gamma^{E}))^{\intercal}\NT X_{-j}b_{-j}\right|\\
&\lesssim \left(\frac{\Mq}{\tau_*} \sqrt{s}\lambda_j+\frac{\|\NT  X_{-j}\gamma^{A}\|_2}{\sqrt{n}}\right) \|\frac{1}{\sqrt{n}}\NT X_{-j}\|_2\|b_{-j}\|_2\\
&\lesssim \left(\frac{\Mq}{\tau_*} \sqrt{s}\lambda_j+\sqrt{\max\left\{1,\frac{p}{n}\right\}\cdot \frac{q \sqrt{\log p}}{\lambda^2_{q}(\Psi_{-j})}}\right)\sqrt{\max\left\{1,\frac{p}{n}\right\}\cdot \frac{q \sqrt{\log p}}{\lambda^2_{q}(\Psi_{-j})}}.
\end{aligned}
\label{eq: bound3-d}
\end{equation}
where the last inequality follows from the upper bound for $\|\gamma^{A}\|_2$ in \eqref{eq: approximation decoup}, the upper bound for $\|b_{-j}\|_2$ in \eqref{eq: b approximation upper}
 together with the property ${\rm (P1)}.$

By \eqref{eq: approximation decoup}, \eqref{eq: variance limit 1} and \eqref{eq: variance limit 2}, we have 
$\frac{|b_j|}{\sqrt{V}}\cip 0$ if 
$
\sqrt{n}\frac{q \sqrt{\log p}}{1+\lambda_{q}^2(\Psi)}\rightarrow 0.$

We now combine the decomposition \eqref{eq: bias 2 decomposition} and the upper bounds \eqref{eq: bound3-a}, \eqref{eq: bound3-b}, \eqref{eq: bound3-c} and \eqref{eq: bound3-d}.
Together with \eqref{eq: variance limit 1} and \eqref{eq: variance limit 2}, we establish $B_{b}\cip 0$ under the condition

\begin{equation*}
\lambda_{q}(\Psi)\geq \lambda_{q}(\Psi_{-j})\gg \sqrt{q}(\log p)^{1/4} \max\left\{\sqrt{\frac{p}{n}},n^{1/4}\right\} \quad \text{and}\quad \sqrt{s}\lambda_j\Mq \rightarrow 0.
\label{eq: dim condition d}
\end{equation*}
Note that the above condition is implied by \eqref{eq: factor condition} and $s\ll n/[\Mq^2\log p]$. 

\subsection{Proof of Lemma \ref{lem: dense confounding}}
\label{sec: proof dense confounding}
We first control the lower bound of $\lambda_{q}(\Psi)$ and the argument for $\lambda_{q}(\Psi_{-j})$ is similar. Note that $\lambda_{q}^2(\Psi)$ is the smallest eigenvalue of $\Psi \Psi^{\intercal}=\sum_{l=1}^{p}\Psi_{\cdot,l} \Psi_{\cdot, l}^{\intercal}.$ Since $\Psi_{\cdot,l}\in \R^{q}$ for $1\leq j\leq p$ are i.i.d. sub-Gaussian random vectors, it follows from (5.26) in \citep{vershynin2010introduction}, with probability larger than $1-p^{-c},$ 
\begin{equation*}
\norm{\frac{1}{p}\sum_{l=1}^{p}\Psi_{\cdot,l} \Psi_{\cdot, l}^{\intercal}-\Sigma_{\Psi}}_2\leq C \lambda_{\max}(\Sigma_{\Psi}) \sqrt{\frac{q+\log p}{p}},
\end{equation*}
for some positive constants $c, C>0$. This gives us that, with probability larger than $1-p^{-c},$
\begin{equation}
\lambda_{q}^2(\Psi)=\lambda_{\min}(\sum_{j=1}^{p}\Psi_{\cdot,l} \Psi_{\cdot,l}^{\intercal})\gtrsim p\left(\lambda_{\min}(\Sigma_{\Psi})-\lambda_{\max}(\Sigma_{\Psi}) \sqrt{\frac{q+\log p}{p}}\right).
\label{eq: inter lower 1}
\end{equation}
Similarly, we establish that, with probability larger than $1-p^{-c},$
\begin{equation}
\lambda_{q}^2(\Psi_{-j})=\lambda_{\min}(\sum_{l\neq j}^{p}\Psi_{\cdot,l} \Psi_{\cdot,l}^{\intercal})\gtrsim (p-1)\left(\lambda_{\min}(\Sigma_{\Psi})-\lambda_{\max}(\Sigma_{\Psi}) \sqrt{\frac{q+\log p}{p}}\right).
\label{eq: inter lower 2}
\end{equation}
In the following, we control $\Psi a$ for $a\in \R^{p}$ by noting that $\E\|\Psi a\|_2^2={\rm Tr}(\Sigma_{\Psi})\|a\|_2^2.$ Hence, with probability larger than $1-\frac{1}{t^2}$, we have 
\begin{equation}
\|\Psi a\|_2^2 \leq t^2{\rm Tr}(\Sigma_{\Psi})\|a\|_2^2\leq t^2 q \lambda_{\max}(\Sigma_{\Psi})\|a\|_2^2.
\label{eq: one-dimension control}
\end{equation}
By taking $a\in \R^{p}$ as $((\Omega_{E})_{1,j},\ldots,(\Omega_{E})_{j-1,j},0,(\Omega_{E})_{j+1,j},\ldots,(\Omega_{E})_{p,j})$, $e_j$ and $(\Omega_{E})_{j,\cdot}$, we establish that with probability larger than $1-\frac{1}{t^2},$
\begin{equation}
\|\Psi_{-j}(\Omega_{E})_{-j,j}\|_2\lesssim t\sqrt{q}\sqrt{\lambda_{\max}(\Sigma_{\Psi})}\|(\Omega_{E})_{-j,j}\|_2
\label{eq: inter upper 1}
\end{equation}
\begin{equation}
\|\Psi_{j}\|_2\lesssim t\sqrt{q}\sqrt{\lambda_{\max}(\Sigma_{\Psi})}
\label{eq: inter upper 2}
\end{equation}
\begin{equation}
\|\Psi(\Omega_{E})_{\cdot,j}\|_2\lesssim t\sqrt{q}\sqrt{\lambda_{\max}(\Sigma_{\Psi})}\|(\Omega_{E})_{\cdot,j}\|_2
\label{eq: inter upper 3}
\end{equation}
The lemma follows from a combination of \eqref{eq: inter lower 1}, \eqref{eq: inter lower 2}, \eqref{eq: inter upper 1}, \eqref{eq: inter upper 2} and \eqref{eq: inter upper 3}.
\subsection{Proof of Lemma \ref{lem: dense confounding general}}
\label{sec: general dense confounding}
The proof is a generalization of that of Lemma \ref{lem: dense confounding} in Section \ref{sec: proof dense confounding}. Note that $\lambda_{q}^2(\Psi)$ is the smallest eigenvalue of $\Psi \Psi^{\intercal}=\sum_{l=1}^{p}\Psi_{\cdot,l} \Psi_{\cdot,l}^{\intercal}$ and $\sum_{l=1}^{p}\Psi_{\cdot,l} \Psi_{\cdot,l}^{\intercal}-\sum_{l\in A}\Psi_{\cdot,l} \Psi_{\cdot,l}^{\intercal}$ is a positive definite matrix.  By the same argument for \eqref{eq: inter lower 1}, we have 
\begin{equation}
\begin{aligned}
\lambda_{q}^2(\Psi)\geq\lambda_{\min}(\sum_{l\in A}\Psi_{\cdot,l} \Psi_{\cdot,l}^{\intercal})&\gtrsim |A|\left(\lambda_{\min}(\Sigma_{\Psi})-\lambda_{\max}(\Sigma_{\Psi}) \sqrt{{q}/{|A|}}\right).
\end{aligned}
\label{eq: general lower 1}
\end{equation}
Similarly, we have 
\begin{equation}
\lambda_{q}^2(\Psi_{-j})\gtrsim |A|\left(\lambda_{\min}(\Sigma_{\Psi})-\lambda_{\max}\Sigma_{\Psi}) \sqrt{{q}/{p}}\right).
\label{eq: general lower 2}
\end{equation}
We establish \eqref{eq: factor condition} by the condition \eqref{eq: confounding number condition} on the set cardinality $|A|$.

Similarly to \eqref{eq: one-dimension control}, we establish that, with probability larger than $1-\frac{1}{t^2},$
\begin{equation*}
\|\Psi a\|_2^2 \lesssim t^2 q \max\{\lambda_{\max}(\Sigma_{\Psi}),C_1\}\|a\|_2^2.
\end{equation*}
Then we can establish \eqref{eq: inter upper 1}, \eqref{eq: inter upper 2} and \eqref{eq: inter upper 3} by replacing $\sqrt{\lambda_{\max}(\Sigma_{\Psi})}$ with $\sqrt{\max\{\lambda_{\max}(\Sigma_{\Psi}),C_1\}}.$ Combined with \eqref{eq: general lower 1} and \eqref{eq: general lower 2}, we establish the lemma.

\section{Additional Simulations}
\label{sec: additional simulations}
We present here some additional simulations to the ones presented in the Section \ref{sec: sim}. We use the same simulation setup where we further vary certain aspects of the data generating distribution or we vary the tuning parameters of the proposed Doubly Debiased Lasso method. 

\begin{figure}[htp]
    \hspace{-0.65cm}
    \includegraphics[width=1.04\linewidth]{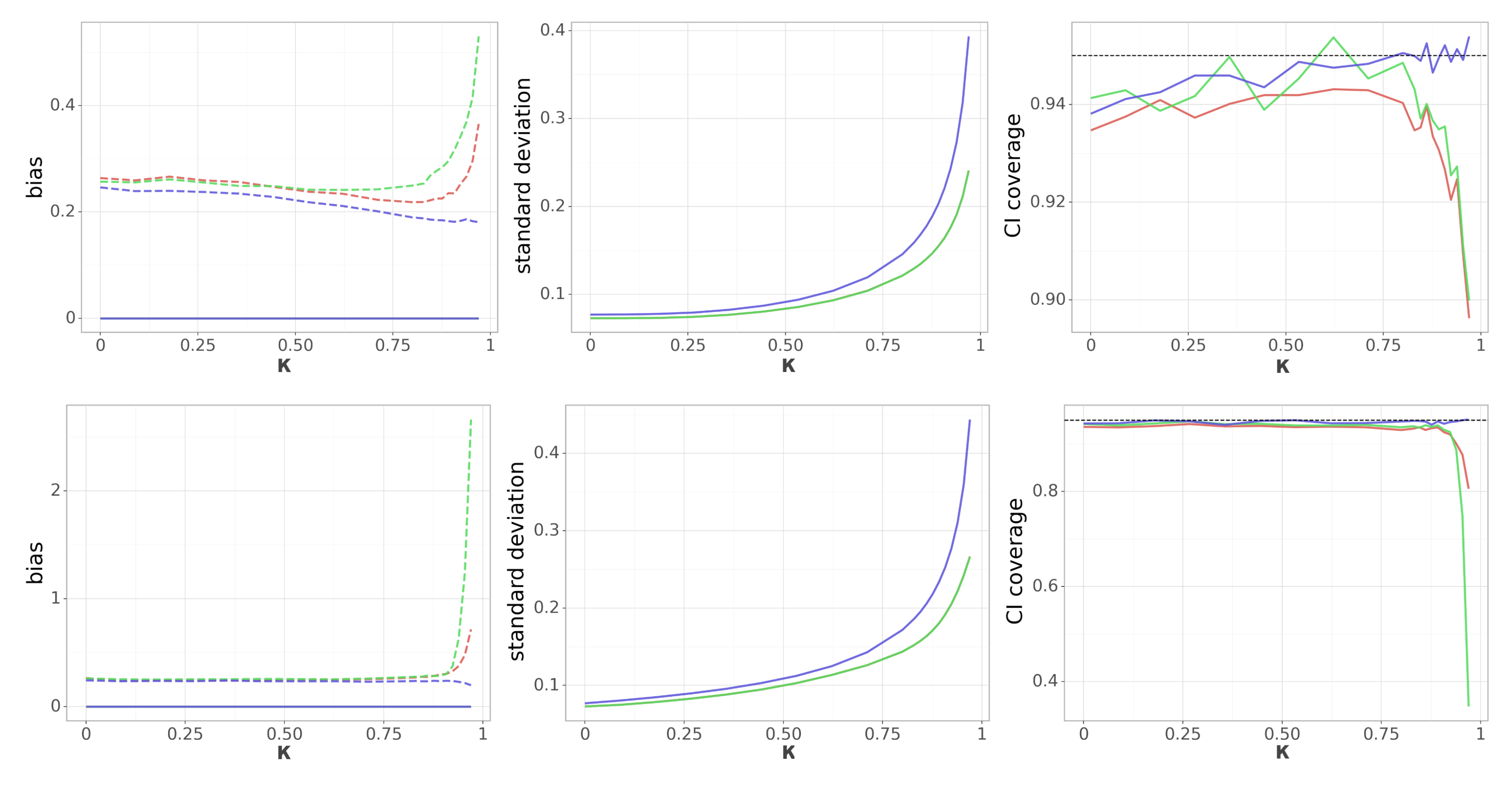}
    \caption{\textit{(No confounding - Toeplitz and Equicorrelation covariance)} Dependence of the (scaled) absolute bias terms $|B_\beta|$ and $|B_b|$ (left), standard deviation $V^{1/2}$ (middle) and the coverage of the $95\%$ confidence interval (right) on the correlation parameter $\kappa$, while keeping $p=1,000, n=300, q=0$ fixed. In the plots on the left, $|B_\beta|$ and $|B_b|$ are denoted by a dashed and a solid line, respectively, but $B_b=0$ since we zero confounders $q=0$. Top row corresponds to the Toeplitz covariance structure $(\Sigma_E)_{i,j} = \kappa^{|i-j|}$, whereas for the bottom row we have equicorrelation covariance matrix where the off-diagonal elements equal $\kappa$. Blue color corresponds to the Doubly Debiased Lasso, red color represents the standard Debiased Lasso and green color corresponds also to the Debiased Lasso estimator, but with the same $\betainit$ as our proposed method. Note that the last two methods have almost indistinguishable $V$.}
    \label{fig: no_bias_toeplitz_equicorrelation}
\end{figure}

\paragraph*{No confounding - Toeplitz and Equicorrelation covariance}
Here we explore further the scenarios where there is no confounding at all, i.e. $q = 0$, similarly as in the bottom part of Figure \ref{fig: no_bias}, but with different covariance structure of $X=E$. We fix $n=300, p=1,000,$ and take the covariance matrix $\Sigma_E$ to be either a Toeplitz matrix, with $(\Sigma_E)_{i,j} = \kappa^{|i-j|}$ for $\kappa \in [0, 1)$, or we take it to be equicorrelation matrix where $(\Sigma_E)_{i,j} = \kappa \in [0,1)$ when $i\neq j$ and $1$ otherwise. In both cases, as the correlation parameter $\kappa$ approaches $1$, the singular values become more spiked and the predictors become more correlated. The results can be seen in Figure \ref{fig: no_bias_toeplitz_equicorrelation}. We see that Doubly Debiased Lasso seems to have much smaller bias $|B_\beta|$ and thus better coverage even in the case when $q=0$, because Trimming large singular values reduces the correlations between the predictors. This difference in bias and the coverage is even more clearly pronounced for the equicorrelation covariance structure, since for the Toeplitz covariance structure $\text{Cor}(X_i, X_j)$ decays as $|i-j|$ gets bigger, whereas for equicorrelation case it is constant and equal to $\kappa$.

\begin{figure}[htp]
    \hspace{-0.65cm}
    \includegraphics[width=1.04\linewidth]{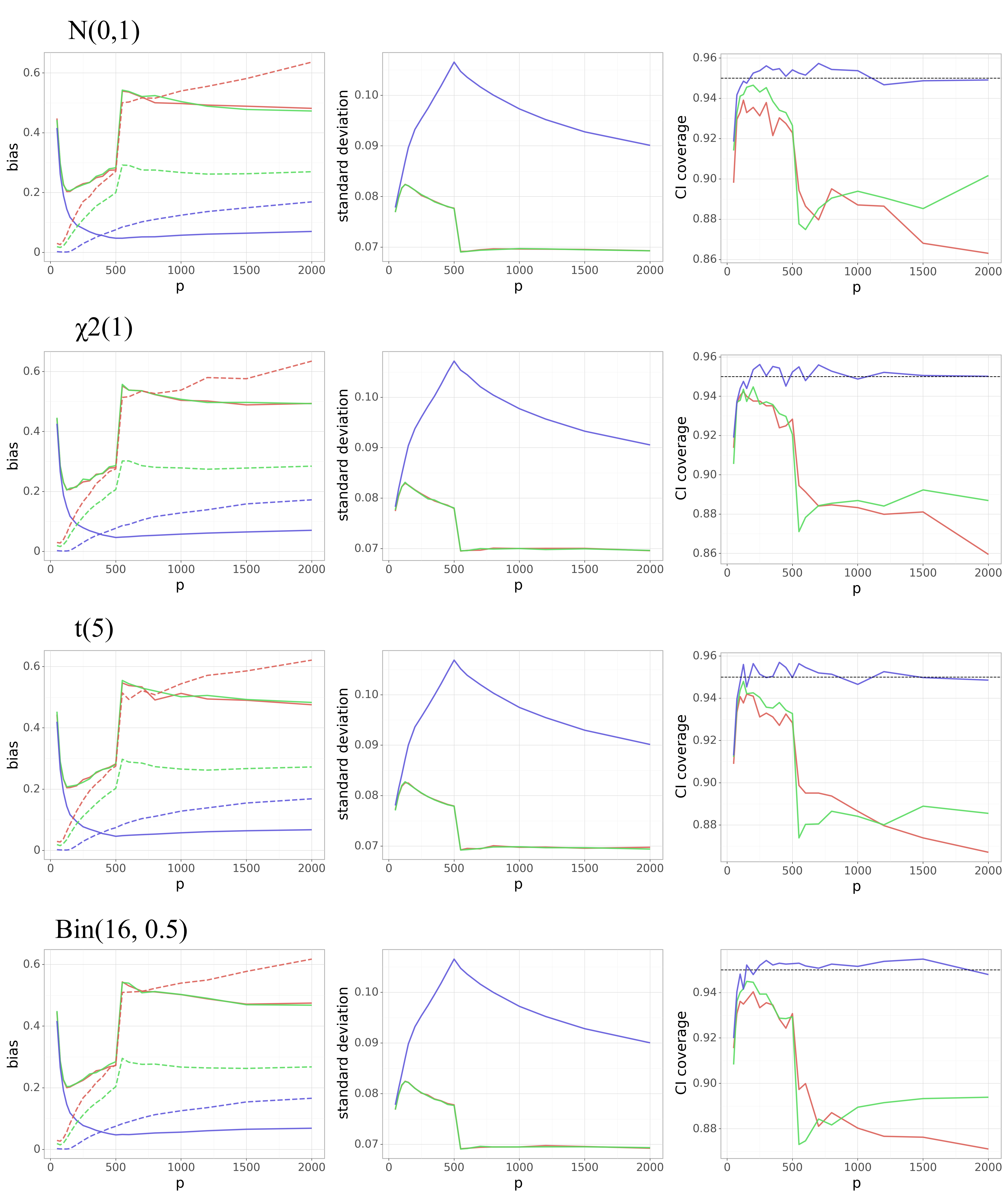}
    \caption{\textit{(Non-Gaussian distribution)} Dependence of the (scaled) absolute bias terms $|B_\beta|$ and $|B_b|$ (left), standard deviation $V^{1/2}$ (middle) and the coverage of the $95\%$ confidence interval (right) on the number of predictors $p$, while keeping $n=500, q=3$ fixed.  On the left side, $|B_\beta|$ and $|B_b|$ are denoted by a dashed and a solid line, respectively. We change the distribution of $H, E, e$ in \eqref{eq: hidden SEM} as described in the text. Each row in the plot corresponds to a different distribution $\P$. We set $\Sigma_E$ to have Toeplitz structure with parameter $\kappa = 0.7$. Blue color corresponds to the Doubly Debiased Lasso, red color represents the standard Debiased Lasso and green color corresponds also to the Debiased Lasso estimator, but with the same $\betainit$ as our proposed method. Note that the last two methods have almost indistinguishable $|B_b|$ and $V$.}
    \label{fig: nongaussian}
\end{figure}

\paragraph*{Non-Gaussian distribution}
The Assumption (A3) in Section \ref{sec: theory} requires that the noise term $\nu_{i,j}=E_{i,j}-E_{i,-j}^{\intercal}\gamma^{E}$ is is independent of $E_{i,-j}$. This condition will automatically hold if $E_{i,\cdot}$ is multivariate Gaussian or $E_{i,\cdot}$ has independent entries. We now test the robustness of Doubly Debiased Lasso method when this assumption is violated. 
In order to examine that, we repeat the simulation setting displayed in Figure \ref{fig: n_p_growing}, where $n=500$ and $p$ varies from $1$ to $2,000$. We change the distribution as follows: Let $\P$ be some real distribution with zero mean and unit variance. The entries of the matrix of the confounders $H$ are generated i.i.d. from $\P$. Furthermore, the unconfounded part of the predictors $E$ is generated as $Z\Sigma_E^{1/2}$, where $Z$ is a $n\times p$ matrix with i.i.d. entries coming from the distribution $\P$ and $\Sigma_E$ is a Toeplitz matrix with $(\Sigma_E)_{i,j} = \kappa^{|i-j|}$ for $\kappa=0.7.$ 
Finally, the noise variables $e_i$ used for generating $Y$ (see Equation \ref{eq: confounder model}) are also generated from $\P$. The results can be seen in Figure \ref{fig: nongaussian}. We take $\P$ to be the following distributions: standardized chi-squared with $1$ degree of freedom, standardized t-distribution with $5$ degrees of freedom and standardized $\text{Bin}(16, 0.5)$. For comparisons of the performance, we also include $N(0,1)$ distribution, but one needs to keep in mind that the obtained plot differs from the one in Figure \ref{fig: n_p_growing} because of different correlation structure of $E$. We can see that there is very little change in the performance of the proposed estimator, thus showing that Doubly Debiased Lasso can be used for a wide range of models.

\paragraph*{Comparison to PCA adjustment}
Here we investigate how the choice of the spectral transformation can affect the performance of the Doubly Debiased Lasso estimator. We focus on the PCA adjustment which maps first $\hat{q}$ singular values to $0$, for some tuning parameter $\hat{q}$, while keeping the remaining singular values unchanged. This transformation is used frequently in the literature because it arises by regressing out the top $\hat{q}$ principal components from every predictor.

\begin{figure}[htp]
    \hspace{-0.65cm}
    \includegraphics[width=1.04\linewidth]{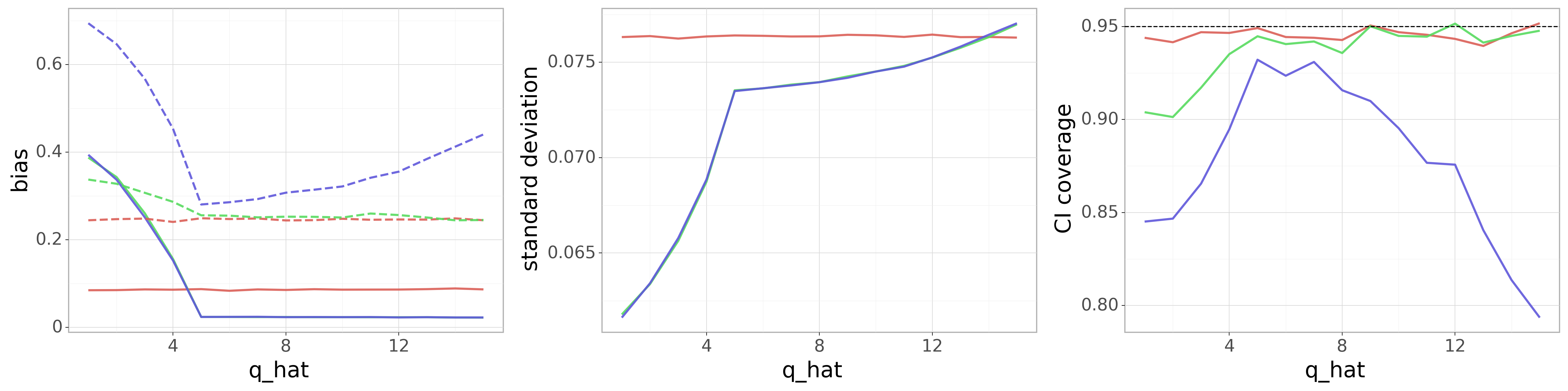}
    \caption{\textit{(Comparison to PCA adjustment)} Dependence of the (scaled) absolute bias terms $|B_\beta|$ and $|B_b|$ (left), standard deviation $V^{1/2}$ (middle) and the coverage of the $95\%$ confidence interval (right) on the correlation parameter $\kappa$, while keeping $p=1,000, n=300, q=3$ fixed. In the left plot, $|B_\beta|$ and $|B_b|$ are denoted by a dashed and a solid line, respectively. We vary the parameter $\hat{q}$ of the PCA adjustment, which maps the first $\hat{q}$ to zero. Red color corresponds to the Doubly Debiased Lasso using Trim transform for both $\NT$ and $Q$, blue color represents the Doubly Debiased Lasso using PCA adjustment for both $\NT$ and $\Q$ and green color corresponds to the Doubly Debiased Lasso estimator using the same default $\betainit$ with $\Q$ being the median Trim transform, but uses PCA adjustment for $\NT$. Note that the last two methods have almost indistinguishable $V$.}
    \label{fig: PCA adjustment}
\end{figure}

We fix $n=300, p=1,000, q=5$ and vary the parameter $\hat{q}$. We compare the estimator using the PCA adjustment for both $\NT$ and $\Q$ with the estimator using the Trim transform with the median rule for both $\NT$ and $\Q$. Finally, we also consider the estimator using the Trim transform for $\Q$ and PCA adjustment for $\NT$, in order to separate the effects of changing the spectral transformation for the initial estimator $\betainit$ and the overall estimator construction. The results can be seen in Figure \ref{fig: PCA adjustment}. 

We see that the performance is very sensitive to the choice of the tuning parameter $\hat{q}$. On one hand, if $\hat{q} < q$, we do not manage to remove enough of the confounding bias $B_b$, which has as a consequence that there is certain undercoverage of the confidence intervals. On the other hand, if $\hat{q} \leq q$, the bias $B_b$ becomes very small, but the variance of our estimator increases slowly as $\hat{q}$ grows. Also, removing too many principal components when computing $\betainit$ can remove too much signal, resulting in the higher bias $B_\beta$. Trim transform has an advantage that we do not need to estimate the number of latent confounders $q$ from the data, which might be a quite difficult task. This is done by trimming many principal components, but not removing them completely. However, this can result in a small increase of the estimator variance compared to the PCA adjustment with the optimal tuning $\hat{q} = q$.

\begin{figure}[htp]
    \hspace{-0.65cm}
    \includegraphics[width=1.04\linewidth]{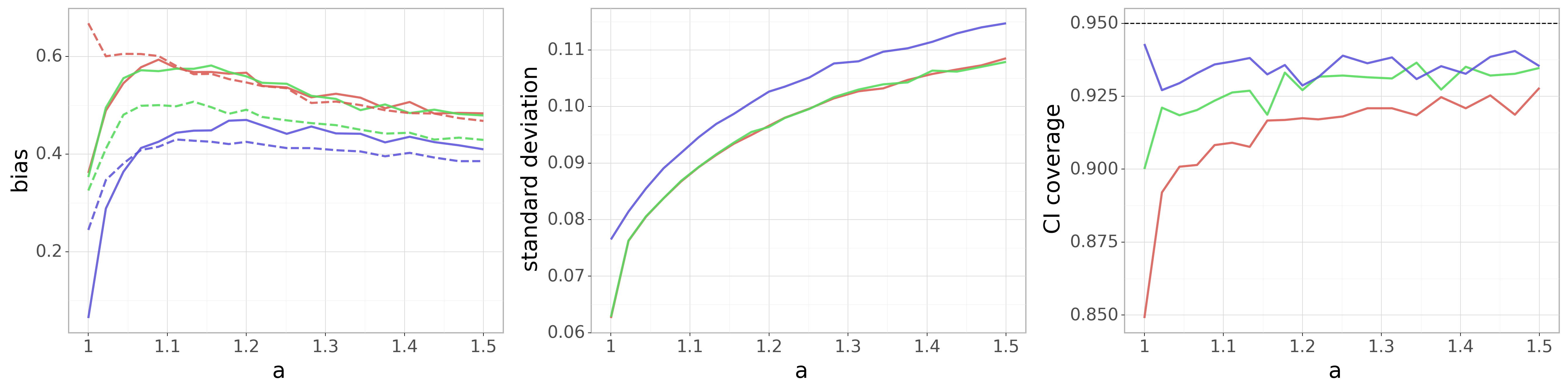}
    \caption{\textit{(Weak confounding)} Dependence of the (scaled) absolute bias terms $|B_\beta|$ and $|B_b|$ (left), standard deviation $V^{1/2}$ (middle) and the coverage of the $95\%$ confidence interval (right) on the loadings decay parameter $a$, while keeping $p=1,000, n=300, q=3$ fixed. In the left plot, $|B_\beta|$ and $|B_b|$ are denoted by a dashed and a solid line, respectively. Blue color corresponds to the Doubly Debiased Lasso, red color represents the standard Debiased Lasso and green color corresponds also to the Debiased Lasso estimator, but with the same $\betainit$ as our proposed method. Note that the last two methods have almost indistinguishable $V$.}
    \label{fig: loadings_decay}
\end{figure}

\paragraph*{Weak confounding}
Here, we explore how the performance of our estimator depends on the strength of the confounding, i.e. how $H$ affects $X$. In Figure \ref{fig: proportion}, we have already explored how the performance of our method depends on the number of affected predictors by each confounder. Here we allow all predictors to be affected, but with decaying strength. This we achieve by generating the entries of the loading matrix $\Psi$ as $\Psi_{ij} \sim N(0, 1/\sigma_i(j)^a)$, where for each of the $q$ rows we take a random permutation $\sigma_i: \{1, \ldots, p\} \to \{1, \ldots, p\}$, and $a \geq 1$ is a tuning parameter describing the decay of the loading coefficients. The values $n=300, p=1,000$ and $q=3$ are kept fixed. The results can be seen in the Figure \ref{fig: loadings_decay}. We see that when $a$ is close to $1$ and the confounding is strong that our proposed estimator is much better that the standard Debiased Lasso estimator. On the other hand, when $a$ is larger, meaning that the confounding gets much weaker, the difference in performance decreases, but Doubly Debiased Lasso still has smaller bias and thus better coverage.

\end{document}